\documentclass[preprint]{aastex}
\usepackage{url}

\shorttitle{Improving Galactic Center Astrometry}
\shortauthors{Yelda, Lu, Ghez, Clarkson, Anderson, Do, Matthews}

\begin{document}

\title{Improving Galactic Center Astrometry by Reducing
       the Effects of Geometric Distortion}
\author{
S. Yelda\altaffilmark{1}, 
J. R. Lu\altaffilmark{2}, 
A. M. Ghez\altaffilmark{1,3},
W. Clarkson\altaffilmark{4}, 
J. Anderson\altaffilmark{5},
T. Do\altaffilmark{6},
K. Matthews\altaffilmark{2}}
\altaffiltext{1}{UCLA Department of Physics and Astronomy, Los Angeles, CA 90095
-1547; syelda, ghez@astro.ucla.edu}
\altaffiltext{2}{Astrophysics, California Institute of Technology, MC 249-17, 
Pasadena, CA 91125; jlu, kym@caltech.edu}
\altaffiltext{3}{UCLA Institute of Geophysics and Planetary Physics,
Los Angeles, CA 90095-1565}
\altaffiltext{4}{Indiana University, Department of Astronomy, Bloomington, IN 47405;
clarkson@astro.ucla.edu}
\altaffiltext{5}{Space Telescope Science Institute, 3700 San Martin Drive, 
Baltimore, MD 21218; jayander@stsci.edu}
\altaffiltext{6}{UC Irvine Department of Physics and Astronomy, Irvine, CA 92697;
tdo@astro.ucla.edu}

\begin{abstract}
We present significantly improved proper motion measurements of the Milky Way's 
central stellar cluster. These improvements are made possible by refining our 
astrometric reference frame with a new geometric optical distortion model
for the W. M. Keck II 10 m telescope's Adaptive Optics camera 
(NIRC2) in its narrow field mode.  For the first time, this distortion model
is constructed from on-sky measurements, and is made available to the public 
in the form of FITS files.
When applied to widely dithered images, it produces residuals in the separations 
of stars that are a factor of $\sim$3 smaller compared to the outcome using 
previous models.  By applying this new model, along with corrections for 
differential atmospheric refraction, 
to widely dithered images of SiO masers at the Galactic center, we improve our 
ability to tie into the precisely measured radio Sgr A*-rest frame. The resulting
infrared reference frame is $\sim$2-3 times more accurate and stable 
than earlier published efforts.  In this reference frame, Sgr A* is localized to 
within a position of 0.6 mas and a velocity of 0.09 mas yr$^{-1}$, 
or $\sim$3.4 km s$^{-1}$ at 8 kpc (1$\sigma$).
Also, proper motions for members of the central stellar cluster are more accurate, 
although less precise, due to the limited number of these wide field measurements.   
These proper motion measurements show that, with respect to Sgr A*, the central 
stellar cluster has no rotation in the plane of the sky to within
0.3 mas yr$^{-1}$ arcsec$^{-1}$, has no net translational motion with respect 
to Sgr A* to within 0.1 mas yr$^{-1}$, and has net rotation perpendicular to the 
plane of the sky along the Galactic plane, as has previously been observed. 
While earlier proper motion studies defined a reference frame by assuming no net 
motion of the stellar cluster, this approach is fundamentally limited by the 
cluster's intrinsic dispersion and therefore will not improve with time.
We define a reference frame with SiO masers and this reference
frame's stability should improve steadily 
with future measurements of the SiO masers in this region ($\propto t^{3/2}$).   
This is essential for achieving the necessary reference frame stability 
required to detect the effects of general relativity and extended mass on 
short-period stars at the Galactic center.
\end{abstract}

\section{Introduction}
High angular resolution astrometry has been a powerful technique for 
studies of the Galactic center (GC).  Over the last decade, it has revealed a 
supermassive black hole \citep{eckart97, ghez98}, a disk of young stars 
surrounding the central supermassive black hole
\citep{levin03,genzel03,paumard06,lu09}, and allowed for measurements of the 
orbit about the GC of the Arches, a massive young star cluster located at a projected 
galacto-centric distance of 30 pc \citep[][Clarkson et al. in prep.]{stolte08}.  
While the speckle imaging work carried out on the Galactic center in the 1990's
had typical centroiding uncertainties of $\sim$1 mas, 
recent deep, adaptive optics (AO) images have improved the precision of stellar 
centroiding by a factor of $\sim$6-7, significantly increasing 
the scientific potential of astrometry at the Galactic center 
\citep{ghez08, gillessen09}. Further gains in astrometric precision 
would allow ultra-precise measurements of the distance to the Galactic 
center (R$_o$), measurements of individual stellar orbits at larger 
galacto-centric radii, and, more ambitiously, measurements of post-Newtonian 
effects in the orbits of short-period stars
\citep[e.g.,][]{jarosz98,jarosz99,salim99,fragile00,rubilar01,weinberg05,
zucker07, kraniotis07,nucita07,will08}. Such gains will also probe the
possibility that the supermassive black hole is moving with
respect to the central stellar cluster, due either to the gravitational 
influence of a massive companion or from a systematic effect
produced by improper alignment of images.  

Two factors that currently limit astrometric measurements 
of stars at the Galactic center are (1) the level to which AO cameras'
geometric distortions are known and (2) differential atmospheric 
refraction (DAR), which has not yet been explicitly corrected for in any
Galactic center proper motion study \citep{ghez08, gillessen09}.  While 
optical distortion from an infrared camera is expected to be static, 
distortion from the AO system and the atmosphere not corrected
by AO, is not.  Initial estimates of the optical distortions for AO cameras
are generally based on either the optical design or laboratory test, which 
do not perfectly match the actual optical distortion of the system. Both 
uncorrected camera distortions and DAR leave $\sim$1-5 mas scale distortions 
over the spatial scales of the SiO masers that are used to
define the Sgr A*-radio rest frame for proper motions 
of stars at the Galactic center \citep[see e.g.,][]{reid07}.
These are significantly larger than the $\sim$0.2-0.3 mas precision achieved
in the relative astrometry of \citet{ghez08} and \citet{gillessen09}.
The impact of all these effects on relative astrometry
has been minimized by mapping the coordinate systems of different epochs
of observations to a reference frame using high order transformations, 
allowing $\sim$0.2-0.3 mas precision in the relative astrometry to be achieved.
However, the full impact of these effects is imposed on astrometric measurements 
in a reference frame that is known to be at rest with respect to Sgr A*-radio 
(henceforth, the Sgr A*-radio rest frame).
Therefore, correcting these effects would have the greatest improvement on 
astrometric measurements in the maser frame.  Relative astrometry would 
also be improved by eliminating these effects before the images, which are obtained 
at different times and occasionally different orientations, are combined.

In this paper, we improve the astrometric accuracy and precision of Keck AO 
measurements of the Galactic center by (1) deriving 
a new, publicly-available distortion solution for the infrared imaging camera 
behind the Keck AO system (NIRC2) and (2) correcting for DAR.
Furthermore, having corrected for these effects, we show that an astrometric 
reference frame for the Galactic center can now be established with Sgr A* at rest to
within 0.09 mas yr$^{-1}$ (3.4 km s$^{-1}$ at the distance to the GC), thereby improving
the stability of the reference frame.
Section \ref{sec:data} presents observations and analysis of the globular 
cluster, M92, that 
were used to derive the first distortion solution for NIRC2 that is based upon 
on-sky measurements, as opposed to NIRC2's internal pinhole mask. We also discuss
the observations and analysis of Galactic center data used to illustrate
the impact of our technical work here. We present 
the results and tests of the distortion solution in \S\ref{sec:newDist}.
In \S\ref{sec:gcRefFrame} we apply this solution, along with corrections for DAR, to 
observations of the GC and report the positions and proper motions of
a set of infrared astrometric standards (N$\sim$10$^3$) in a Sgr A*-radio rest frame.
In \S\ref{sec:secondary}, we measure the motion of the stellar cluster in this 
reference frame and show that these stars exhibit significant net motion in 
the plane of the Galaxy.
Finally, we consider the implications of this work for 
measuring relativistic and extended mass effects on short period stars.
While this work has been carried out in the context of the 
Galactic center, the new distortion solution also benefits a wide array of 
other science that is currently being carried out with NIRC2, including
astrometric studies of extrasolar planets \citep{marois08}, brown dwarf binaries 
\citep{quinn07,liu08,dupuy08}, compact objects \citep{cameron07},
and external galaxies \citep[e.g.,][]{max05}.

\section{Observations \& Analysis}
\label{sec:data}
In this section, we report all the new observations and analysis carried out
for this paper. Section \ref{sec:m92} discusses the observations of M92 that are 
used to create a new NIRC2 distortion model presented in \S\ref{sec:newDist}.
Section \ref{sec:gcKeck} describes the observations of the Galactic center that
are used to test the new distortion model in \S\ref{sec:testNew} and to generate
a new IR astrometric reference frame at the Galactic center in \S\ref{sec:gcRefFrame}.

\subsection{M92}
\label{sec:m92}
To characterize the optical distortion in the NIRC2 camera, it is
ideal to compare the measured set of stellar positions to those 
in a distortion-free reference frame.  As this idealized 
reference frame does not exist, we choose observations of M92 
(NGC 6341; $\alpha$ = 17 17 07.27, $\delta$ = +43 08 11.5) made with 
the well-characterized Advanced Camera for Surveys Wide Field Channel 
(ACS/WFC) on the Hubble Space Telescope (HST), which has a 
plate scale $\sim$49.9933 $\pm$ 0.0005 mas pix$^{-1}$ and position angle 
offset = -0.0006$^o$ $\pm$ 0.0023$^o$ \citep{vanderMarel},
as our reference frame. The static distortion in 
this camera has been corrected down to the $\sim$0.01 pix 
($\sim$0.5 mas) level \citep{anderson05,anderson06,anderson07} and 
is therefore a useful reference for our purposes given the level of 
distortion in the NIRC2 camera.  While several clusters
were considered during the planning phase of this project, M92 was chosen
because it had been extensively observed with ACS/WFC, was observable in the 
northern hemisphere during the summer, was sufficiently crowded, and had an 
isolated natural guide star (NGS) available.
The HST observations of M92 used for this analysis were made on 
2006 April 11 with both the 
F814W ({\em I}) and F606W ({\em V}) filters as part of the ACS Survey
of Globular Clusters (GO-10775, PI: A. Sarajedini).  The details of the
observations and data reduction can be found in \citet{anderson08}, 
and the catalog of positions were provided in advance of publication 
by J. Anderson.  We used the \citet{anderson07} correction for the
linear skew in ACS to ensure that our reference frame was free of skew.

Observations of M92 were made from 2007 June to 2009 May using the AO
system on the W. M. Keck II 10 m telescope with the facility 
near-infrared camera NIRC2 (PI: K. Matthews). Aside from the 2007 July
data set, these observations were obtained upon completion of our
primary science program for the night (Galactic center astrometry),
or when conditions were not optimal for the primary science program
(e.g., clouds were present or seeing was relatively poor). All images were taken 
with the narrow field camera, which maps the 1024$\times$1024 pix 
array into $\sim$10$\arcsec\times$10$\arcsec$ field of view, and
through the K' ($\lambda_0$=2.12 $\mu$m, $\Delta\lambda$=0.35 $\mu$m) 
band-pass filter.  While the Natural Guide Star adaptive optics (NGSAO)
system was used to obtain the majority of the data, the Laser Guide Star
(LGS) AO system was used for one run in 2008 June.  The NGSAO atmospheric 
corrections and the LGSAO low-order, tip-tilt corrections were made 
using visible observations of USNO-B1.0 1331-0325486 (R = 8.5 mag). 
The resulting image point spread functions (PSFs) had Strehl ratios
of $\sim$0.55 and FWHM of $\sim$50 mas, on average.

\begin{figure}
\begin{center}
\begin{minipage}[b]{1.00\linewidth}
\centering
\fbox{\includegraphics[scale=0.2]{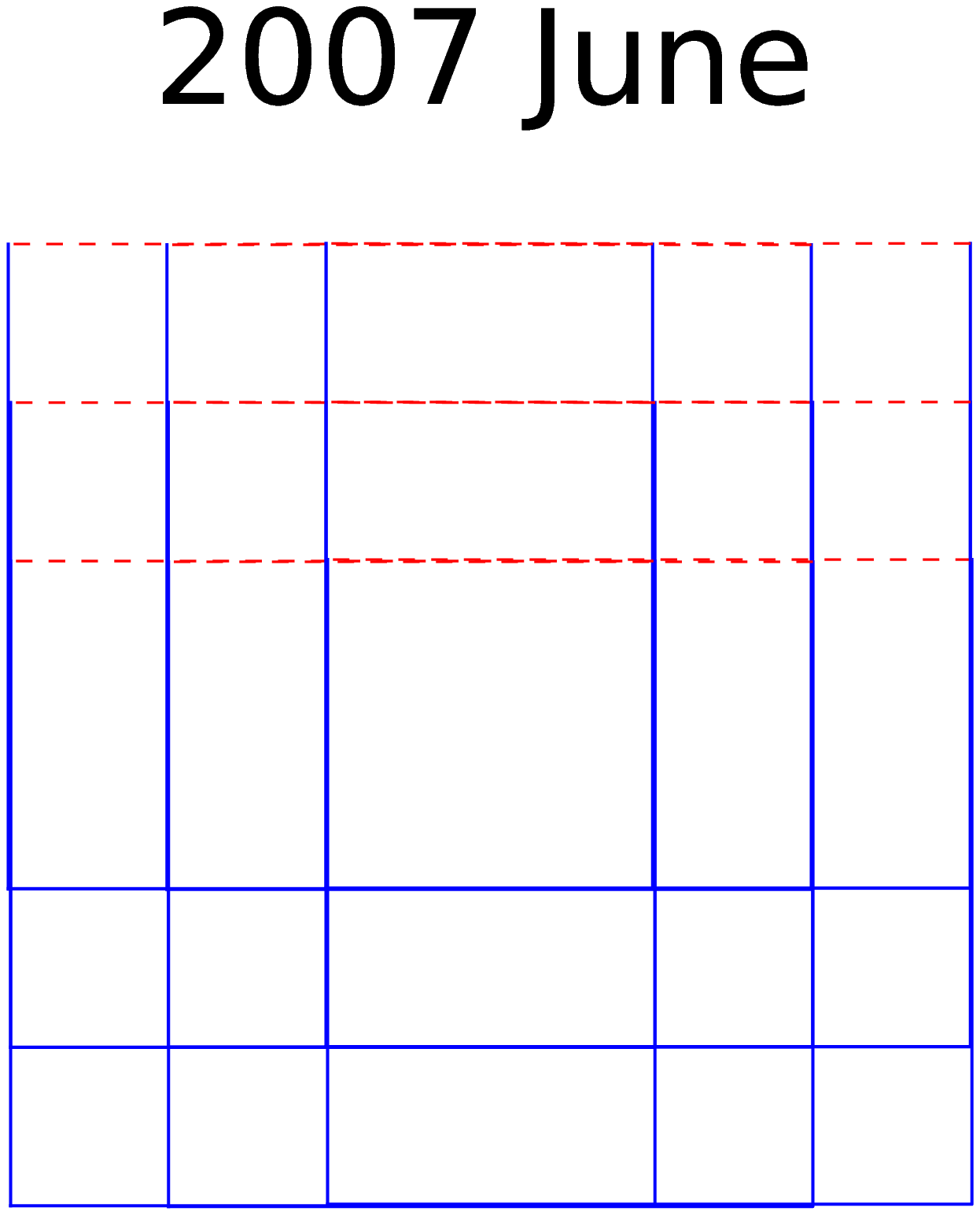}}
\fbox{\includegraphics[scale=0.2]{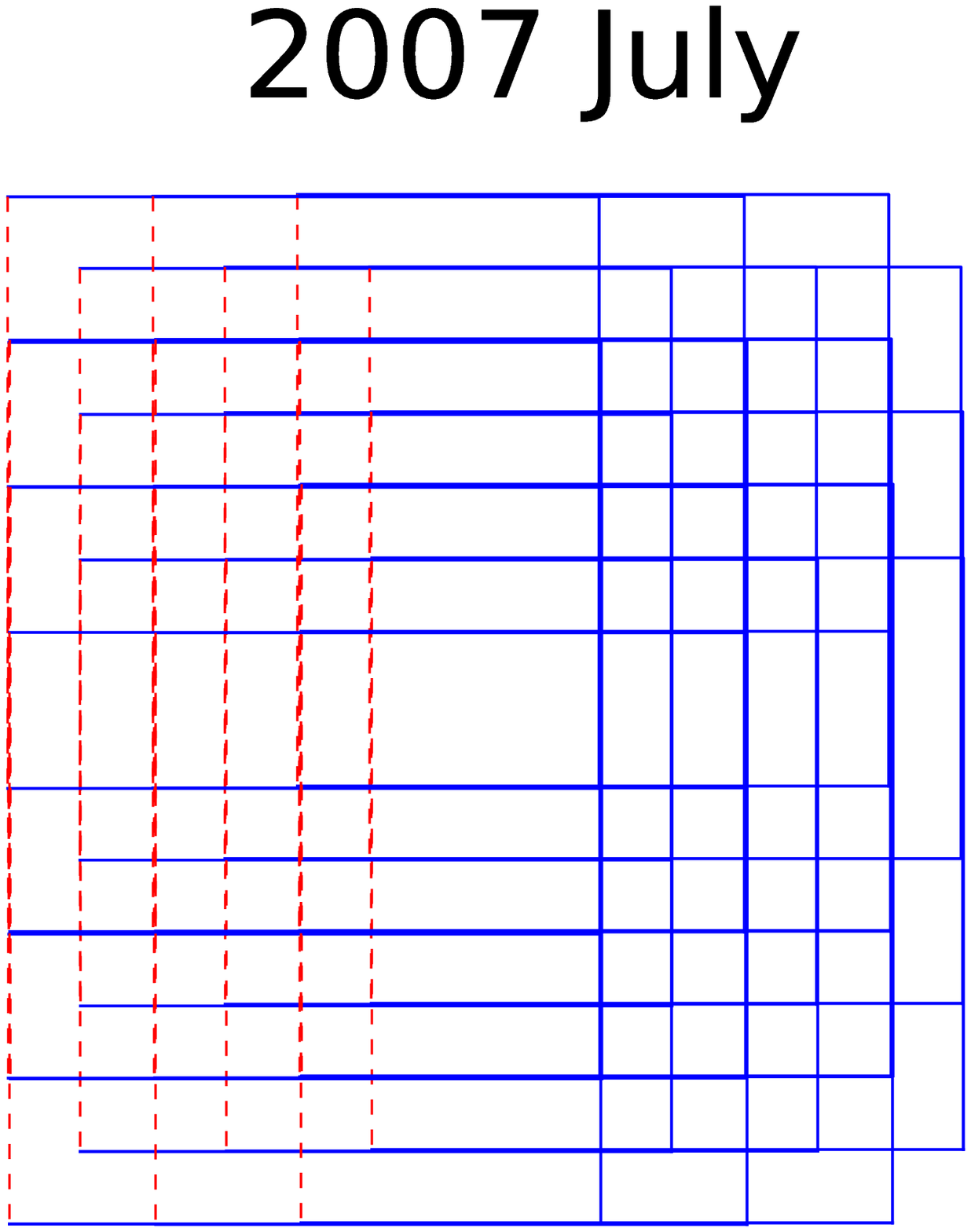}}
\fbox{\includegraphics[scale=0.2]{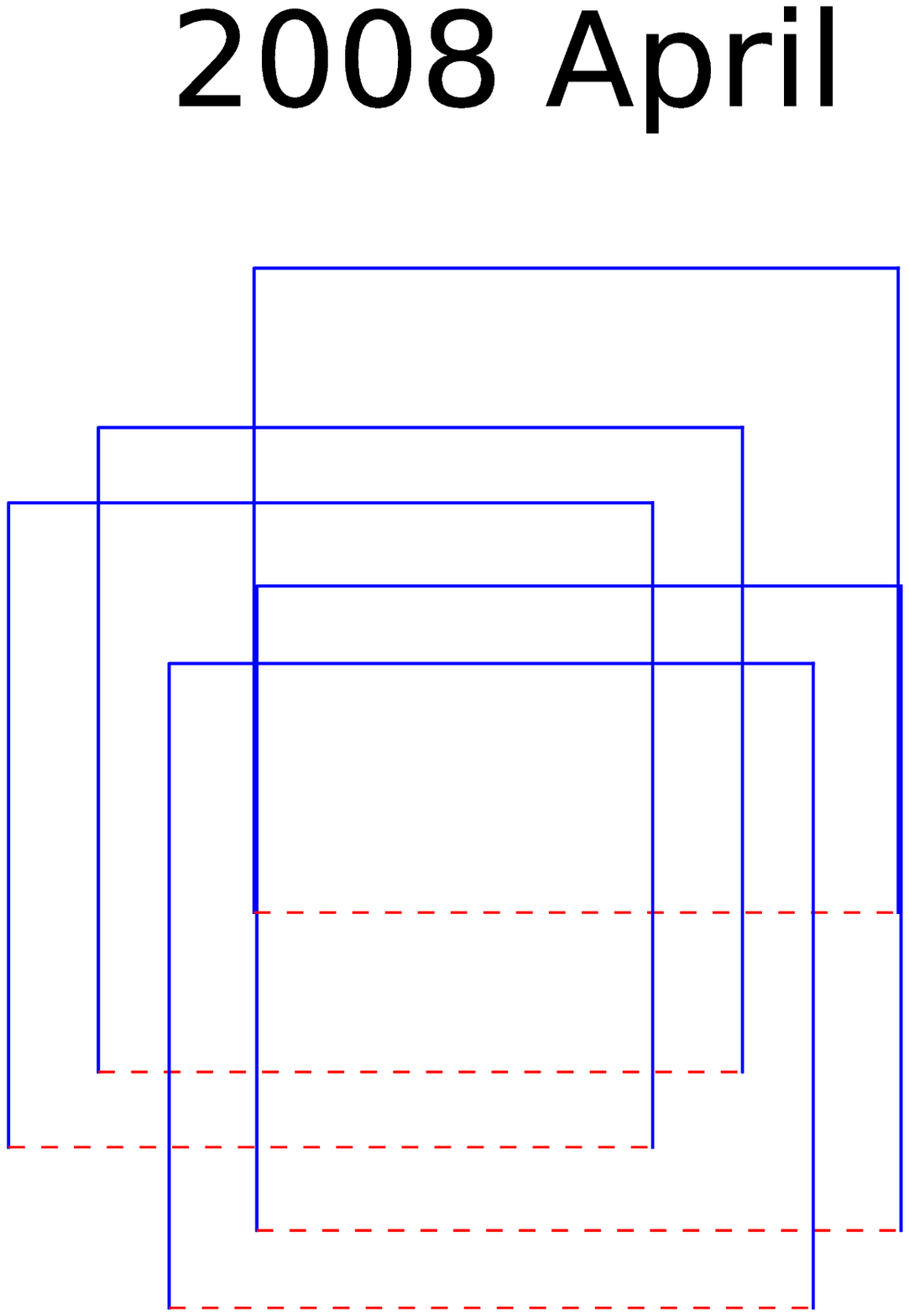}}
\end{minipage}\\
\vspace{4pt}
\begin{minipage}{1.00\linewidth}
\centering
\includegraphics[scale=0.65]{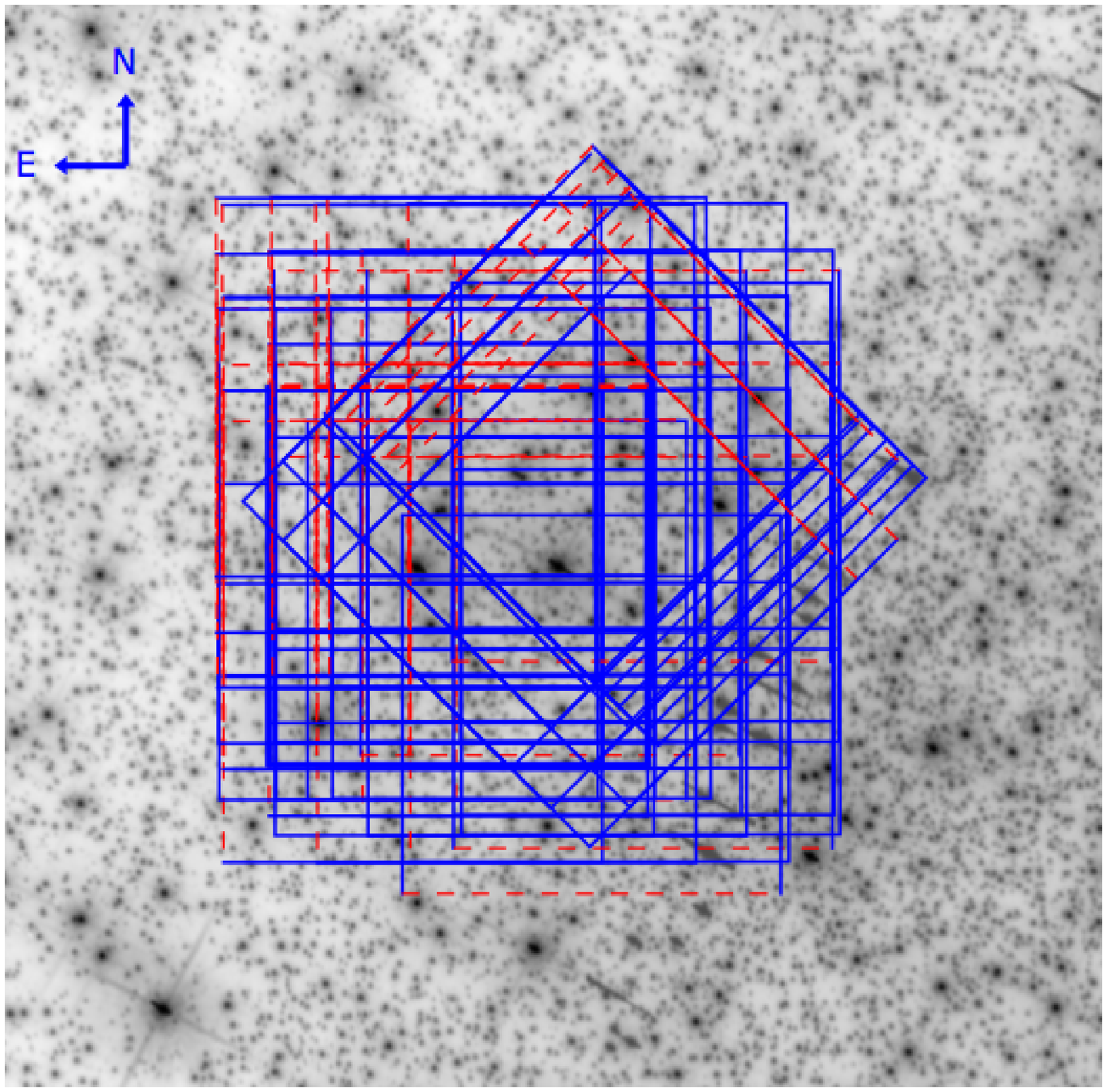}
\end{minipage}\\
\vspace{4pt}
\begin{minipage}[b]{1.0\linewidth}
\centering
\fbox{\includegraphics[scale=0.2]{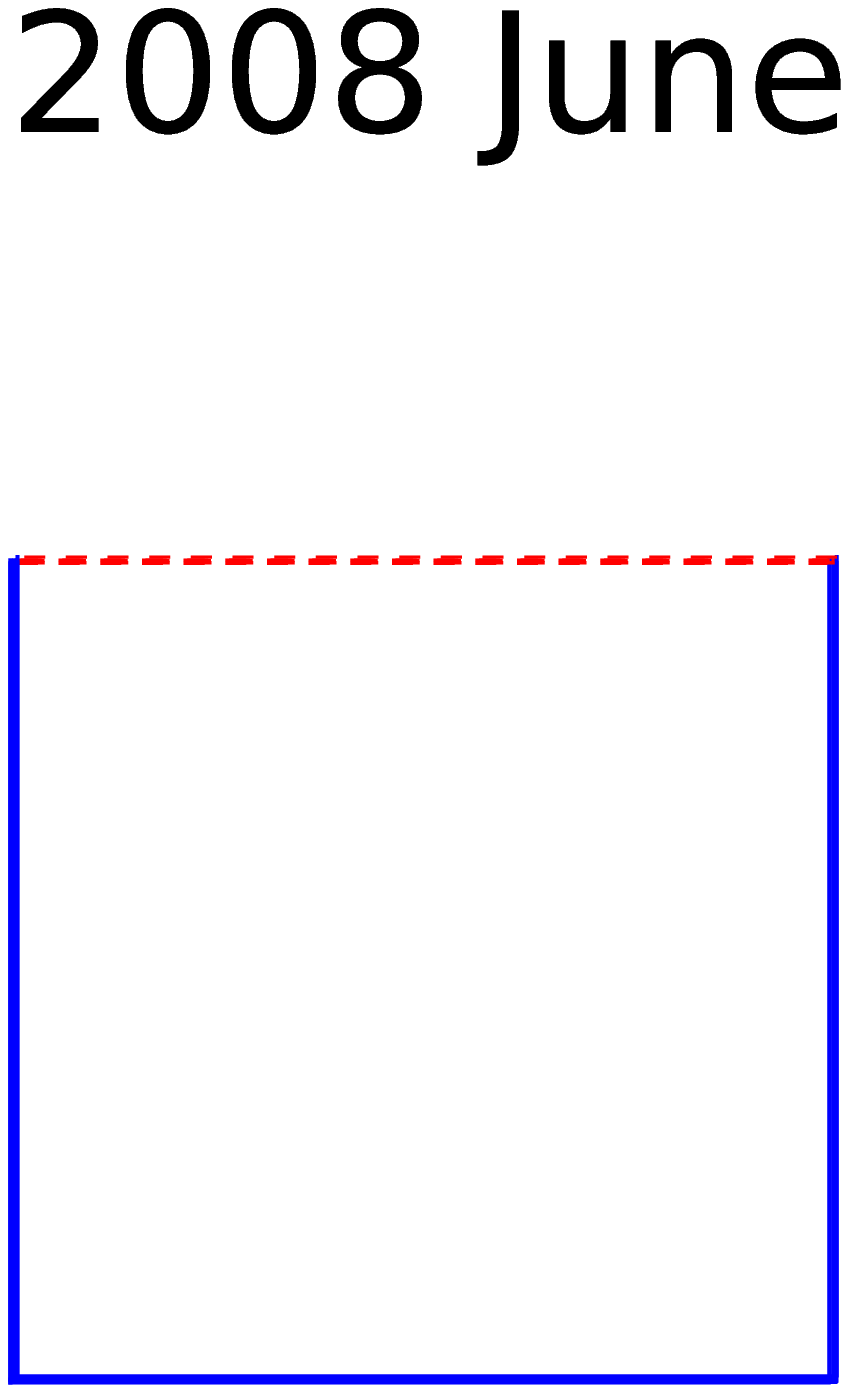}}
\fbox{\includegraphics[scale=0.2]{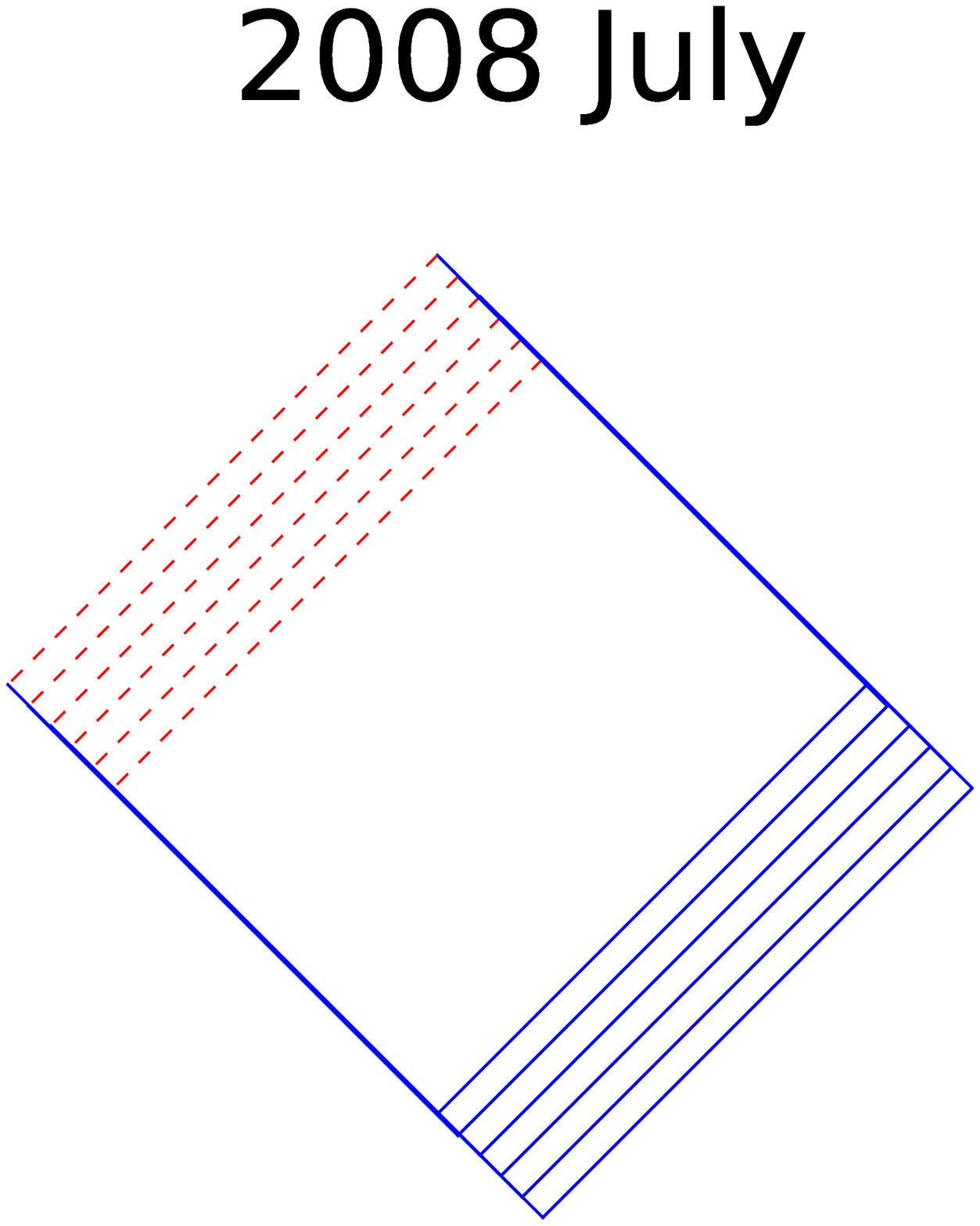}}
\fbox{\includegraphics[scale=0.2]{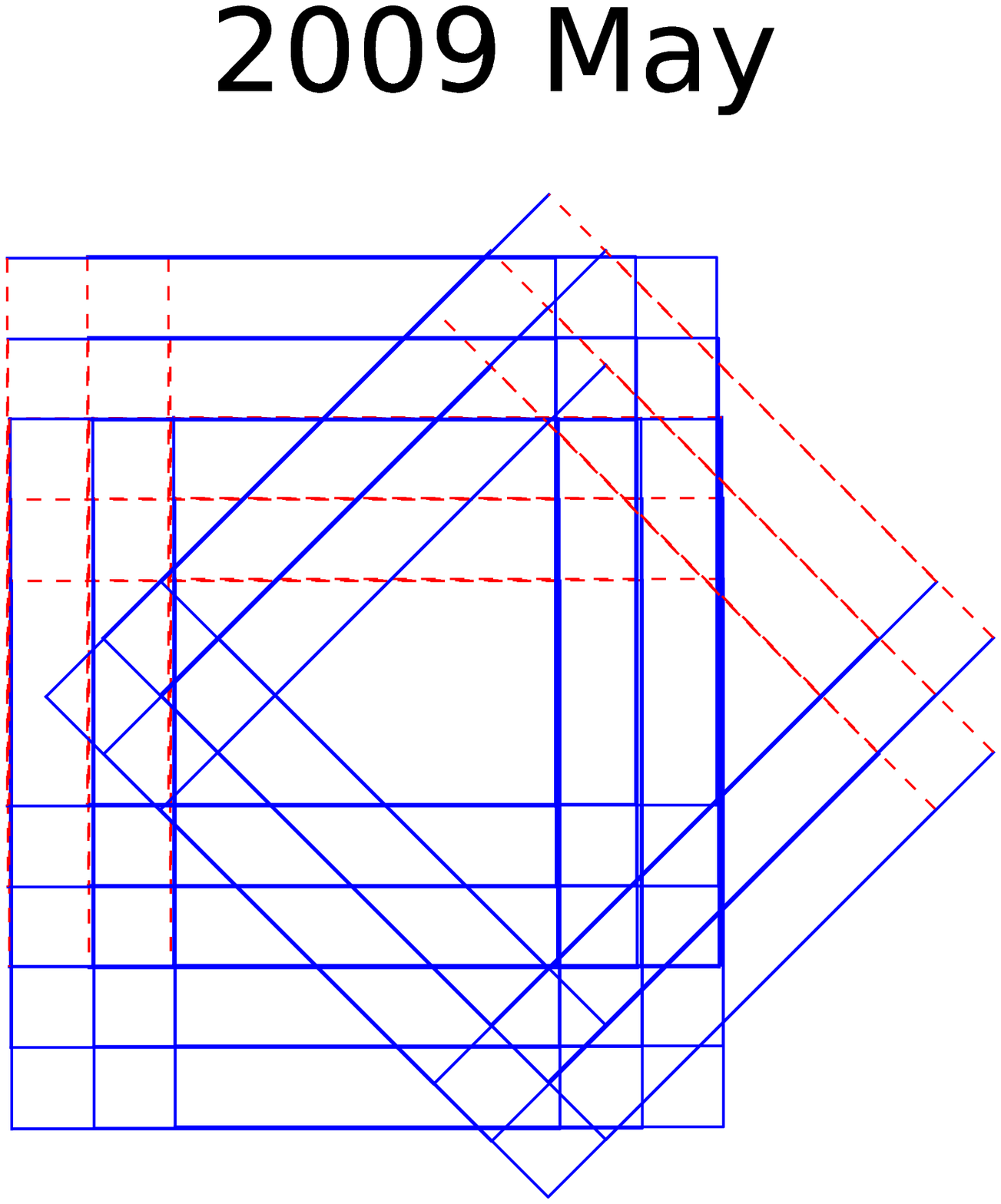}}
\end{minipage}
\caption{ACS/WFC image with the 79 NIRC2 pointings.  The red dashed side
of each NIRC2 box denotes the top of the detector's field of view.
Each NIRC2 field is 10$\arcsec\times$10$\arcsec$, while the ACS image shown is
$\sim$30$\arcsec\times$30$\arcsec$. The pattern for the individual epochs' exposures are 
shown in the insets.
}
\label{fig:m92_acs_obs}
\end{center}
\end{figure}

\begin{figure}
\epsscale{0.8}
\plotone{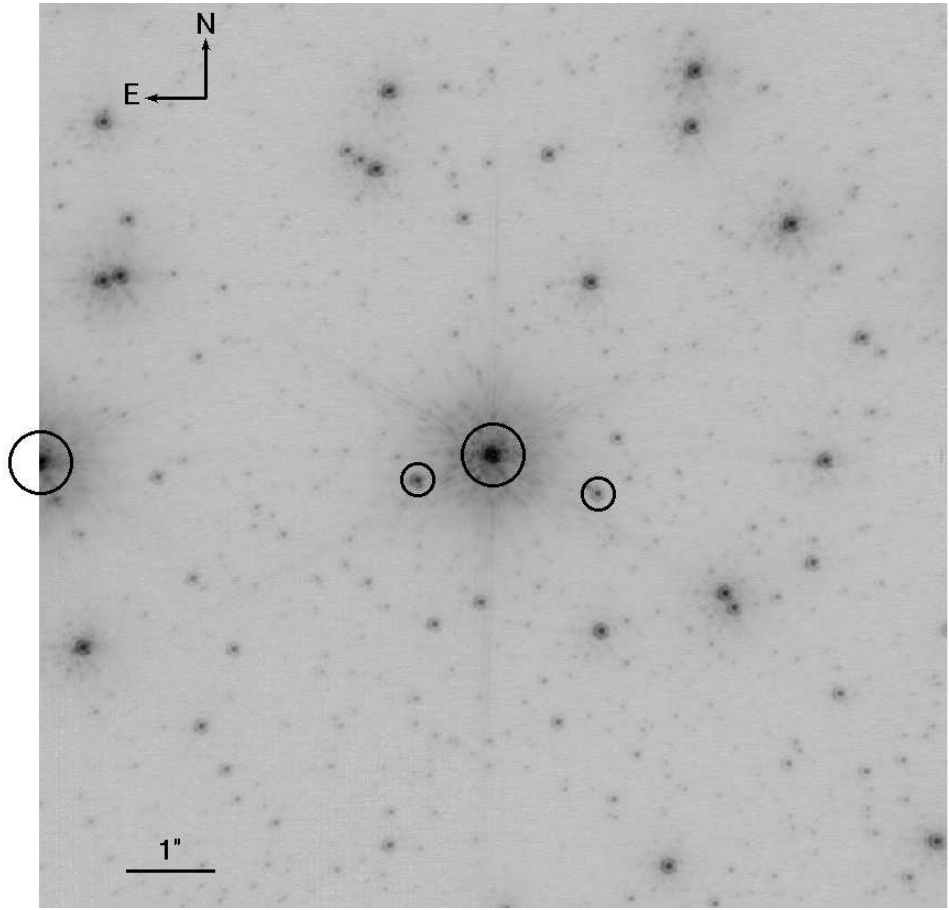}
\caption{
Diffraction-limited NGSAO NIRC2 image of one of the M92 fields used to 
characterize the optical distortion in the NIRC2 camera. 
The circled stars at the center of the image, the NGS and two fainter stars, 
are present in most of the M92 NIRC2 observations and are used to register
the images, each of which had a different position/orientation
on the sky.  The NGS and the circled star $\sim$5$\arcsec$ to its east were almost 
always (i.e., when clouds weren't present) detected at levels that saturated 
the detector and were therefore removed from the analysis (see \S\ref{sec:data}).  
}
\label{fig:m92}
\end{figure}

M92 was observed at 79 different combinations of position angles (PAs) and
offsets (see Figure \ref{fig:m92_acs_obs}), with three identical exposures taken at 
each pointing. This allowed for a given star to fall on several different parts
of the detector over the course of the observations.  
We note for clarity that the reported PA value is the angle (eastward)
of the camera's columns with respect to North.
The field of view of NIRC2's narrow camera contained the 
Natural Guide Star (NGS) in each pointing, and in most cases two other 
nearby stars, which are circled in Figure \ref{fig:m92}; this facilitated the 
process of combining the positional information from all of the different pointings. 
Table \ref{tab:m92_obs_all} provides the details of the NIRC2 M92 observations.

The M92 images are calibrated and stellar positions are measured from 
these images using standard techniques.  Specifically, the images are
first dark- and sky-subtracted, flat-fielded, and bad-pixel and cosmic ray 
corrected.  The images are then run through the point spread function 
(PSF) fitting program {\em StarFinder} \citep{stf}, which is 
optimized for adaptive optics observations of crowded stellar 
fields to identify and characterize stars in the field of view.  
{\em StarFinder} iteratively constructs a PSF from a set 
of bright stars in the field, which have been pre-selected by the 
user.  For M92, a total of 16 stars spread out across the 
detector are used to obtain a PSF that is representative 
of the entire field. The resulting PSF is then cross-correlated with 
the image and detections with a correlation peak of at least 0.7 are 
considered candidate stars. Relative astrometry and photometry are 
extracted by fitting the PSF to each candidate star. 
This results in a star list, which contains the NIRC2 pixel coordinates for
the detected stars, for each of the 237 images.

\begin{figure}
\epsscale{0.7}
\plotone{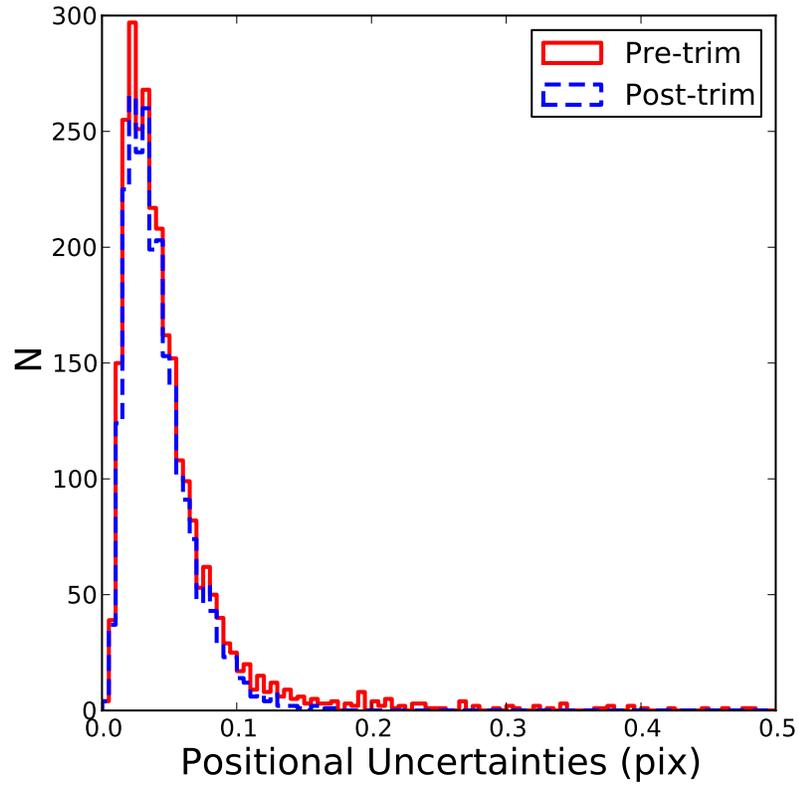}
\caption{Histogram of NIRC2 (plate scale $\sim$10 mas pix$^{-1}$) positional 
uncertainties for stars matched to the ACS/WFC star list
before ({\em solid red}) and after ({\em dashed blue}) removing all outliers (see text). 
The uncertainties are calculated from the RMS error of the positions obtained
from three images taken at the same position on the sky. The distributions 
peak at $\sim$0.02 pix.
}
\label{fig:n2posErr}
\end{figure}

Final star lists for each pointing are produced by combining each set
of three star lists from images with the same observational setup.  Positions
are taken from the first of the three images and the centroiding uncertainties 
are estimated empirically using the three images at each pointing and computing
the RMS error of each star's 
position\footnote{Choosing positions from the first of the three images was
unintentional, but should not affect the results since the centroiding
uncertainties are smaller than the level of distortion in the images.}.
The median centroiding uncertainty is $\sim$0.035 pix 
($\sim$0.35 mas; see Figure \ref{fig:n2posErr}). Two
initial criteria are used to trim out false or problematic source detections.
First, only stars detected in all three images are kept.
Second, we remove the two brightest stars (the NGS and a comparably bright star 
$\sim$5$\farcs$1 to the east that appears in the images of 147 out of 237 pointings) 
and any other source identified within a 60-pixel ($\sim0\farcs6$) radius 
of these stars (see Figure \ref{fig:m92}). These two sources are $\sim$1 mag 
brighter than any other detected star and are often detected at levels that 
saturate the detector.  Saturation leads to poor PSF matching with the 
empirical PSF estimate, and consequently poor positional estimates for these 
two stars, as well as $\sim$20-50 false detections in their
halos.  With these selection criteria, the
79 final star lists contain a combined total of 3846 stellar position 
measurements of more than 150 independent stars. 

\subsection{Galactic Center}
\label{sec:gcKeck}
Two types of Galactic center observations were obtained with the NIRC2 narrow
camera and the LGSAO system at Keck.  First, two sets of deep Galactic center
observations centered roughly on Sgr A* offer a test of the new distortion
model because something is changed in each.  In a data set from 2007 May,
previously reported in \citet{ghez08}, 103 frames were obtained with a camera
orientation of PA=0$^o$ (May 17) and another 20 frames were collected at
PA=200$^o$ (May 20).  In a new data set collected on 2008 May 15, the first 
22 images were obtained with the LGSAO system and the remaining 112 images were
taken with the NGSAO system; the Strehl ratio was 31\% and 22\% for the
LGSAO and NGSAO data that evening, respectively.  Second, three new epochs
of observations, designed to measure the relative positions of seven IR-bright
SiO masers, were carried out in 2008 May, 2009 June, and 2010 May, bringing the
total number of such observations to six.  These observations are primarily to
generate an astrometric reference frame in which Sgr A* is at rest, but
are also used as an additional test of the new distortion solution.  The 2008
May data set is identical to those reported in detail in 
\citet[][Appendix C]{ghez08} and consists of a total of 27 images, which were
obtained in a widely dithered (6$\arcsec\times$6$\arcsec$) 9-point box pattern 
with three images
at each of the nine pointing positions.  The 2009 June data set differed in that we 
repeated the box pattern three times, resulting in a deeper image by 
$\sim$0.5 magnitude, and the 2010 May data set was similar in total exposure time to
the initial mode, but made a trade of less coadds for more recorded images in an
attempt to compensate for the shorter atmospheric coherence times that evening.
We note that these observations were generally carried out under poorer seeing
conditions that our other GC observations (see below), since the quality of the
seeing is not the limiting factor in measuring the positions of the SiO masers.
These observations are summarized in Table \ref{tab:maserObs}.
USNO 0600-28577051, which is offset by 9.4$\arcsec$ E and 16.9$\arcsec$ N
from Sgr A*, served as the tip-tilt star for all of the LGSAO observations and
as the natural guide star for the NGSAO observation.

All the GC images are calibrated in a similar manner to what was carried out
with the M92 data sets, with a few exceptions.  First, all images are corrected
for differential atmospheric refraction (see \S\ref{sec:newDist}) and the
geometric optical distortion using the new model derived in \S\ref{sec:newDist}.
Second, all the images from each epoch (and each configuration in the case of
the deep central 10$\arcsec\times$10$\arcsec$ observations) are combined into a 
final average 
map as described in \citet{ghez08}.  In addition, three subset images, each
containing 1/3 of the data, are created for estimating centroiding uncertainties.

Astrometry was extracted using {\em StarFinder} \citep{stf} in a similar
manner as in \citet{ghez08}, with a few minor modifications.
Images from the individual pointings were analyzed with a correlation 
threshold of 0.9 in order to minimize spurious detections.  The maser
mosaics and the deep central 10$\arcsec\times$10$\arcsec$ average maps were run at a 
correlation threshold of 0.8, but used an improved 
algorithm to minimize spurious detections, which is described in Appendix
\ref{app:starfinder}.

\section{A New Distortion Model for NIRC2's Narrow Camera}
Ground-based astrometric observations are subject to rapidly varying effects
(such as instantaneous changes in the spatial pattern of PSF variation around
the detector), which lead to measurable nonlinear residuals between positions 
even when comparing frames within a night \citep[e.g.,][]{luPhd08}.
In this work, we seek to characterize the static component of the distortion,
which may be dominated by distortion within NIRC2 itself. Residuals between
observed stellar positions in NIRC2 and their counterparts in a nominally
distortion-free frame are represented as a single residual surface, which when
smoothed, forms our distortion model.  The result is a model for the
time-averaged distortion felt by the telescope and detector system.

\subsection{Constructing the Model}
\label{sec:newDist}
\begin{figure}
\begin{center}
\epsscale{1.0}
\plotone{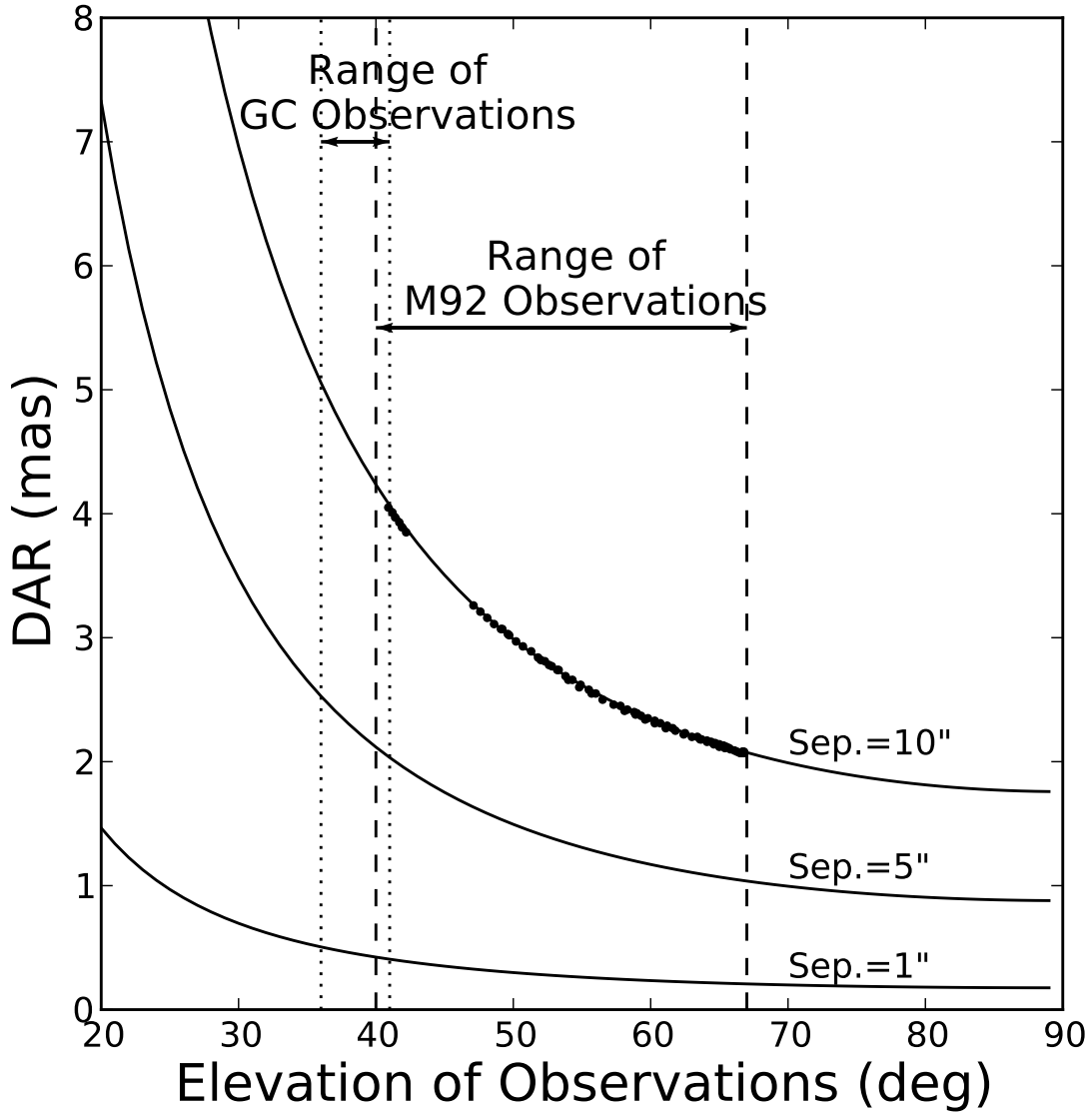}
\end{center}
\caption{
The predicted achromatic differential atmospheric refraction at a range of 
elevation angles for {\em typical} observing conditions at Keck. DAR 
causes the separation of two stars to appear smaller along the zenith
direction and the 
change in the separation is shown for three pairs of stars separated 
by 1'', 5'', and 10''.  The black dots show the amount of DAR over 
the 10'' field for each of the M92 observations used in the distortion 
solution. These are slightly offset from the predicted curve because
the atmospheric conditions differed slightly from the reference conditions 
used to generate the curves. The range of our GC and M92 observations are 
also shown. In all of our analyses, we apply DAR corrections to each 
individual image based on the conditions at the time.
}
\label{fig:keckDAR}
\end{figure}

To find the best fit model for NIRC2's geometric optical distortion
from the M92 observations, one must account for the fact that 
the ACS/WFC data do not suffer from differential atmospheric refraction
(DAR), while the NIRC2 data come 
from ground-based observations and therefore will be affected by 
the earth's atmosphere (see Figure \ref{fig:keckDAR}). 

Differential atmospheric refraction will compress an image along the zenith
direction, causing the
apparent separation between a pair of stars to be smaller than their true
separation.  Since the stellar positions are first geometrically distorted
by the atmosphere and then the telescope/instrument, it is best to "undo" 
these effects in the reverse direction.  Assuming one has a distortion
solution for NIRC2 at hand, to convert observed positions to their
counterparts in rectilinear space, the distortion solution should first
be applied to the observed positions, and then corrected for DAR.  When
attempting to {\it find} the distortion solution, however, the positions 
observed at NIRC2 cannot be corrected for DAR to compare with HST because
we do not know the distortion-corrected positions this process requires.
Instead, DAR is applied to the HST positions to produce a set of reference
positions that should correspond to distortion-free positions as observed
through the atmosphere.
Because the effects of DAR depend on the elevation and, to a much lesser 
extent, the atmospheric conditions of the observations, it is necessary 
to create a separate DAR-transformed ACS/WFC star list for each NIRC2 
image and associated star list.  To account for DAR, we follow the
prescription for DAR given in \citet{gublerTytler98}.  The stellar positions
are only corrected for achromatic DAR, as the error from chromatic DAR is
negligible ($<$0.2 mas) relative to the residual distortion in ACS/WFC ($\sim$0.5 mas).
Neglecting chromatic effects, 
the DAR term ($\Delta R$) depends on (1) the observed zenith 
angle of star 1, (2) the wavelength of the observations, (3) the observed 
zenith separation of star 1 and star 2, (4) the temperature at the 
observatory, (5) the pressure at the observatory, and (6) the relative 
humidity at the observatory.
The atmospheric parameters of interest are downloaded from an archive
maintained at the Canada-France-Hawaii Telescope 
(CFHT)\footnote{\url{http://kiloaoloa.soest.hawaii.edu/archive/wx/cfht/}}
for the night of each observation.
These values are recorded every five minutes, allowing us to find the 
appropriate atmospheric conditions on Mauna Kea within three minutes of 
the observation \citep{luPhd08}.
As shown in Figure \ref{fig:keckDAR}, the magnitude of the achromatic effect 
over the range of elevations for 
the M92 observations is expected to be $\sim$2-4 mas across 
NIRC2's 10$\arcsec$ field of view, along the elevation axis.  

Each of the NIRC2 star lists described in \S\ref{sec:m92} is then used as a 
reference coordinate system into which the ACS/WFC star list of positions 
is transformed. In this process, the ACS/WFC star list is transformed by 
minimizing the error-weighted (NIRC2 positional errors) net displacement 
for all the stars, allowing for translation, rotation, and a global 
plate scale (i.e., a four-parameter transformation model).  
This process is described in greater detail in 
\citet{ghez08} and \citet{lu09}. Only sources that are cross-identified 
in both the NIRC2 and ACS star lists are used in the remaining analysis.  
From the 79 separate alignments, a total of 2743 matches in stellar positions
are obtained for a total of 150 independent stars.  The differences in the
matched positions, or deltas, are a result of the optical distortion in NIRC2.

\begin{figure}
\epsscale{0.6}
\plotone{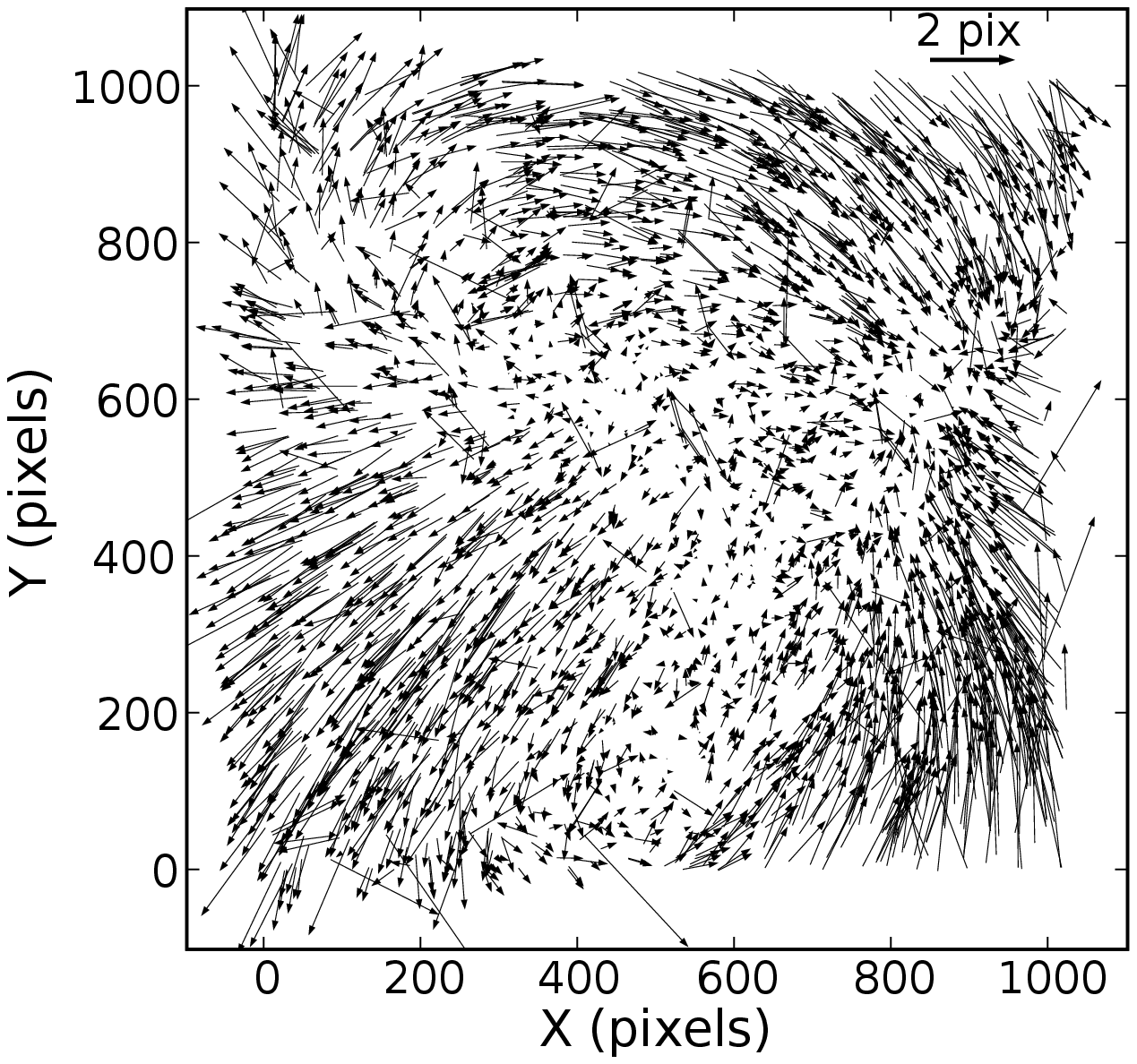}
\plotone{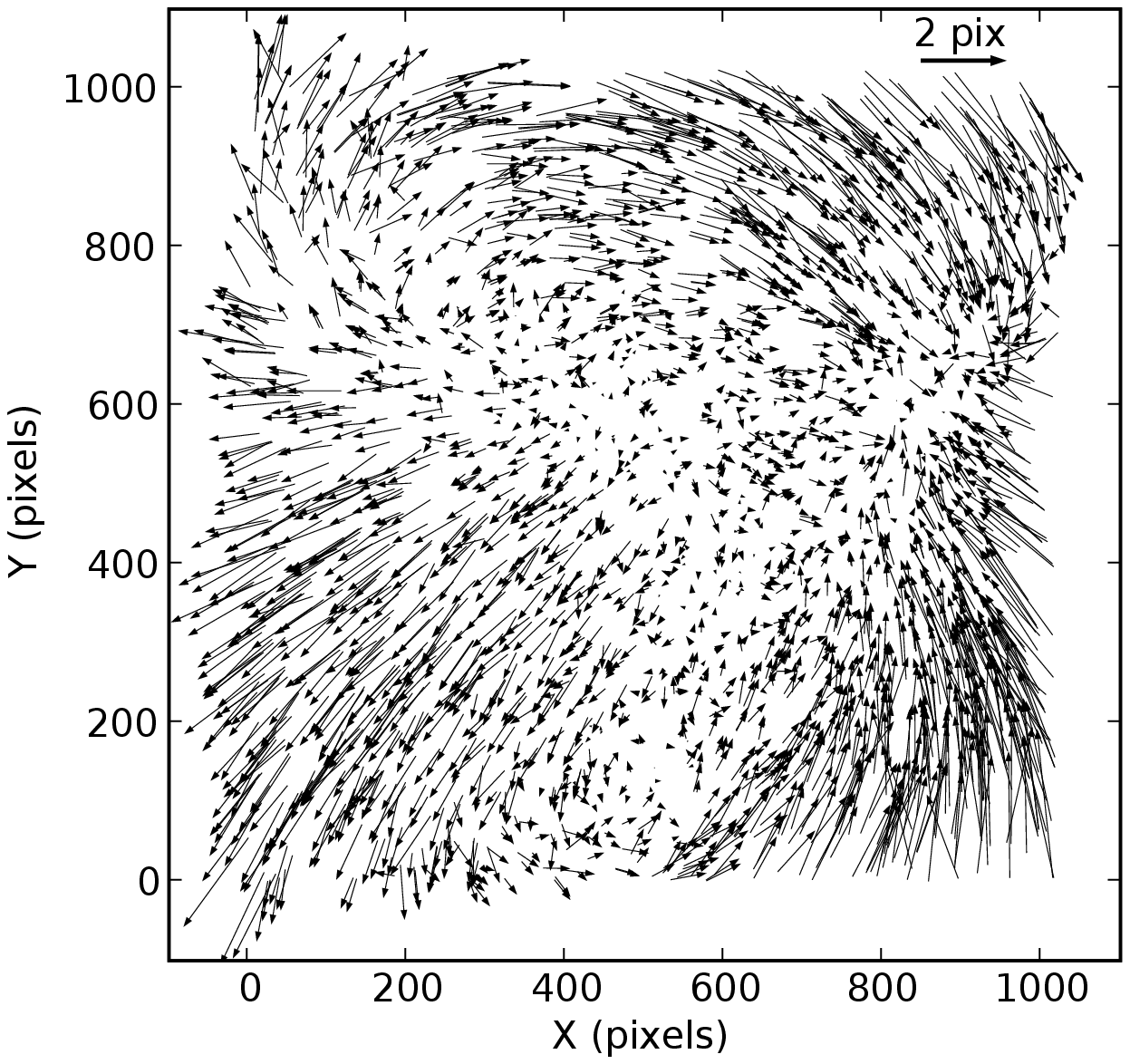}
\caption{
Optical distortion in the NIRC2 camera obtained from positional 
measurements of stars in the globular cluster M92.
Arrows indicate the difference between measurements made with 
NIRC2 ({\em arrow tail}) and ACS/WFC ({\em arrow head}), which 
has a well characterized distortion solution to the $\sim$0.5 mas 
level \citep{anderson06,anderson07}. The two figures show pre- 
({\em top}) and post- ({\em bottom}) trimming. 
}
\label{fig:distM92}
\end{figure}

The mapping of ACS positions to NIRC2 positions shows clear spatial structure 
across the detector, as expected from optical distortion 
(see Figure \ref{fig:distM92}). However, some deltas are 
inconsistent with those in their immediate surroundings.
These outliers are found by examining the 
vector deviations in 205$\times$205 pixel bins and determining the average 
and standard deviation. Any 3$\sigma$ outliers in either the X or Y 
direction are removed.  A total of 75 deltas are removed based
on this criterion.  An additional cut ($>$3$\sigma$) in each of 
these bins is made on NIRC2 positional uncertainties, as they 
may vary with respect to detector position.  This cut removes 73 data 
points, four of which were also eliminated by the first cut.
These bins are examined a second time for vector 
outliers, as they often show a rather wide distribution. The average 
and standard deviation in each bin are recalculated 
and the vector outliers ($>$3$\sigma$) are removed once again.
This resulted in an additional loss of 26 deltas. 

Many of the eliminated measurements come from common stars or images.
We therefore remove all measurements of the 9 out of 150 stars and of the
8 out of 79 images that were eliminated more than 20\% of the time by the
sigma-clipping process. Many of these problematic stars have close neighbors
($<$ 0$\farcs$2) that are not resolved or not well measured in the
lower-resolution ACS observations (FWHM$\sim$70 mas for the F814W
observations).
The majority of the rejected frames have exposure times less 
than 10 sec, while the remaining frames are at least 30 sec.  This results 
in significantly higher centroiding
uncertainties, residual atmospheric effects, and fewer stars detected.
We note that our trimming criteria mentioned above also results in the 
exclusion of all data from the 2008 June epoch (t$_{int}$ = 9 sec), which 
coincidentally was the only M92 data set taken in LGSAO mode. 
Although the 2008 April data set had relatively long exposure times 
(t$_{int}$ = 48 sec), the observations were heavily impacted 
by clouds and the AO system was often unable to remain locked on the NGS.
These rejected frames show a value of -1.0 in the last column of 
Table \ref{tab:m92_obs_all}.
Our final data set consists of 2398 positional deviations between 
ACS and NIRC2, with median centroiding uncertainty for the NIRC2 images 
of 0.035 pix ($\sim$0.35 mas).
The vector plot for this cleaned sample is shown in the bottom of 
Figure \ref{fig:distM92}.  

\begin{figure}
\begin{center}
\epsscale{1.0}
\plottwo{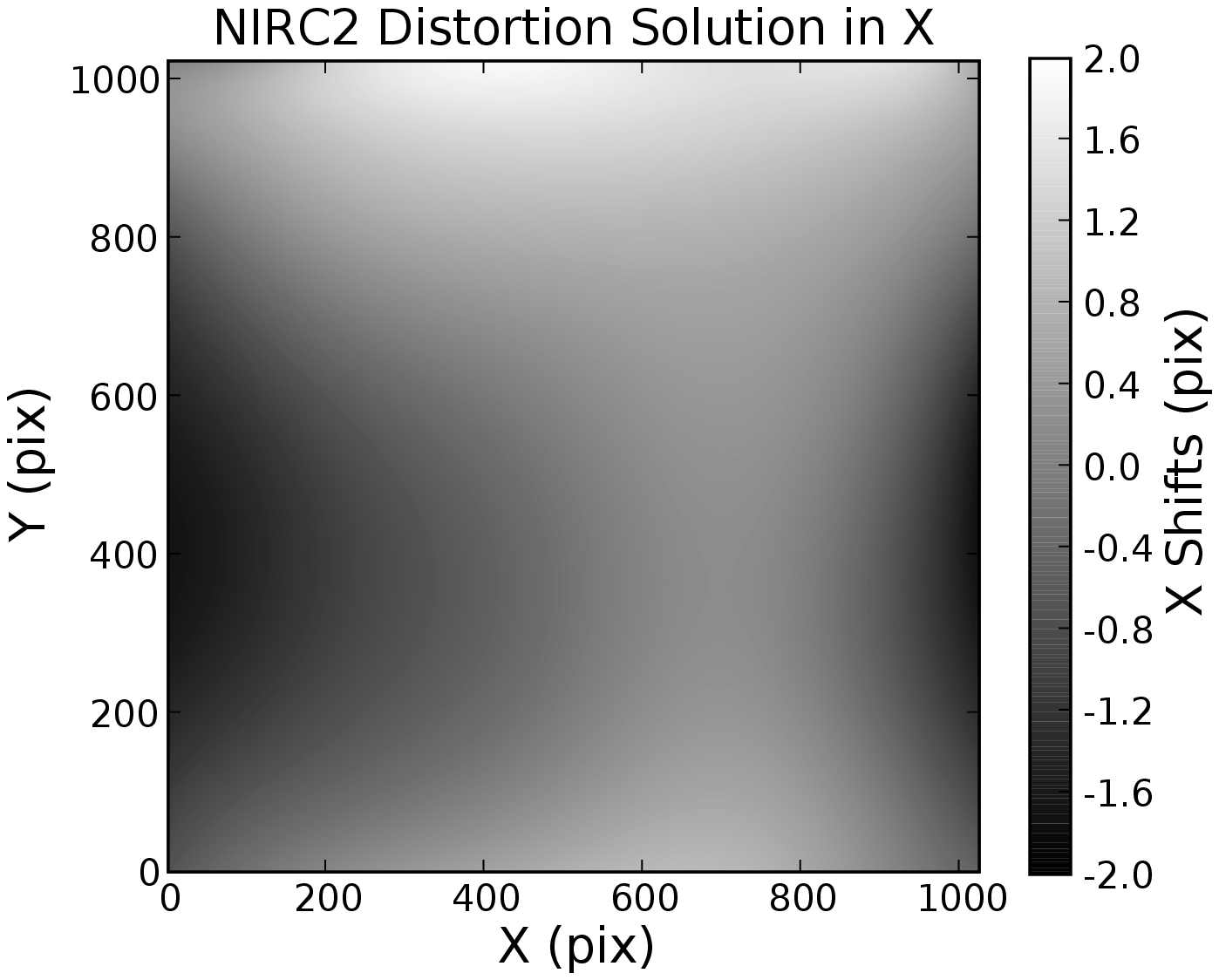}{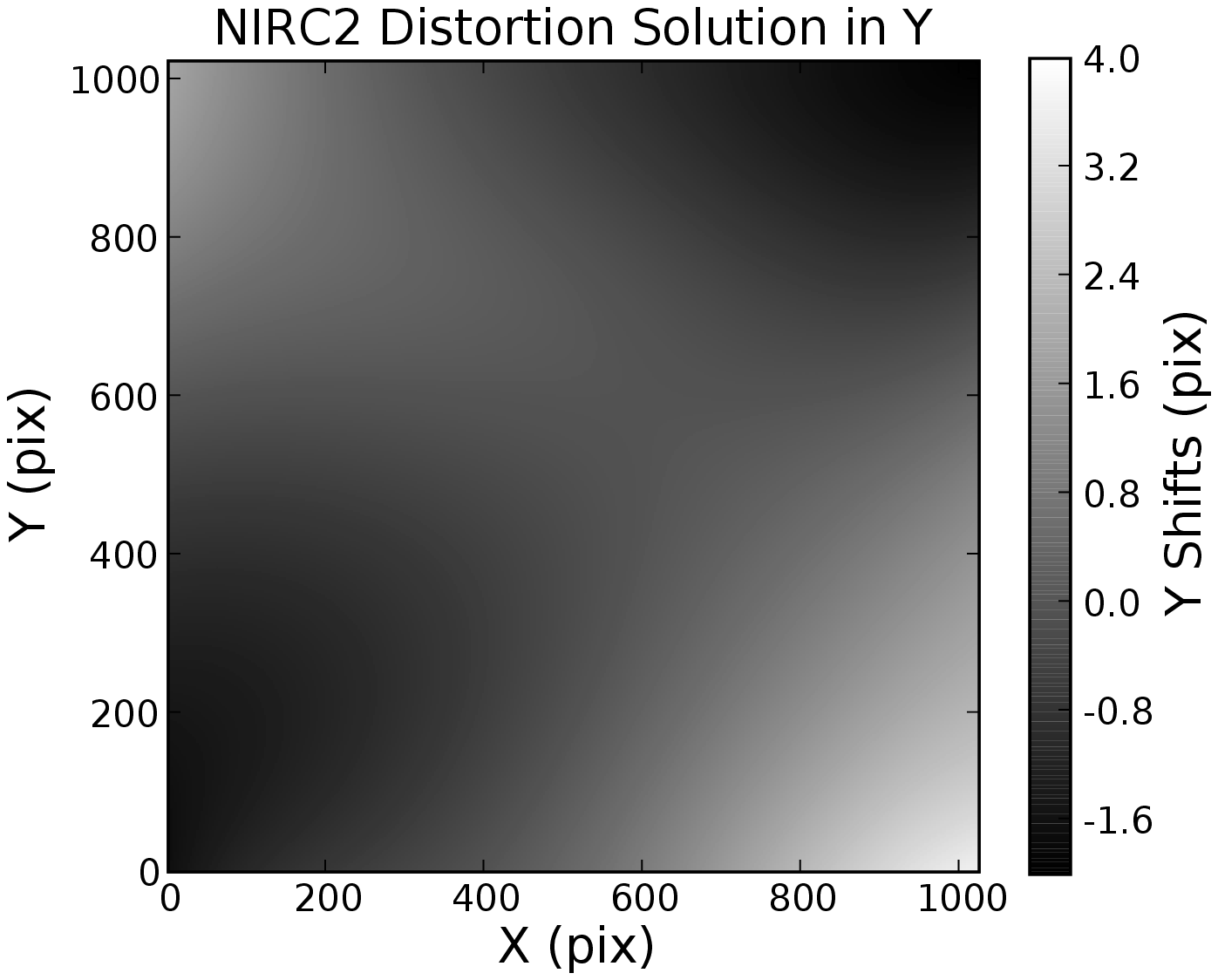}
\plottwo{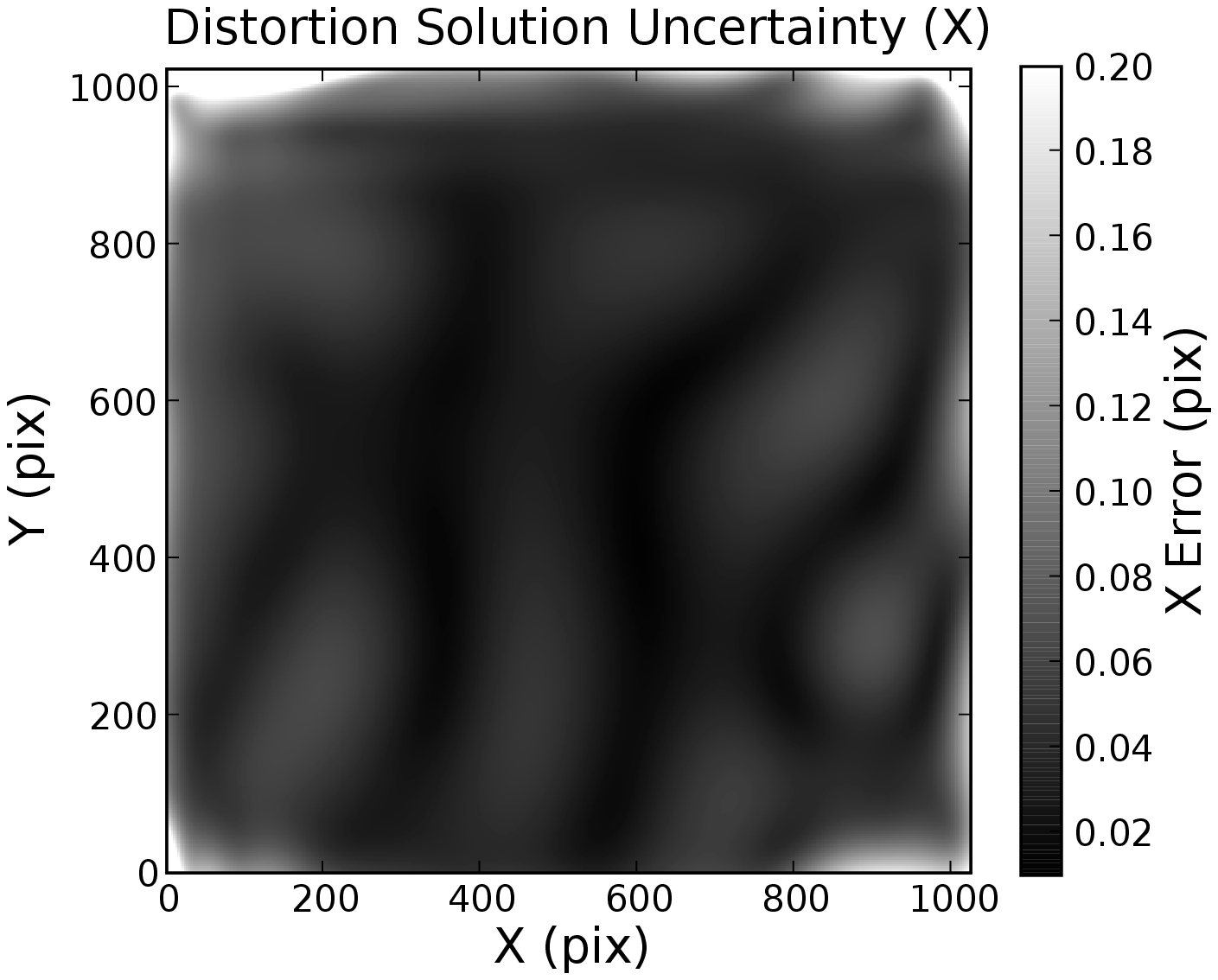}{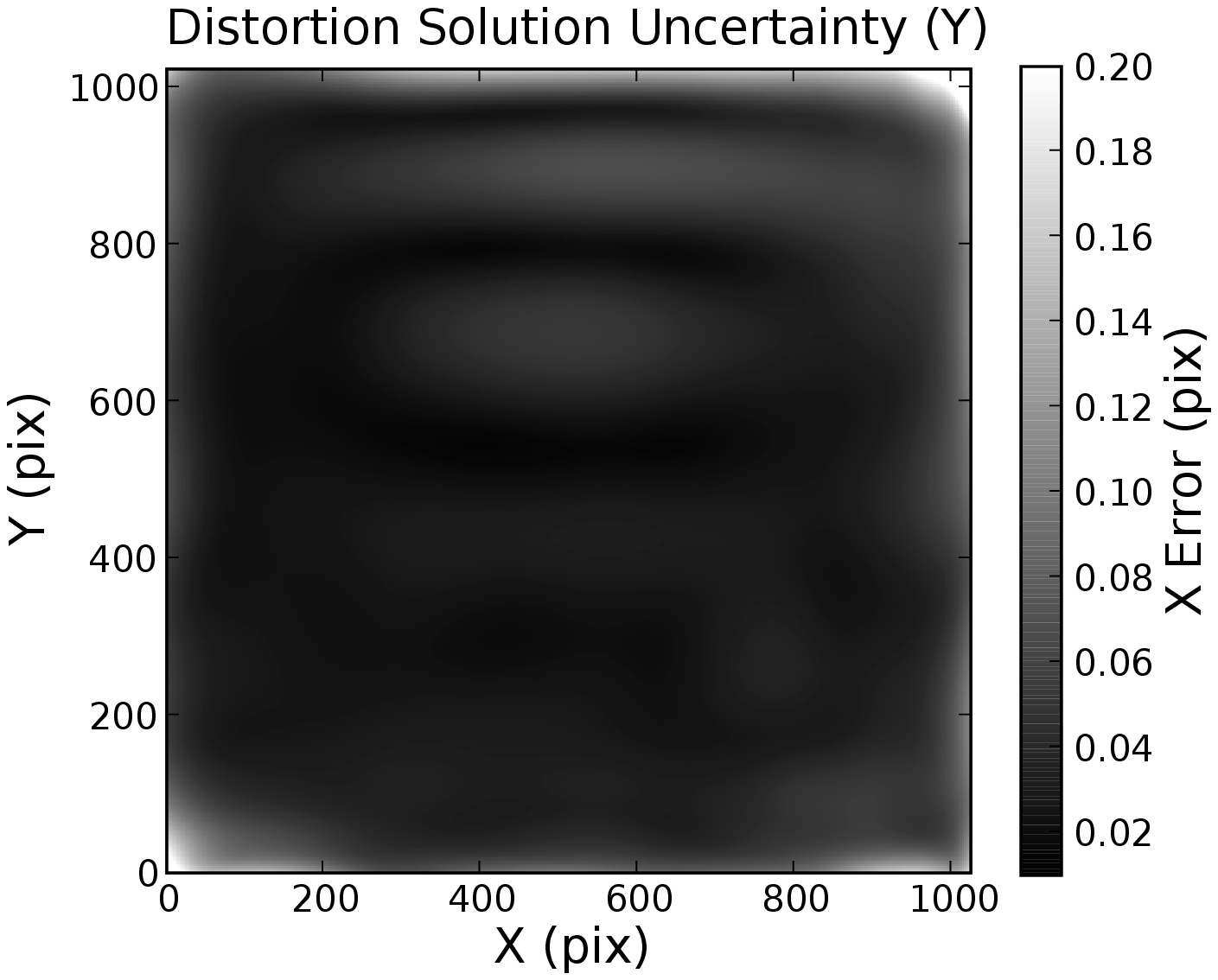}
\caption{
({\em Top}) Distortion solution in the form of a look-up table for X ({\em left}) 
and Y ({\em right}). The tables give the X and Y values for each 
pixel required to remove the optical distortion from NIRC2 images.  
This was generated by fitting a surface to the distortion map in the 
bottom of Figure \ref{fig:distM92}. 
({\em Bottom}) RMS error of the 1000 simulations of the distortion solution 
based on M92 data (\S\ref{sec:newDist}) for X ({\em left}) and Y ({\em right}).  
The images are shown in linear stretch.  The average errors in X and Y are 
($\sigma_X$, $\sigma_Y$) = (0.05, 0.04) pix $\sim$ ($\sim$0.5, $\sim$0.4 mas), 
respectively. Note an additional error of 0.1 pix is required to fully
describe the uncertainty in the optical distortion (see \S\ref{sec:otherSrcs}).
}
\label{fig:distFits}
\end{center}
\end{figure}

\begin{figure}
\begin{center}
\epsscale{1.0}
\plottwo{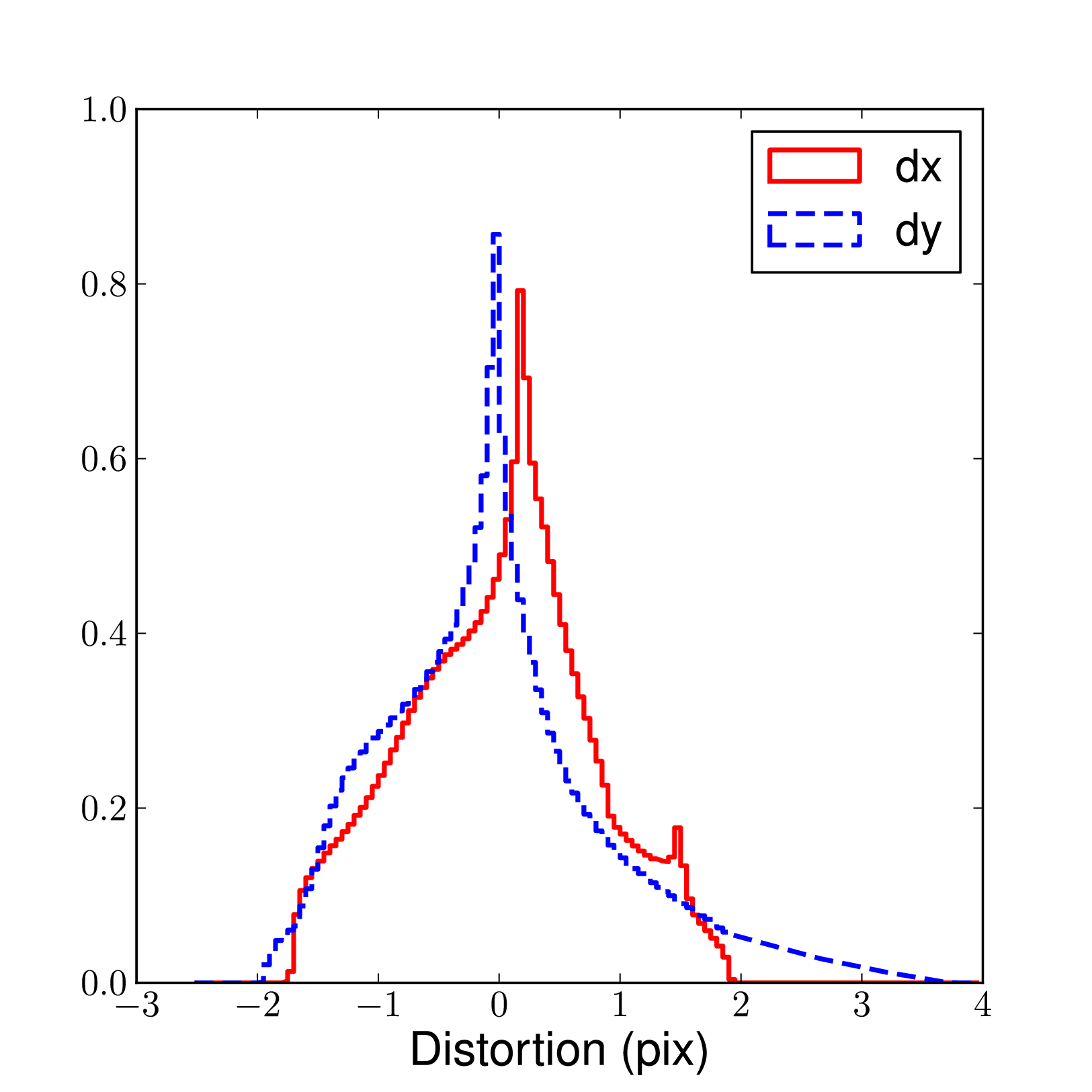}{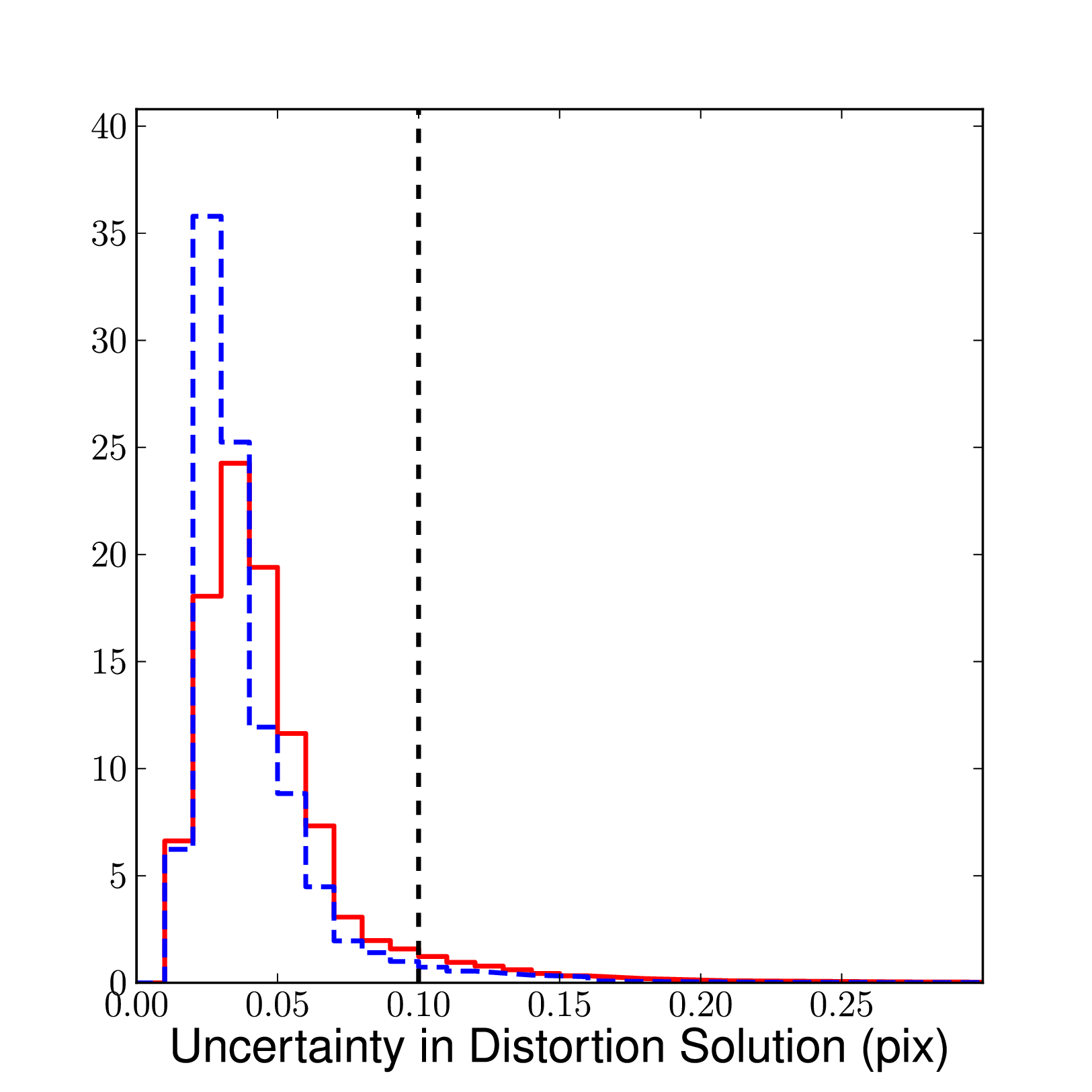}
\caption{
({\em Left}) Distribution of the shifts in the distortion solution look-up table
over all NIRC2 pixels for X ({\em solid red}) and Y ({\em dashed blue}).  
({\em Right}) Distribution of the RMS uncertainties from the 1000 simulations 
of the distortion solution. The average errors in X and Y are 
0.05$\pm$0.04 pix and 0.04$\pm$0.02 pix, respectively.
The vertical dashed line represents the additive error that is found when
the models are tested with Galactic center data (see \S\ref{sec:testNew}).
}
\label{fig:monteHist}
\end{center}
\end{figure}

A bivariate B-spline is fit to the distortion map (Figure \ref{fig:distM92})
using the SciPy package {\em interpolate}, and a look-up table sampled at each of 
the 1024$\times$1024 NIRC2 pixels is subsequently produced. The effect of 
the smoothing factor ($f$; which is related to the number of nearest-neighbor
measurements used to calculate the smoothing) used in the interpolation 
routine was investigated extensively in order to find a good compromise 
between the closeness of fit and the smoothness of fit.  The residuals 
between the original distortion vectors in the bottom panel of Figure 
\ref{fig:distM92} and the computed shift at the nearest pixel 
(from the smoothed look-up table) were measured.  The median deviation is 
found to increase until $f{\sim}150$, where it plateaus at a value of 
$\sim$0.27 pix.  
We choose for our interpolation the smoothing factor that gave nearly 
the lowest median deviation, $f=135$. Although the deviations were lower for 
distortion solutions created with smaller smoothing factors,
the edge effects were prominent in the look-up tables and the distribution 
of deviations was much larger \citep[for details on surface fitting and
the choice of smoothing factors, see][]{dierckx}. The resulting
look-up tables for shifts in X and Y are shown in Figure \ref{fig:distFits},
and are produced in the form of FITS files that may 
be fed into the IRAF routine, {\em Drizzle} \citep{drizzle}, to correct
for the optical distortion.  Figure \ref{fig:monteHist} (left) shows a 
histogram of these values.

Statistical uncertainties in the distortion solution were computed by running 
a bootstrap analysis with 1000 trials. In each trial, we generated
a random set of data pulled from the observed data, 
allowing for replacement after each data point was sampled, and then derived
a distortion model from this resampled data set.
The RMS error with respect to the distortion solution (i.e., the actual 
distortion solution was taken as the average) was calculated at each pixel 
and the results are shown in the bottom of Figure \ref{fig:distFits}. 
The average errors in X and Y are ($\sigma_X$, $\sigma_Y$) = 
(0.05, 0.04) pix $\sim$ (0.5, 0.4 mas), respectively. 
We can see the uncertainties are highest near the edge of 
the detector, where the spline algorithm is less robust. 
The uncertainties are also shown 
in the form of a histogram in Figure \ref{fig:monteHist} along with a
histogram of the distortion solution itself.

To solve for the global plate scale and orientation that results from this
new solution, we re-reduce the raw NIRC2 observations of M92 from all epochs,
and apply corrections for distortion and DAR to these images.
The distortion correction and DAR correction are applied to each image
at the same time in the form of look-up tables using the {\em Drizzle} 
algorithm as implemented in IRAF \citep{drizzle}. The look-up
tables are specified in {\em Drizzle} using the {\em xgeoim} and {\em ygeoim}
keywords and are FITS files of the same dimensions as the science image.
Because DAR depends on the zenith angle and atmospheric conditions, both
of which vary in time, the look-up tables are created by first including 
the distortion solution and then applying the necessary DAR correction. 
Two FITS files, one for shifts in X and one for shifts in Y, are created 
for each NIRC2 observation and contain the shifts to be applied to each 
pixel in the image. From these distortion- and DAR-corrected NIRC2 images, 
star lists were generated and aligned with the original ACS starlist 
(without DAR) as described above. The weighted average of the plate scale is
${\langle}s{\rangle}$ = 9.950 $\pm$ 0.003$_{stat}$ $\pm$ 0.001$_{abs}$ 
mas pix$^{-1}$. The difference between
the orientation given in the header of the NIRC2 
images\footnote{The NIRC2 FITS header keyword for the 
position angle, ROTPOSN, includes a +0.7$^o$ offset (given by header
keyword INSTANGL), the observatory value for the angle offset of NIRC2. 
The nominal position angle for NIRC2 in our analysis is taken 
as (ROTPOSN - INSTANGL).} 
and the measured orientation is on average (weighted)
0.254$^o$ $\pm$ 0.014$^o_{stat}$ $\pm$ 0.002$^o_{abs}$.  We use 
the RMS errors of the average values from each epoch as the statistical 
uncertainties, and
the absolute errors are the RMS errors in the ACS/WFC plate scale 
and orientation angle \citep{vanderMarel}.  The results from each epoch
of M92 data are shown in Table \ref{tab:plateScaleAngleCmpr}.

The new distortion solution and its errors are made public and may be obtained
in the form of FITS files at 
\url{http://www.astro.ucla.edu/~ghezgroup/distortion}.

\subsection{Testing the Model}
\label{sec:testNew}
There are two parts to the error in the distortion model when applied to a
real data set: static distortion error (hereafter "residual distortion",
discussed here) and time-varying effective distortion (discussed in 
\S\ref{sec:otherSrcs}).  The residual distortion map on the detector is
unknown, but is likely to be highly spatially correlated.  The spurious 
position-shift due to residual distortion when comparing two measurements is
a function both of the size of the residual distortion itself, and the
difference $\Delta R$ in location on the detector between the two measurements. 

To estimate the size of the residual distortion from our model, we consider
two cases.  In the first, sets of images are taken at two very different
position angles so that the distance $\Delta R$ between two measurements
of the same object (and therefore the degree to which the residual distortion
varies between measurements) is a strong function of position on the detector.
In the second, images are taken at the same position angle, so that $\Delta R$
is constant over the image, but are widely dithered (60\% of the detector 
side-length) so that the residual distortion is sampled at widely separated
detector locations for all objects.

In both cases, we compare our new distortion model with two previous solutions,
which we refer to as "pre-ship" and "PBC". The pre-ship 
solution\footnote{\url{http://www2.keck.hawaii.edu/inst/nirc2/preship_testing.pdf}},
which is known to $\sim$4 mas, was found using a pinhole mask, and
is in the form of a 3rd-order polynomial.
The more recent solution by P. B. Cameron, also from a pinhole mask, is a 
4th-order polynomial and improves upon the former solution mainly along the X 
axis\footnote{\url{http://www.astro.caltech.edu/~pbc/AO/distortion.pdf}}.

\begin{figure}
\begin{center}
\includegraphics[scale=0.5]{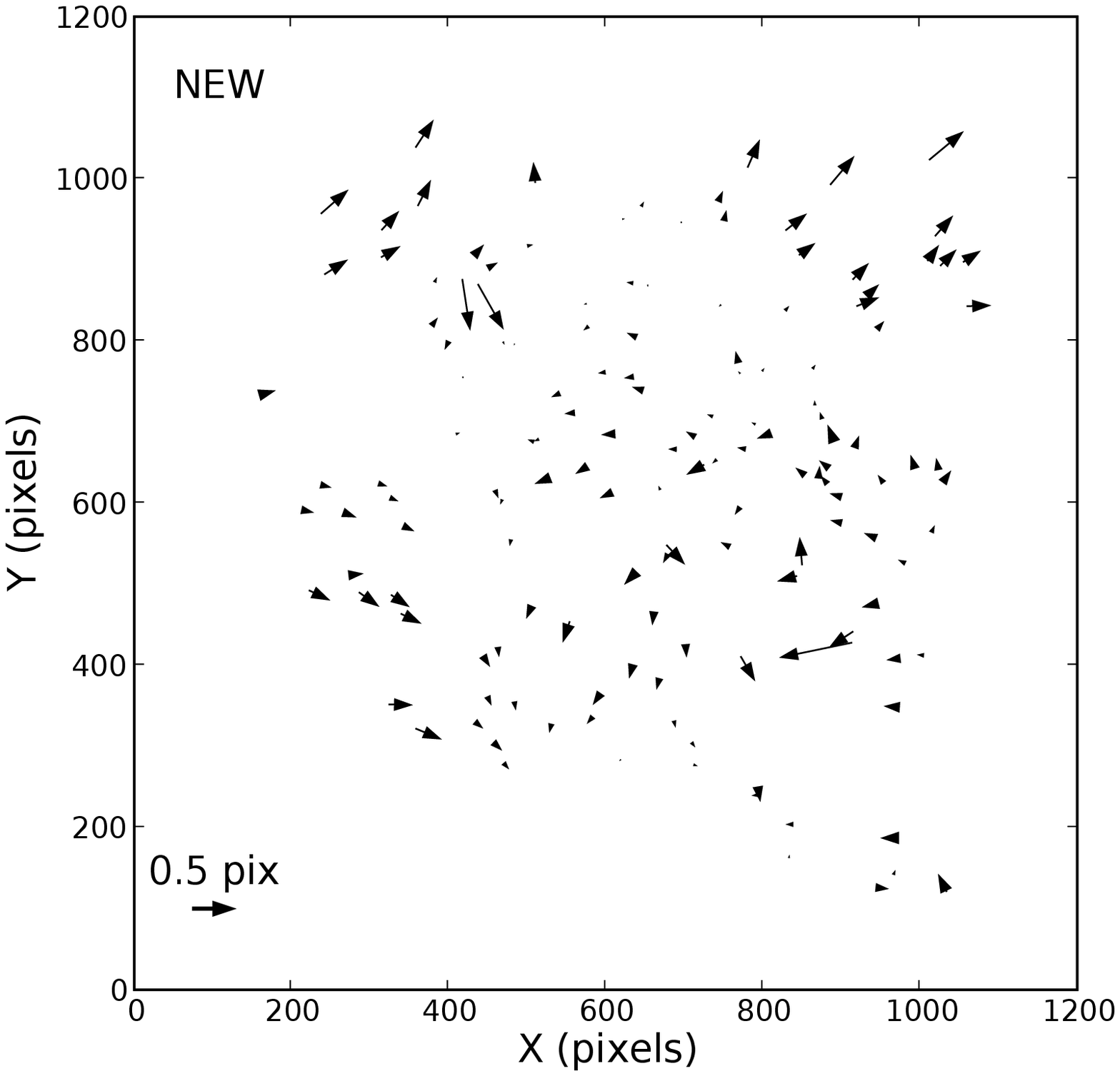}
\end{center}
\includegraphics[scale=0.5]{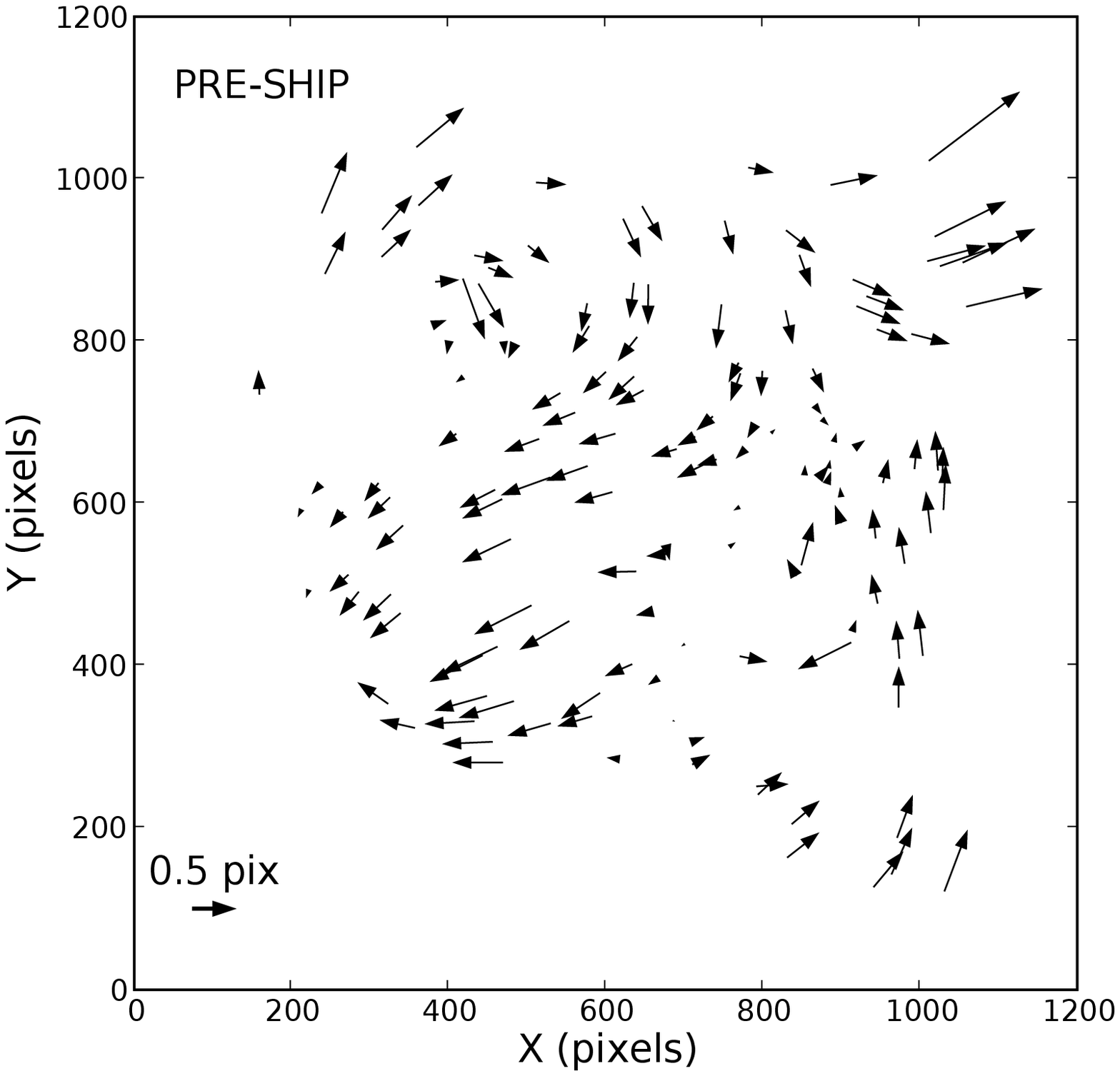}
\includegraphics[scale=0.5]{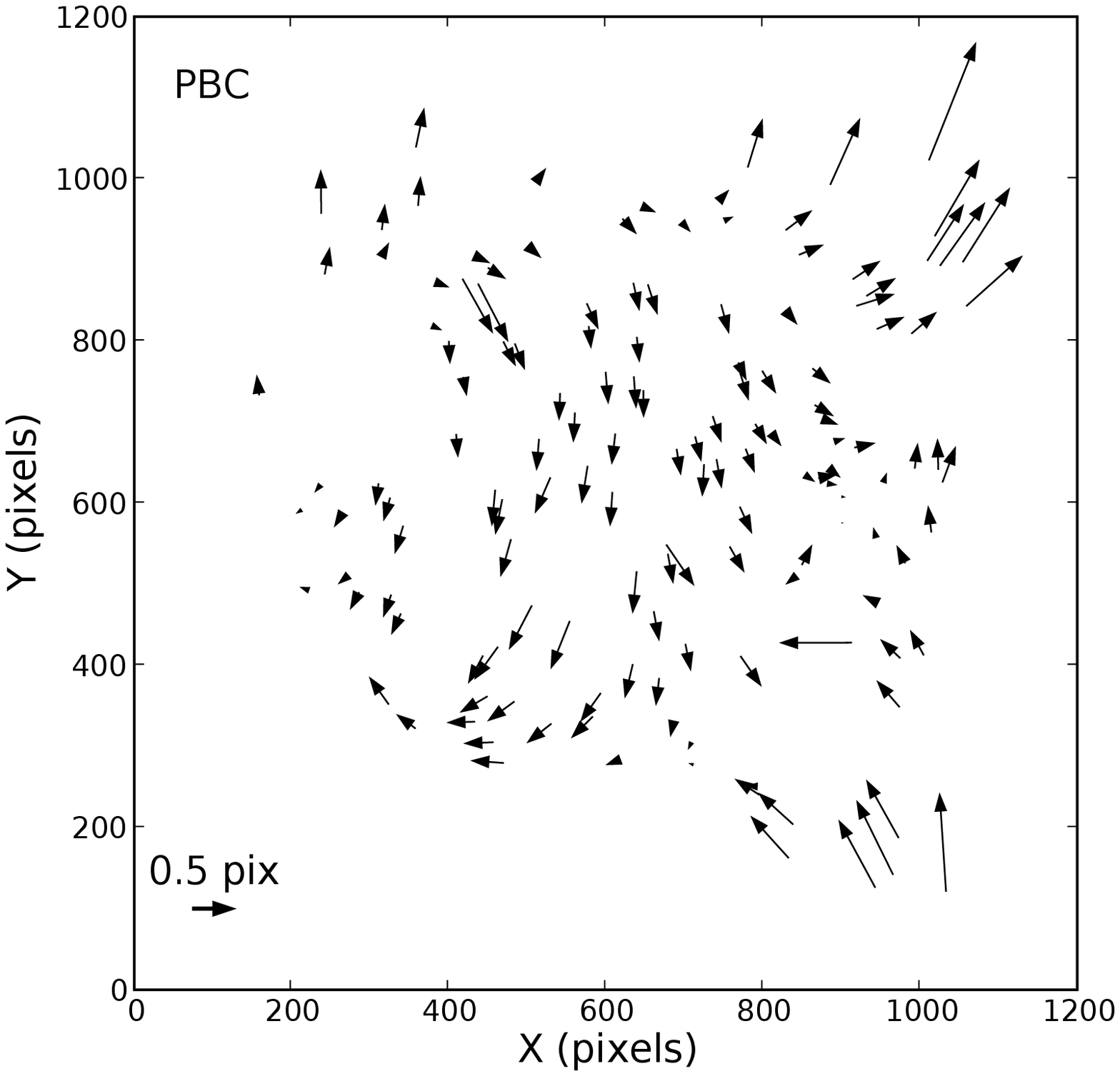}
\caption{
Differences between stellar positions in Galactic center images taken 
at PA=200$^o$ ({\em arrow tail}) and PA=0$^o$ ({\em arrow head}) 
after applying the new ({\em top}), pre-ship ({\em bottom left}), and PBC 
({\em bottom right}) distortion solutions. While some residual distortion 
remains, much of the structure seen after using the pre-ship solution 
is removed with the new solution. The residual distortion remaining
is ($\sigma_{x,0}$, $\sigma_{y,0}$) = (0.12, 0.11) pix, (0.17, 0.28) pix, 
and (0.27, 0.22) pix for the new, PBC, and pre-ship solutions, respectively.
}
\label{fig:diffPAs}
\end{figure}

First, we use the two high precision data sets taken of the central 
10$\arcsec\times$10$\arcsec$ on 2007 May 17 and May 20 at two different PAs 
(0$^o$ and 200$^o$) with roughly the same central position 
(\S\ref{sec:gcKeck}).
The typical NIRC2 positional uncertainties (the standard deviation, 
$\delta_{pos}$) for the PA=0$^o$ and PA=200$^o$ images are $\sim$0.013 pix and 
$\sim$0.018 pix, respectively.
The PA=200$^o$ image was transformed into the PA=0$^o$ image's coordinate 
system, again allowing for translation, rotation, and global plate scale.  
The differences in the aligned positions of stars with K$<$14.5 are shown in 
Figure \ref{fig:diffPAs}. 
This analysis gives an average residual distortion by comparing the 
positions of a star at two distinct locations on the detector.
Our new solution shows significantly less residual structure than the
previous solutions. To estimate the magnitude of the residual distortion 
($\sigma$), we compute the RMS error of the offsets ($\Delta$) between the 
positions in the two images in the X and Y directions 
separately\footnote{These residuals are
measured in the PA=0$^o$ image's coordinate system, as this was the reference
onto which the PA=200$^o$ image was transformed.}, 
and correct for the positional measurement error from both images ($\delta$): 
\begin{equation}
\sigma_{x} = \sqrt{\frac{1}{2}\displaystyle\sum_{i}^{N_{stars}}\frac{(\Delta_{x,i} - 
\langle\Delta_x\rangle)^2}{(N_{stars} - 1)}
- \frac{1}{2}(\delta^2_{pos,0^o} + \delta^2_{pos,200^o})}
\end{equation}
where $N_{stars}$ is the number of stars matched across the two images, 
and $\delta_{pos,0^o}$ and $\delta_{pos,200^o}$ are the positional uncertainties
(quoted above) for stars brighter 
than K=14.5 in the PA=0$^o$ and PA=200$^o$ images, 
respectively.  Average positional uncertainties are subtracted, as opposed
to each star's individual uncertainty since most stars brighter than K=14.5 have
similar centroiding uncertainties.
We compute $\sigma_{y}$ similarly.  We note that the division 
by 2 is necessary to determine the residual distortion incurred {\em per} NIRC2 
image. This results in estimates of the residual distortion of 
($\sigma_{x,0}$, $\sigma_{y,0}$) = (0.12, 0.11) pix, (0.17, 0.28) pix, and 
(0.27, 0.22) pix for the new, PBC, and pre-ship solutions, respectively.
Thus, the new solution results in smaller residuals by a factor of 
$\sim$2-2.5 over both of the previous solutions. The uncertainty 
in the distortion models has not been removed
from these values, as the uncertainty in the PBC and pre-ship solutions are
unknown. We can, however, remove the average uncertainty in our new model in
quadrature (Figure \ref{fig:distFits}, bottom) to obtain a final measure of
the residual distortion: ($\sigma_{x,0}$, $\sigma_{y,0}$) = (0.11, 0.10) pix.

\begin{figure}
\begin{center}
\includegraphics[scale=0.5]{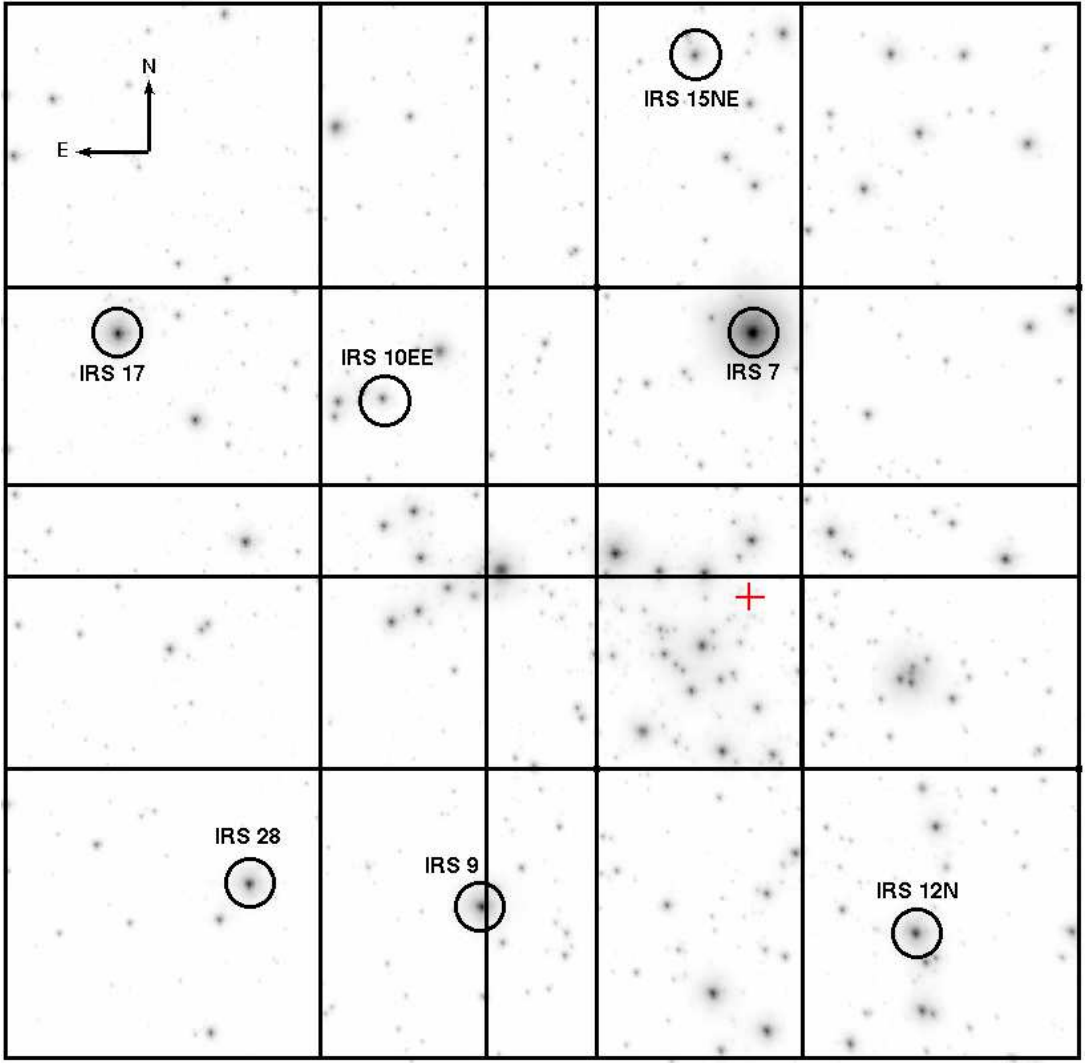}
\end{center}
\caption{
NIRC2 K' mosaic image of the Galactic center scaled to show the bright
stars.  The full field is 
22$\arcsec\times$22$\arcsec$. The black boxes show the nine dither positions 
making up the mosaic, with each box corresponding to the 
10$\arcsec\times$10$\arcsec$ NIRC2 field of view. The 7 SiO masers used in the
construction of the Sgr A*-radio reference frame are circled, and 
Sgr A* is marked with a red cross.
The four images that make up the SW corner of the mosaic were used in
\S\ref{sec:testNew} to determine the quality of the distortion solution.
}
\label{fig:msrFlds}
\end{figure}

\begin{figure}
\epsscale{1.2}
\plottwo{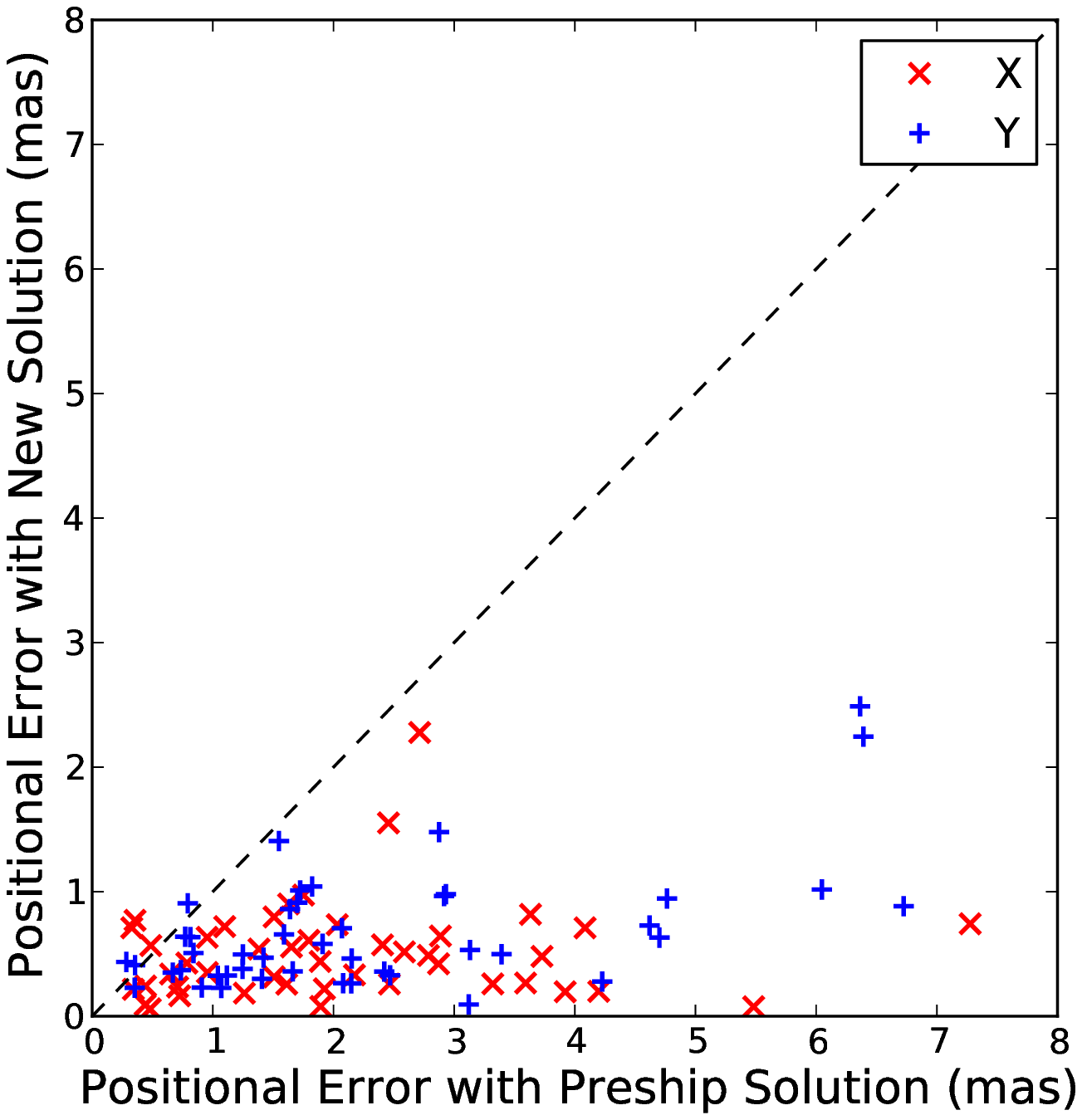}{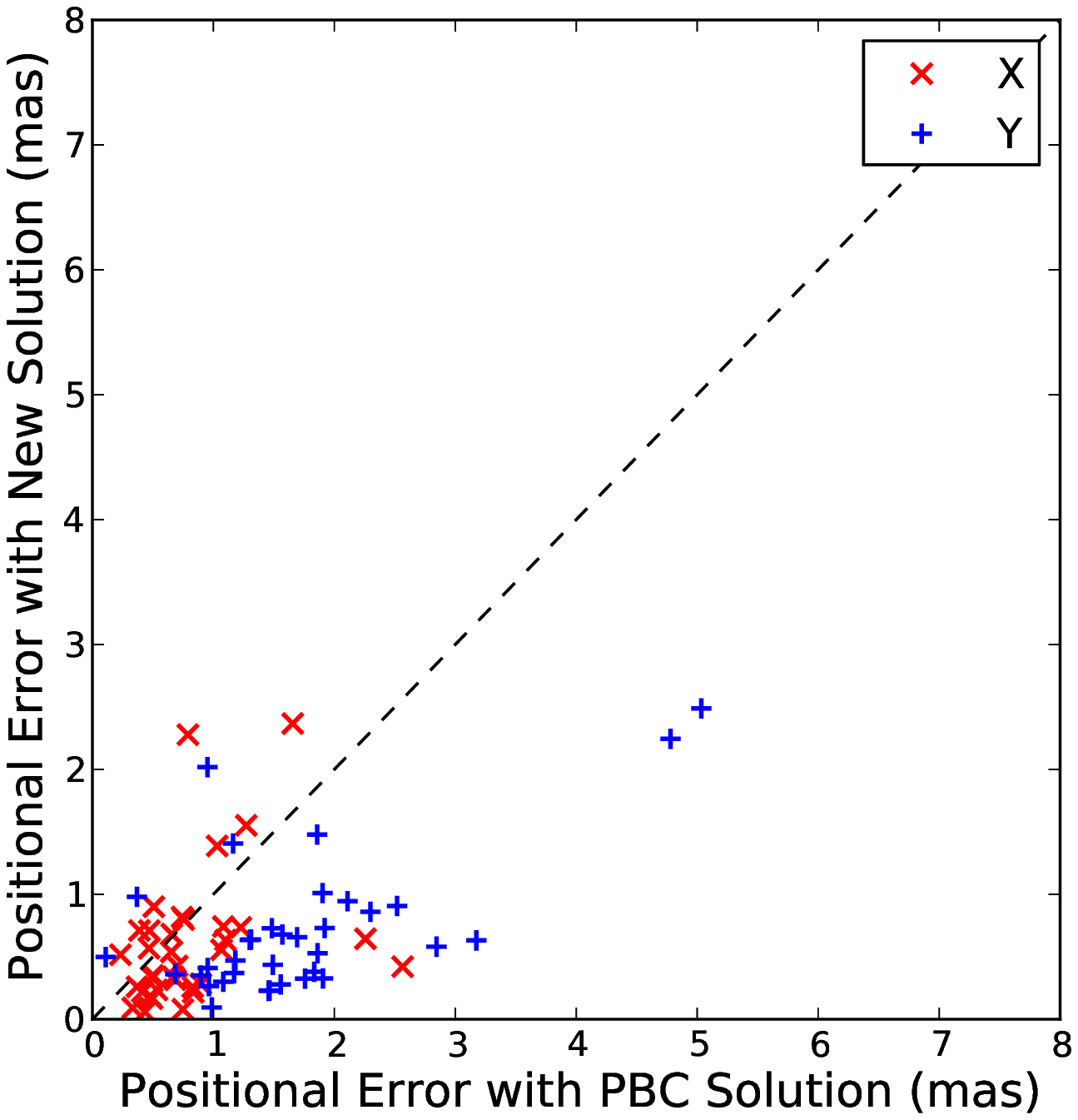}
\caption{
Pair-wise analysis on widely-dithered Galactic center data taken 
in 2008 May.  The RMS error of the positional offsets from IRS 16SW-E 
are plotted.  The plots compare the RMS error values from images 
corrected with the new versus the pre-ship distortion solution 
({\em left}) and the new versus the PBC distortion 
solution ({\em right}). The new solution is a factor of $\sim$3-4
improved in both X ({\em red crosses}) and Y ({\em blue plus signs}) 
over the pre-ship solution and in Y over the PBC solution.
}
\label{fig:pwise}
\end{figure}

As an additional check on the distortion solution, we use the images from 
widely-dithered (6$\arcsec$) 2008 May data set taken at PA=0$^o$ (\S\ref{sec:gcKeck}), 
which, unlike the test just described, maintain the independence of the X and Y 
axes as they are shifted relative to one another by only a translation.  
Largely-dithered data sets are essential in testing the distortion
solution because stars are placed on very different locations on the detector
and therefore provide a sensitive test of residual distortion.
These data have an average RMS error on the positions of 
0.05 pix. Only four overlapping fields, each of which was 
imaged three times and whose centers are the corners of a 6$\arcsec\times$6$\arcsec$ box, were 
examined from this data set (Figure \ref{fig:msrFlds}). Only stars 
detected in at least 6 of the 12 images (and therefore at least two of 
the four overlapping fields) were kept in the analysis.  The stars had to
also be detected in all three exposures at each dither position, and the
average of the positions in these three exposures was taken as the
position at the corresponding dither position.
The error ($\sigma_{x,i}$, $\sigma_{y,i}$) from residual optical distortion of 
{\em each star's} offsets (${\Delta}x$, ${\Delta}y$) from 
IRS 16SW-E (which was in each of the four fields) was computed as:
\begin{equation}
\sigma_{x,i} = \sqrt{\frac{1}{2}
\displaystyle\sum^{N_{fields}}_j\frac{({\Delta}x_j - \langle{\Delta}x_j\rangle)^2}{N_{fields} - 1}
- \frac{1}{2}\frac{1}{N_{fields}}
\displaystyle\sum^{N_{fields}}_j\bigg(\frac{\delta_{pos,IRS16SWE}^2}{N_{exp} - 1} + 
\frac{\delta_{pos,i}^2}{N_{exp} - 1}\bigg)}
\end{equation}
and likewise for $\sigma_{y,i}$,
where we divide by the number of overlapping fields in which a star was 
detected ($N_{fields}$), and we correct for the NIRC2 positional measurement 
error (the standard deviation, $\delta_{pos}$) per exposure ($N_{exp}$) for both 
IRS 16SW-E and star $i$. The factor of 2 in the denominator accounts for the fact 
that the distortion affects both stars, IRS 16SW-E and star $i$. These errors
from the residual optical distortion are shown in Figure \ref{fig:pwise} for
all three solutions. The median values ($\sigma_x$, $\sigma_y$) are (0.05, 0.06) pix,
(0.07, 0.15) pix, and (0.18, 0.17) pix, for the new, PBC, and pre-ship solutions,
respectively. The new solution was found to significantly improve positional 
measurements overall as compared to both of the previous solutions and in particular,
it is a factor of 3 better in the Y direction over the more recent PBC solution.
As mentioned above, the uncertainties in the distortion models have not been
removed from these values, as they are unknown for the two previous solutions.
For the new distortion solution, however, the RMS offsets are consistent with the
average uncertainty in the distortion model itself.

The errors computed using the pairwise analysis are approximately half the size
of those reported using the GC observations at two position angles.
We note that the pairwise analysis uses measurements that have uncertainties
that are a factor of 3 larger than our other test and is therefore more
sensitive to the removal of the measurement bias term.
We take as the final residual
error term for the new NIRC2 distortion solution the value from the first test:
($\sigma_{x,0}$, $\sigma_{y,0}$) = (0.11, 0.10) pix.

\subsection{Additional Sources of Uncertainty}
\label{sec:otherSrcs}
While the new distortion solution represents a significant step forward
in our astrometric capabilities, it still leaves $\sim$0.1 pix or 
$\sim$1 mas residual distortion in LGSAO images that are widely
dithered or taken at different position angles. 
The residual distortion is twice as large as the estimated uncertainties
in the distortion solution
($\sim$0.05 pix; Figure \ref{fig:monteHist}), and must come from sources of 
uncertainty that are not accounted for in our analysis. Below we consider
two possibilities, time-variable distortion, and the difference between
NGSAO and LGSAO observations.

To test the stability of the camera's distortion, we created a 
distortion solution with data points from 2007, the year with the most
data (N=1711).  A smoothing factor of $f=120$ was used for the spline fitting 
and was determined in the same manner as our new distortion solution.  
As the number of data points from each of the years 2008 (N=253) and 
2009 (N=489) was not
sufficient to make separate distortion solutions for these years, we take the 
differences between the 2007-only distortion solution and the actual measured 
data from each of the individual years (see Table \ref{tab:m92_obs_all}).  
We find no significant differences (0.05$\pm$0.30 pix and 
0.09$\pm$0.29 pix for 2008 and 2009, respectively), suggesting that the 
distortion solution is stable within our measurement uncertainties. 
As a check, comparing the data from 2007 to the 2007-only solution gives
an average difference of 0.01$\pm$0.22 pix, which has a smaller RMS error
since it was the data set used to create the single-year model. Based on this
analysis, we conclude that there is no evidence for time-dependent changes.

While we tested our distortion solution on LGSAO data, the model itself
was computed using only NGS data, as the six LGS
frames from 2008 June were thrown out based on the cuts mentioned in 
\S\ref{sec:newDist}.  To test the possibility that the NGS and LGS AO
systems have different distortion solutions, we compare Galactic center data 
taken in both LGS and NGS modes, but otherwise the same setup and in the same 
night in 2008 May. The data were reduced using the usual data reduction
steps \citep[see ][]{ghez08}, and final LGS- and NGS-only images of the Galactic 
center were produced. The astrometric precision for each of these images was 
0.007 pix (NGS) and 0.012 pix (LGS) for stars with K$<$15.
The NGS image was transformed into the LGS image's coordinate system 
allowing only for translation between the two frames. Differences between
the transformed positions would indicate a possible difference in the
distortion between the LGS and NGS observing modes.
The RMS difference in the aligned positions, corrected for measurement
error bias, is only ($\Delta_x,\Delta_y$) = (0.06, 0.05) pix 
(1$\sigma$) and is therefore comparable to the error in the distortion model
($\sim$0.05 pix, Figure \ref{fig:monteHist}).
Thus, given the uncertainties in the distortion solution, we do not see a 
difference in the astrometry from images taken in NGS or LGS mode and conclude
that this is a negligible contribution to the residual distortion.

While we have not identified the source of residual distortion, we can 
functionally include it by adding a constant term of 0.1 pix (1 mas) in quadrature
with the error map of the distortion (see \S\ref{sec:newDist}) when analyzing 
astrometric data. We note that while these error terms are important for our 
localization of Sgr A*-radio in our infrared reference frame (\S\ref{sec:gcRefFrame}),
they do not come into consideration for relative proper motion measurements based 
on data using similar observational setups.

\section{Application to the Galactic Center}
\label{sec:GC}
Here we apply the new geometric optical distortion model and
DAR corrections from \S\ref{sec:newDist} to Keck/NIRC2 observations of the Galactic 
center in order to construct a new IR reference frame (\S\ref{sec:gcRefFrame})
that is significantly more accurate and stable than 
those that have been made in the past (\S\ref{sec:cmprFrames}).  We also measure
the motion of the nuclear cluster in this well-defined reference frame and
generate a set of secondary infrared astrometric standards that are helpful
for doing astrometry over much smaller fields of view (\S\ref{sec:secondary}).

\subsection{Construction of an Infrared, Sgr A*-Radio Rest Frame}
\label{sec:gcRefFrame}
Measurements of seven SiO masers that are detectable in the radio and infrared
wavelengths are used to transform the IR maser mosaics into a Sgr A*-radio
rest frame.  At radio wavelengths, each of these masers has well measured
positions and velocities with respect to Sgr A* \citep[see, e.g.,][]{reid03,reid07}.
In this analysis, we use radio positions and proper motions from M. Reid
(private communication), who has improved the values compared to what is
published in \citet{reid07} by adding one more epoch of observations and by 
applying a correction for the effects of differential nutation. The radio
maser positions were propagated using the radio proper motion measurements
to create a star list at the epoch of each IR mosaic.
Each of the six infrared mosaics was aligned with a four-parameter model
(two-dimensional translation, rotation, and a single pixel scale) to the 
radio maser star list by minimizing the error-weighted net displacements, $D$, 
for the masers, where the infrared positional errors include the positional 
RMS errors (from the 3 subset images; see \S\ref{sec:gcKeck}), as well as 
errors from the distortion model (see Appendix \ref{app:distErr_mosaic}).
The net displacement and the weighting
scheme used are described in Appendix A of \citet{ghez08}. Errors in the 
transformation to the Sgr A*-radio rest frame in each epoch were determined using a
jack-knife sampling technique, in which one maser at a time is excluded from
the alignment. The various sources of error in
our astrometry are broken down in Table \ref{tab:error_breakdown}.
Distortion errors generally dominate the individual IR positional uncertainties of 
these masers, with the exception of IRS 12N and IRS 28, each of which are in only 
one pointing and are furthest from the tip-tilt star, where the AO corrections 
are the poorest. All the transformed IR positions
agree with the radio positions to within $\sim$1$\sigma$ of each other, suggesting
that our uncertainties are well characterized and that we are not
missing large systematic error sources.

The NIRC2 pixel scale and orientation values obtained from the SiO 
maser alignment are similar to those obtained from the M92 
study (Table \ref{tab:plateScaleAngleCmpr}).  We note, however, that the 
RMS scatter shows a larger variation between the epochs
than the uncertainty inferred from the jack-knife analysis of each epoch.
We therefore take the RMS values as our estimates of the uncertainties 
for our average pixel scale and orientation angle
given in Table \ref{tab:plateScaleAngleCmpr}.
The weighted average NIRC2 plate scale and angle offset from the IR to 
radio alignments are 9.953 $\pm$ 0.002 mas pix$^{-1}$ and 0.249 $\pm$ 0.012$^o$, 
respectively. We average the results from the two methods 
(SiO masers and M92) to obtain our 
final values for the NIRC2 pixel scale and orientation angle,
9.952 $\pm$ 0.002 mas pix$^{-1}$ and 0.252 $\pm$ 0.009$^o$, respectively.
Thus, 0.252$^o$ must be added to the presumed PA (ROTPOSN - INSTANGL)
in order to get the true PA of a NIRC2 image (i.e., the NIRC2 columns
must be rotated eastward of North by 0.252$^o$).

\begin{figure}
\epsscale{1.0}
\plotone{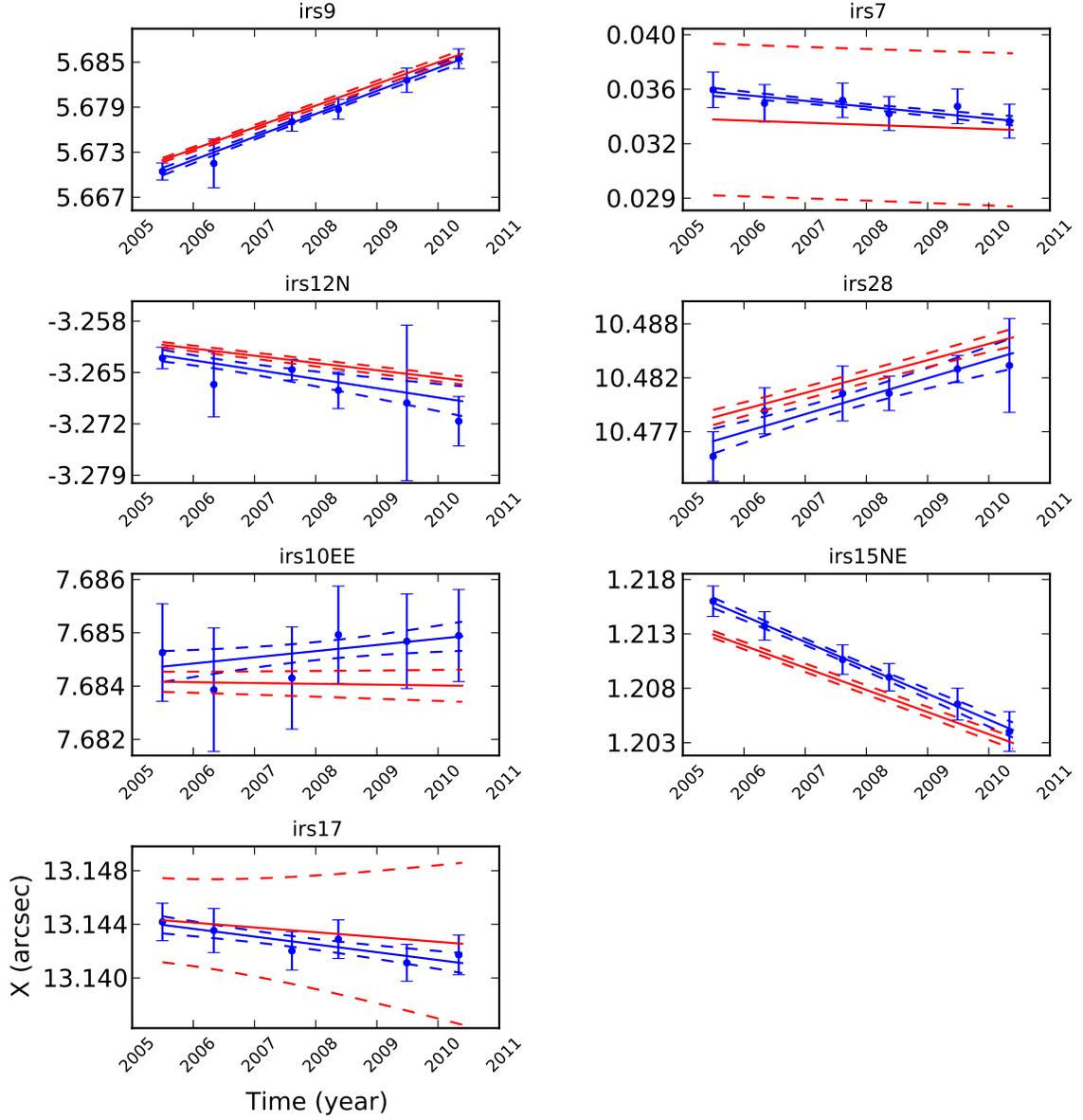}
\caption{
Absolute X positions (where X increases to the east) of Galactic center 
SiO masers in the infrared as a
function of time and the velocity model fit ({\em blue}) and the proper 
motion model for the radio ({\em red}). The 1$\sigma$ errors on the line 
fits are shown as dashed lines. The IR positional errors shown include
centroiding and alignment errors. Radio proper motion measurements are taken
from M. Reid (private communication). 
}
\label{fig:maserXvel}
\end{figure}

\begin{figure}
\epsscale{1.0}
\plotone{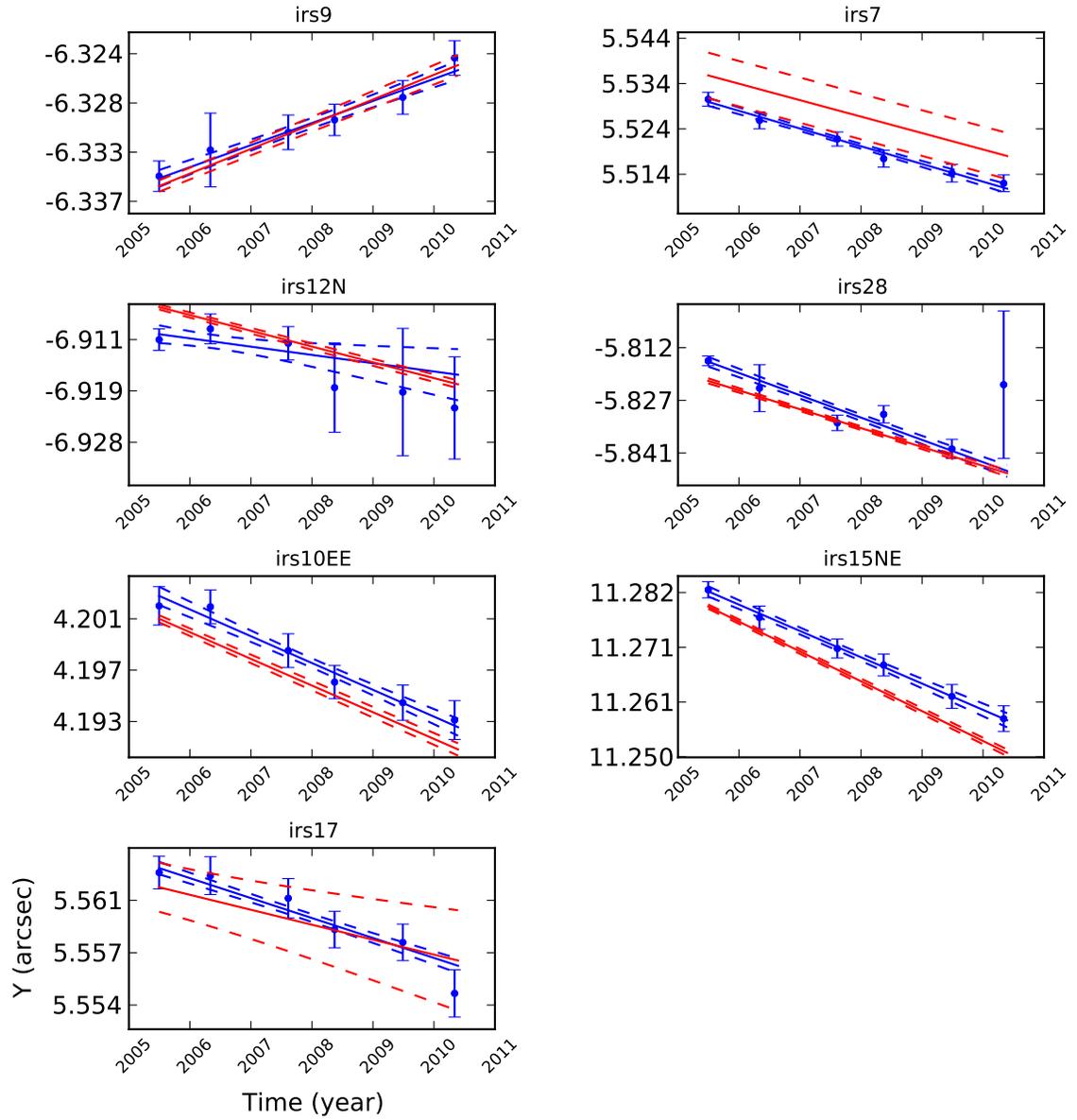}
\caption{
Same as Figure \ref{fig:maserXvel} but for Y positions (where Y increases to the
north).
}
\label{fig:maserYvel}
\end{figure}

Our transformed IR positions from the mosaic images provide a calibrated 
astrometric reference 
frame in which Sgr A* is at rest at the origin.  Comparison of the SiO 
masers as measured in the IR and radio provide estimates of how well we 
can localize the position and velocity of Sgr A*-radio within this reference 
frame. For each maser, a linear motion model is obtained by fitting a line
to the star's transformed infrared positions as a function of time. 
In this initial step, the positional uncertainties include only the 
centroiding and alignment errors; the distortion uncertainty 
(see Appendix \ref{app:distErr_mosaic}) is omitted 
here, since it is correlated across all epochs. 
Furthermore, the model fit is calculated with respect to $T_{0,IR}$, which is
the average time of the IR positional measurements, weighted by the average of
the X and Y positional uncertainties. Figures \ref{fig:maserXvel} and 
\ref{fig:maserYvel}
show the resulting fits, which have an average reduced $\chi^2$ value of
0.62 (see Table \ref{tab:maser_polyfit}), and the uncertainties in the
resulting fit parameters were determined from the covariance matrix.  
In these figures, and in all other cases, X and Y increase to the east
and north, respectively.
While the reduced $\chi^2$ values suggest that the uncertainties may be
overestimated (possibly due to IRS 7; see Appendix \ref{app:irs7}), we err
on the conservative side and do not re-scale our positional measurements.
The velocities measured in the infrared are statistically consistent with 
the radio proper motion values\footnote{We note that using the latest radio 
values reduces the uncertainty in tying the IR and radio measurements, 
reported in Table \ref{tab:maser_polyfit}, by 40\% compared to the same 
analysis carried out using the radio values from \citet{reid07}.}.
Based on the weighted average of the velocity differences between the 
infrared and radio reference frames, we conclude that Sgr A* is at rest to within 
$\sim$0.09 mas yr$^{-1}$ (compared to $\sim$0.03 mas yr$^{-1}$ in the radio reference 
frame).

While increasing the time baseline and depth of the infrared measurements will
improve upon this precision, we note that four out of 14 one-dimensional relative 
velocity measurements are already limited by the radio measurements.  
Further improvements in the radio will ultimately be required to create a 
reference frame that is more stable than 0.03 mas yr$^{-1}$ (the current limits from the
radio measurements alone).

\begin{figure}
\epsscale{1.0}
\plotone{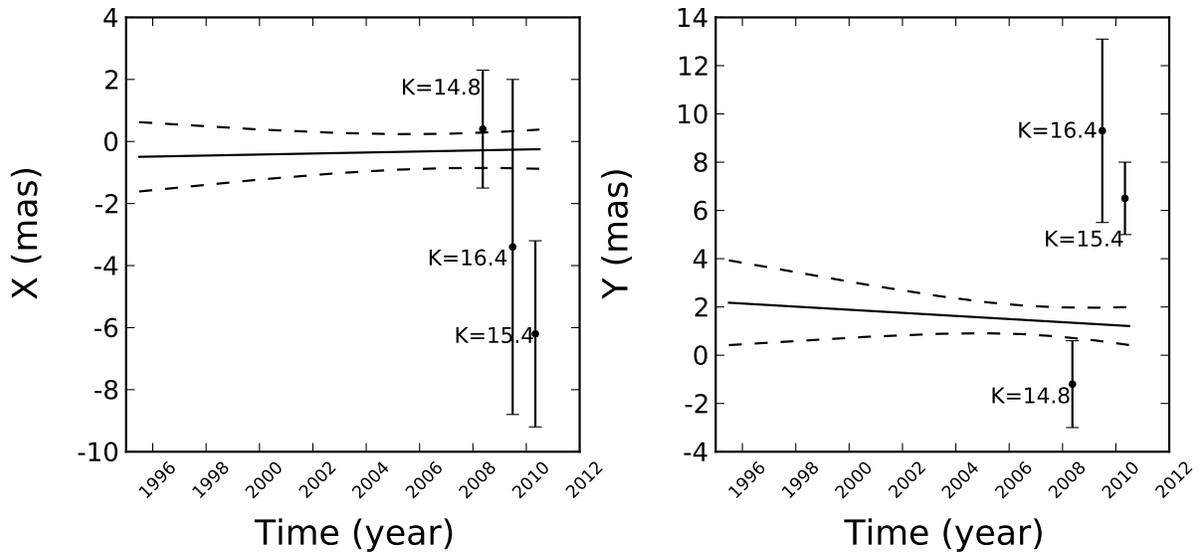}
\caption{
Position of Sgr A*-radio versus time in the IR reference frame based on
analysis in \S\ref{sec:gcRefFrame} and values in Table \ref{tab:maser_polyfit}
(dotted lines show 1$\sigma$ uncertainties).
Detected positions of Sgr A* in the infrared maser mosaics from 2008 May,
2009 June, and 2010 May are overplotted along with the magnitude of Sgr A*-IR.  
We note that the fainter detections may suffer from larger astrometric
biases from underlying sources. Nonetheless, all three IR positions agree
with the radio position of Sgr A*. In the IR reference frame, 
Sgr A*-radio is consistent with being at rest, at the origin.
}
\label{fig:sgraPosT}
\end{figure}

In order to compare the IR and radio positional measurements, two 
additional steps are required. First we determine the time,
$T_{0,(IR+radio)}$, at which the positional difference is expected to have
the smallest uncertainty for each maser,
\begin{equation}
\label{eq:t0}
T_{0,(IR+radio)} = \frac{\sigma^2_{v,IR}T_{0,IR} + \sigma^2_{v,radio}T_{0,radio}}
{\sigma^2_{v,IR} + \sigma^2_{v,radio}}
\end{equation}
where $\sigma_{v,IR}$ and $\sigma_{v,radio}$ are the velocity errors in the 
IR and radio, respectively. We take the average of these seven times, 2006.9,
and find the IR and radio positions and uncertainties at this common epoch. 
Then the correlated distortion error (including both the uncertainty in 
the model and the residual distortion, $\sim$0.1 pix) for each maser is added in 
quadrature to the formal uncertainty from the infrared fit.
Comparison with the radio positions indicates that the position of Sgr A*-radio
is known to within $\sim$0.57 mas in year 2006.9 in our infrared reference frame. 
We note that the localization of Sgr A*-radio in the IR reference frame is
time-dependent (Figure \ref{fig:sgraPosT}).
Decreasing the impact of uncertainties in the IR distortion model, with either
more highly dithered measurements or a better distortion model, would improve
this precision.  Overall, our current measurement uncertainties for both the 
position and velocity of Sgr A* are a factor of 3-4 better than earlier 
measurements - either those of \citet{ghez08}, when treated in the same
manner\footnote{We reported errors from a half-sample
bootstrap in \citet{ghez08}.  To compare values we reran our analysis with the 
half-sample bootstrap, which overestimates the uncertainties since half 
the sample is removed.}, or those reported by \citet{gillessen09} using their
"maser system" method, which is comparable to the method used 
here\footnote{We note that \citet{gillessen09} derive their astrometry using
two distinct methods, and adopted the positional errors from one method (the
"cluster system") and velocity errors from the second method (the "maser system").}.

Sgr A* is detected in three of the IR maser mosaics and its position is consistent
with Sgr A*-radio in our IR reference frame (Figure \ref{fig:sgraPosT}). 
In 2008 May, Sgr A* was as bright
as K=14.8 mag, which is one of the brightest IR detections of Sgr A* 
\citep[see e.g.,][]{do09sgra,sabha10}. In 2009 June and 2010 May, it was detected with
K=16.4 and K=15.3, respectively. The magnitude of Sgr A* in 2009 is more in line
with the faint end of what is observed for this highly variable source. The later 
detections were possible because the mosaics were deeper than the previous mosaics.
No other mosaics show Sgr A* since these observations are composed of very short
exposures to avoid saturation on the infrared-bright SiO masers.  
Figure \ref{fig:sgraPosT} shows that all three detections are consistent 
within 3$\sigma$ in X and Y with the position of Sgr A*-radio in the IR reference 
frame. Furthermore, the IR position is more consistent with the radio
position (within 1$\sigma$) when Sgr A*-IR is in a relatively bright state and
less prone to astrometric biases from underlying sources 
\citep[e.g.,][]{ghez08,gillessen09}. This independent comparison confirms
that we have a well constructed reference frame.  

By carrying out the same analysis in polar coordinates (as opposed to cartesian),
we find that the infrared and radio Sgr A*-rest coordinate systems
show no net relative expansion ($V_r$) nor rotation ($V_t/R$) to within
0.12 mas yr$^{-1}$ and 0.11 mas yr$^{-1}$ arcsec$^{-1}$ (1$\sigma$), respectively 
(Table \ref{tab:maser_polyfit_polar}).

\subsection{Comparison of Sgr A*-Rest Reference Frame vs. Cluster-Rest
Reference Frame}
\label{sec:cmprFrames}
The Sgr A*-rest reference frame generated in \S\ref{sec:gcRefFrame} is
more stable than the cluster-rest reference frame 
that has been used as the principle coordinate system for all previous 
proper motion studies.  Here we are quantifying frame stability as the 
uncertainty in the velocity of the object that is defined to be at rest.  
Since the cluster-rest frame previously used is defined by assuming that a 
set of reference stars has no net motion, the translational stability of this 
reference frame is limited to be $\sigma / \sqrt{N}$, where $\sigma$ is the 
intrinsic dispersion of the stars and $N$ is the number of reference stars used. 
With a dispersion in the plane of the sky of roughly 2.6 mas yr$^{-1}$ 
\citep[][see also \S\ref{sec:secondary}]{trippe08,schodel09}, $\sim$850 reference 
stars would be needed in order to match the stability of our current Sgr A*-rest 
frame. This is a factor of $\sim$2 to 10 more than have been used in earlier 
studies \citep[e.g.,][]{trippe08,ghez08,lu09,gillessen09,schodel09}. 

\begin{figure}
\epsscale{1.0}
\plotone{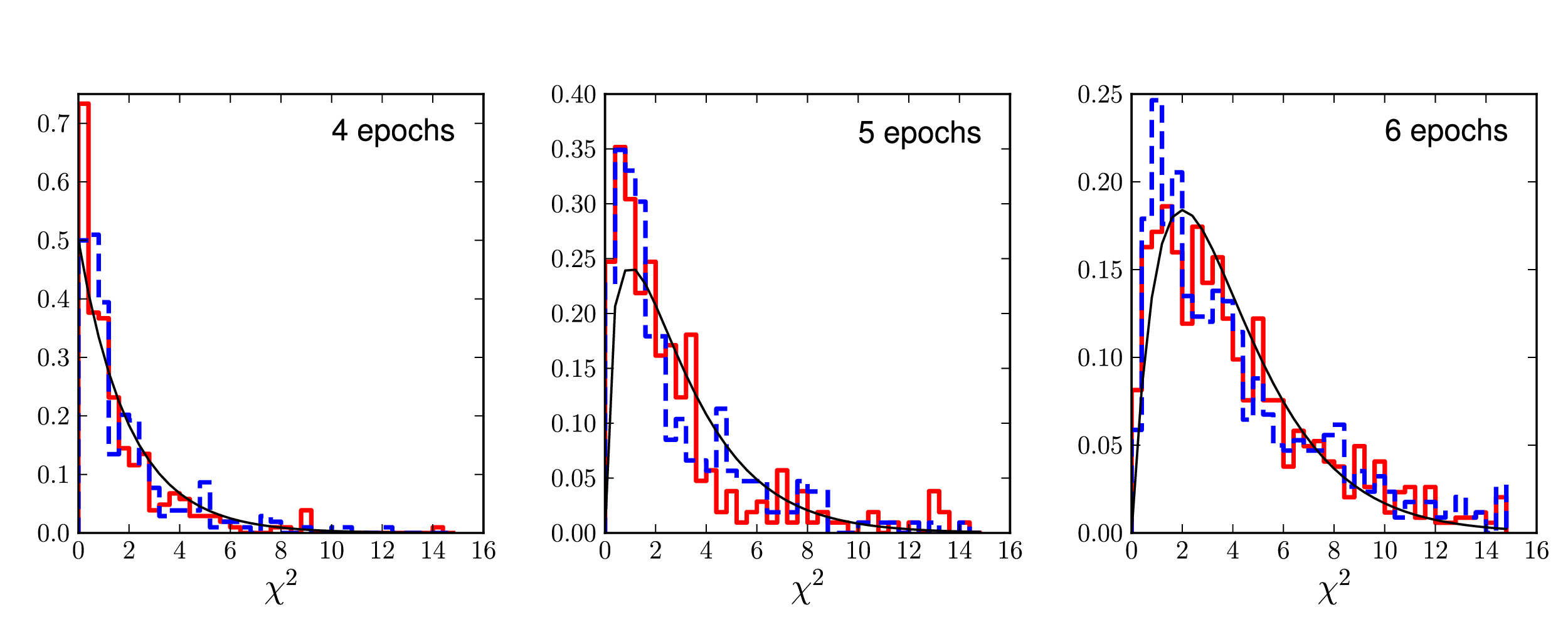}
\caption{
Histogram of velocity $\chi^2$ values in X ({\em solid red}) and 
Y ({\em dashed blue}) for the stars detected in four (N=263), 
five (N=269), and six (N=912) maser mosaics. The theoretical
$\chi^2$ distributions for the corresponding degrees of freedom are 
overplotted as thin black curves for comparison.  We find that the velocities are 
well-behaved, as indicated by the similarity between the observed and 
theoretical distributions.
}
\label{fig:chi2hists_msr}
\end{figure}

\begin{figure}
\epsscale{1.0}
\plotone{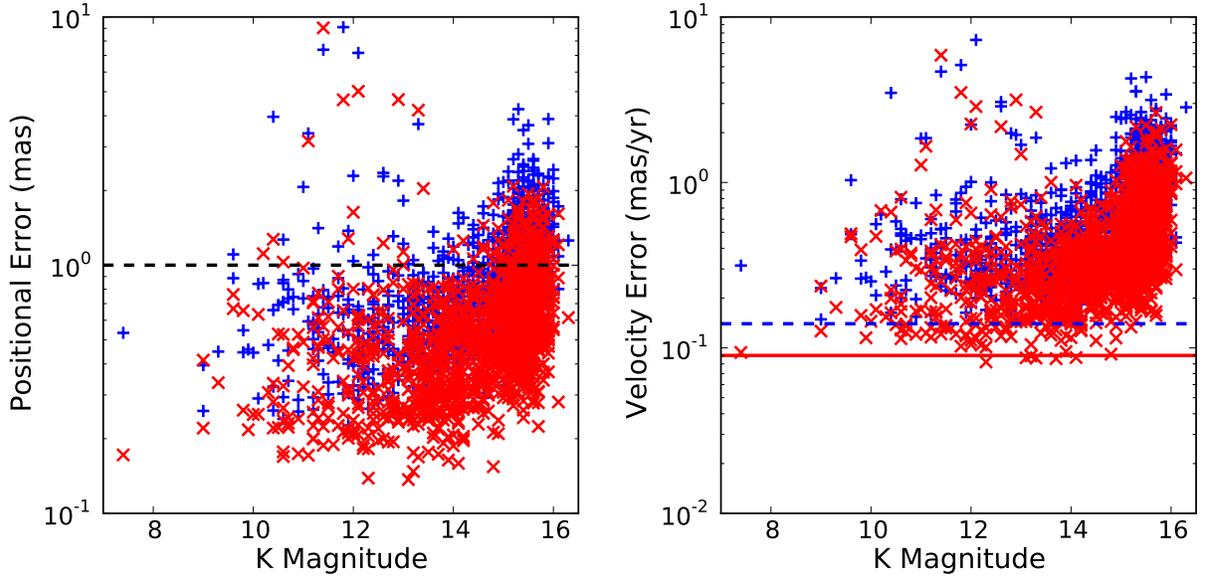}
\caption{
Positional ({\em left}) and
velocity uncertainties ({\em right}) for X ({\em red crosses}) and 
Y ({\em blue plus signs}) versus K magnitude for all stars detected in
at least four of the six maser mosaics (N=1445).
We select stars with velocities less than 10 mas yr$^{-1}$ and velocity errors 
less than 1.5 mas yr$^{-1}$ in both the X and Y directions as our infrared 
astrometric standards, resulting
in a total of 1279 stars. The dashed line in the left-hand plot shows the
level of residual distortion (1 mas) which is added in quadrature
to the positional errors in Table \ref{tab:secondary_short}.  The dashed lines 
in the right plot indicate the level at which the velocity of Sgr A* is known 
in the X ({\em solid red}) and Y directions ({\em dashed blue}), based on the 
IR to radio offsets of the masers 
found in \S\ref{sec:gcRefFrame} and Table \ref{tab:maser_polyfit}.
}
\label{fig:errs_msr}
\end{figure}

\begin{figure}
\epsscale{1.0}
\plottwo{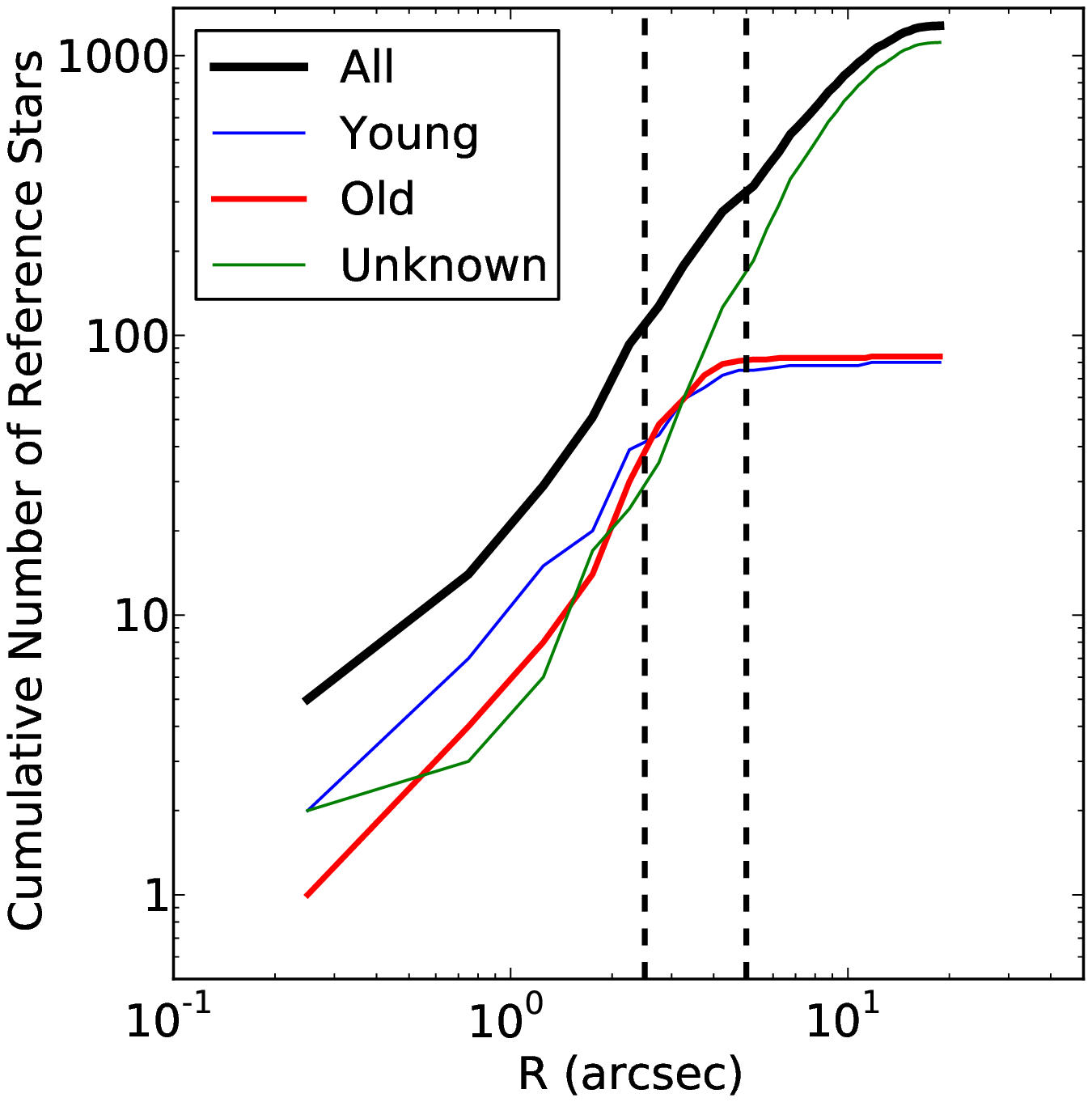}{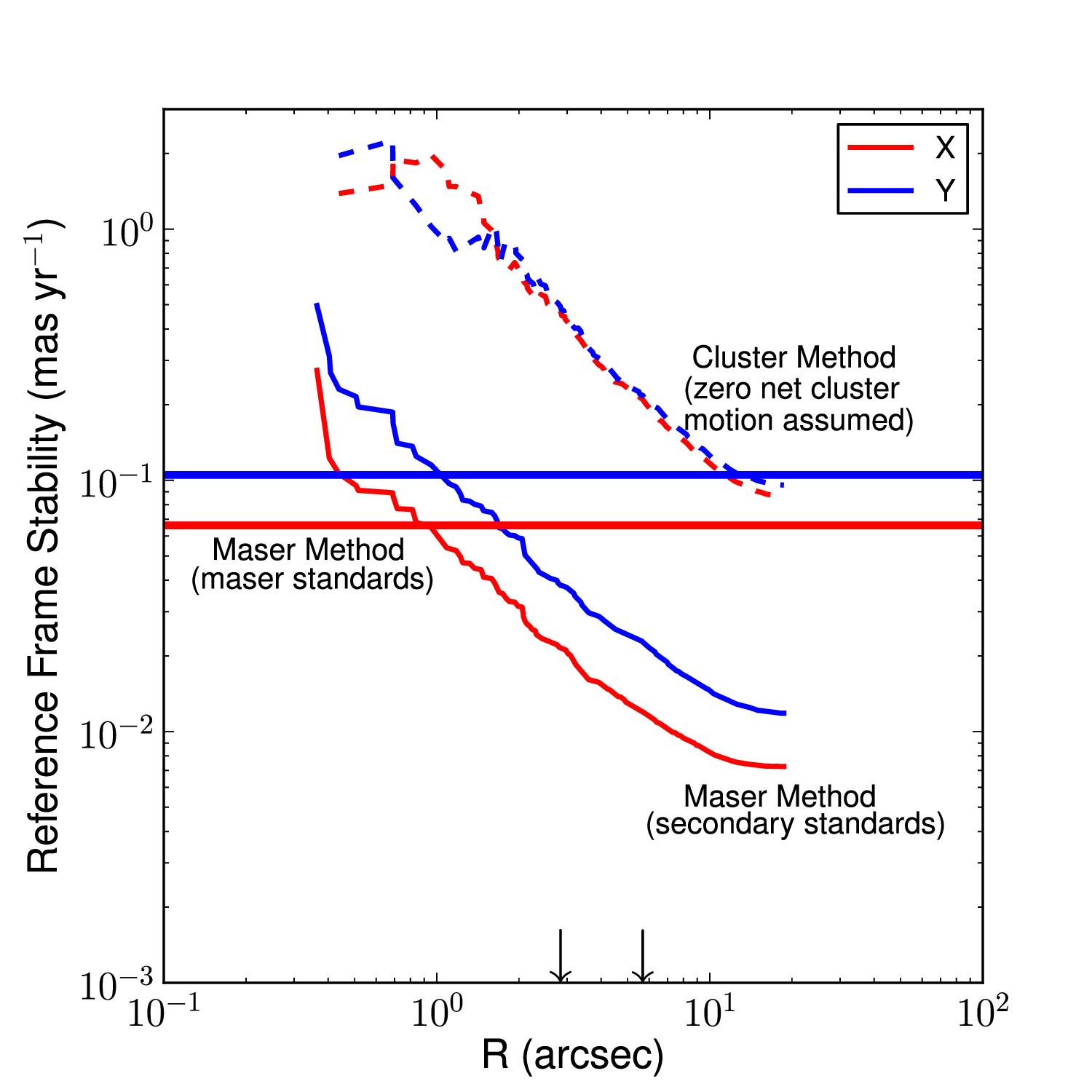}
\caption{
{\em (Left)} Cumulative radial distribution of all 1279 astrometric reference 
stars ({\em black}), as well as the distribution of stars based on
spectral identification:
young (N=77; {\em blue}), old (N=73; {\em red}), and
unknown (N=1129; {\em green}).
The field of view for both the speckle (R=2.5$\arcsec$) and the adaptive 
optics (R=5$\arcsec$) cameras are indicated ({\em dashed lines}).
The speckle data (1995-2005) are dominated by young stars, and thus, 
using them as reference stars in the coordinate transformations
is critical.  Stars with unknown spectroscopic identifications dominate at
larger radii.
{\em (Right)} Error on the weighted average velocity of the cluster stars
in X ({\em red}) and Y ({\em blue}) as a function of distance from 
Sgr A* as measured using two distinct methods. 
The curves show the improvement in the errors as more
reference stars are included at larger radii.  Using the 'cluster
method' and excluding young stars ({\em dashed curves}), the assumption 
of no net cluster motion is made, resulting in a reference frame that 
is stable to $\sim$0.2 mas yr$^{-1}$ at a radius of 5$\arcsec$ 
(corresponding to the field of view of Keck AO data). 
In a reference frame where Sgr A* is at rest, the 'maser method', the 
error on the weighted average velocity of the cluster stars is $\sim$0.02 
mas yr$^{-1}$ over the extent covered by the AO data ({\em solid curves}).
The total error in the weighted average velocity from the method described
in \S\ref{sec:cmprFrames}, which includes contributions from the cluster stars,
as well as the masers in both the radio and infrared ({\em horizontal lines}),
represents the stability of our reference frame (0.09 mas yr$^{-1}$).
}
\label{fig:ref_radial}
\end{figure}

The stability of reference frames for observations made with fields of view that 
are too small to tie into the masers directly can be significantly improved by 
using secondary astrometric standards generated by the proper motion measurements 
for infrared stars other than the masers from the measurements presented in 
\S\ref{sec:gcRefFrame}. To create these secondary standards, we use the same linear 
motion modeling done for the masers and estimate the positions and proper motions 
of stars that are detected in at least four of the 6 maser mosaics (N=1445). 
These stars have K' magnitudes that are brighter than 16 mag and a $\chi^2$ 
distribution that is consistent with their positional uncertainties and degrees of 
freedom (Figure \ref{fig:chi2hists_msr}). From these, we select the
1279 stars that have velocities less than 10 mas yr$^{-1}$ (to exclude mismatches) 
and velocity errors less than 1.5 mas yr$^{-1}$ in both the X and Y directions 
(Figure \ref{fig:errs_msr}). The positions and proper motions of these stars are 
reported in Table \ref{tab:secondary_short}, and the left panel of 
Figure \ref{fig:ref_radial} shows the cumulative distribution as a function 
of radius for the entire sample, as well as for those known to be old and young.  
Reference frames defined based on these secondary astrometric standards 
(which can be young or old), as opposed to one defined on the premise that the 
old stars have no net motion, are significantly more stable, and the exact 
advantage depends on the field coverage; for the Keck 
speckle and 10$\arcsec\times$10$\arcsec$ AO data \citep[see, e.g.,][]{ghez08,lu09}, 
the translational stability is expected to be a factor of 17.5 and 12.5 times 
better, or 0.03 and 0.02 mas yr$^{-1}$, respectively (see right panel of Figure 
\ref{fig:ref_radial}). 
Similarly the rotational stability, quantified as the uncertainty in the average 
rotational velocity, is expected to be a factor of 20 and 9 times better, for the
Keck speckle and AO data, or roughly 0.03 and 0.02 mas yr$^{-1}$ arcsec$^{-1}$, 
respectively.
This improvement has been seen and will be presented in a separate forthcoming 
paper \citep[see also][conference proceedings showing these results]{yelda10}.

\subsection{Motion of the Central Stellar Cluster in a Sgr A* Rest Frame}
\label{sec:secondary}
\begin{figure}
\epsscale{1.0}
\plottwo{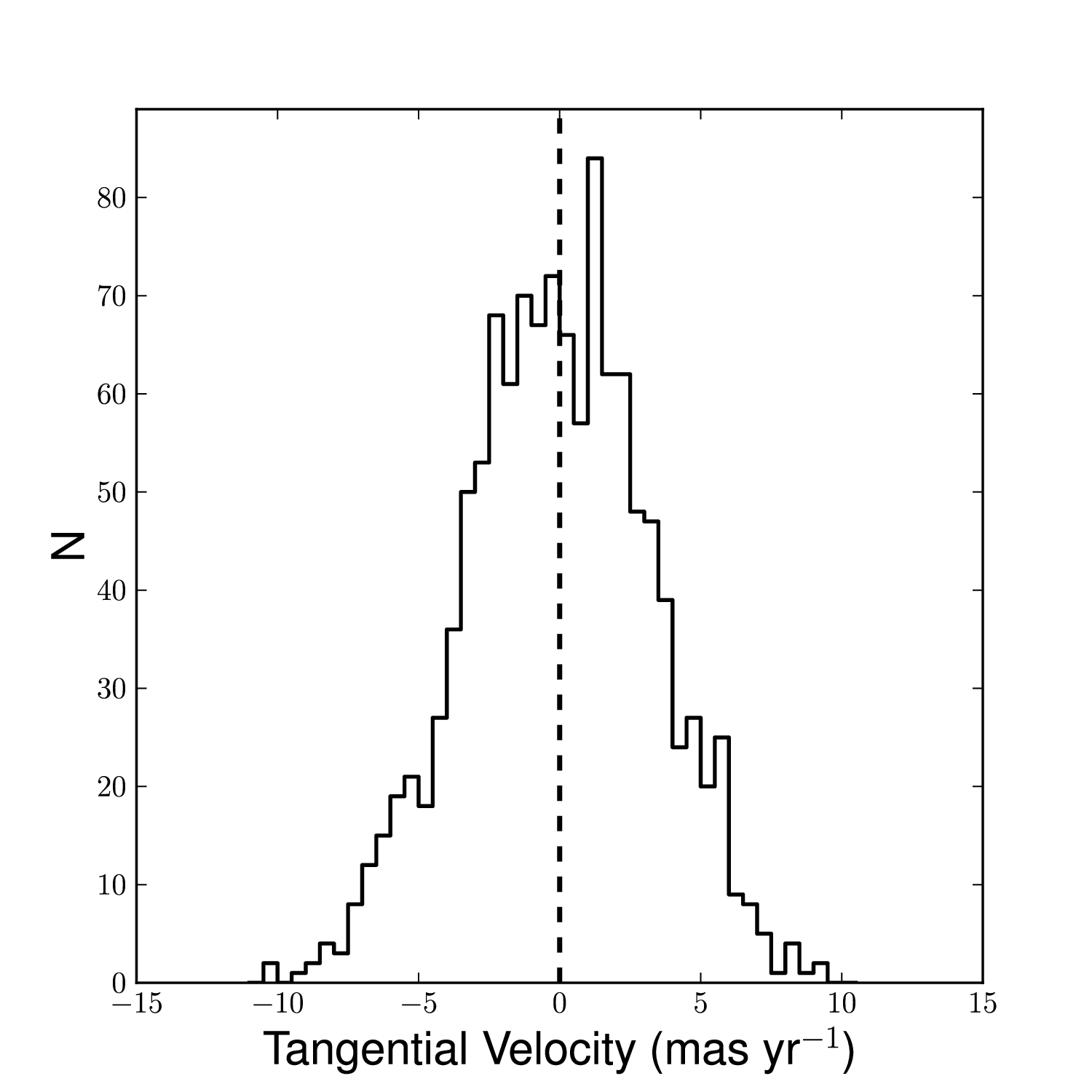}{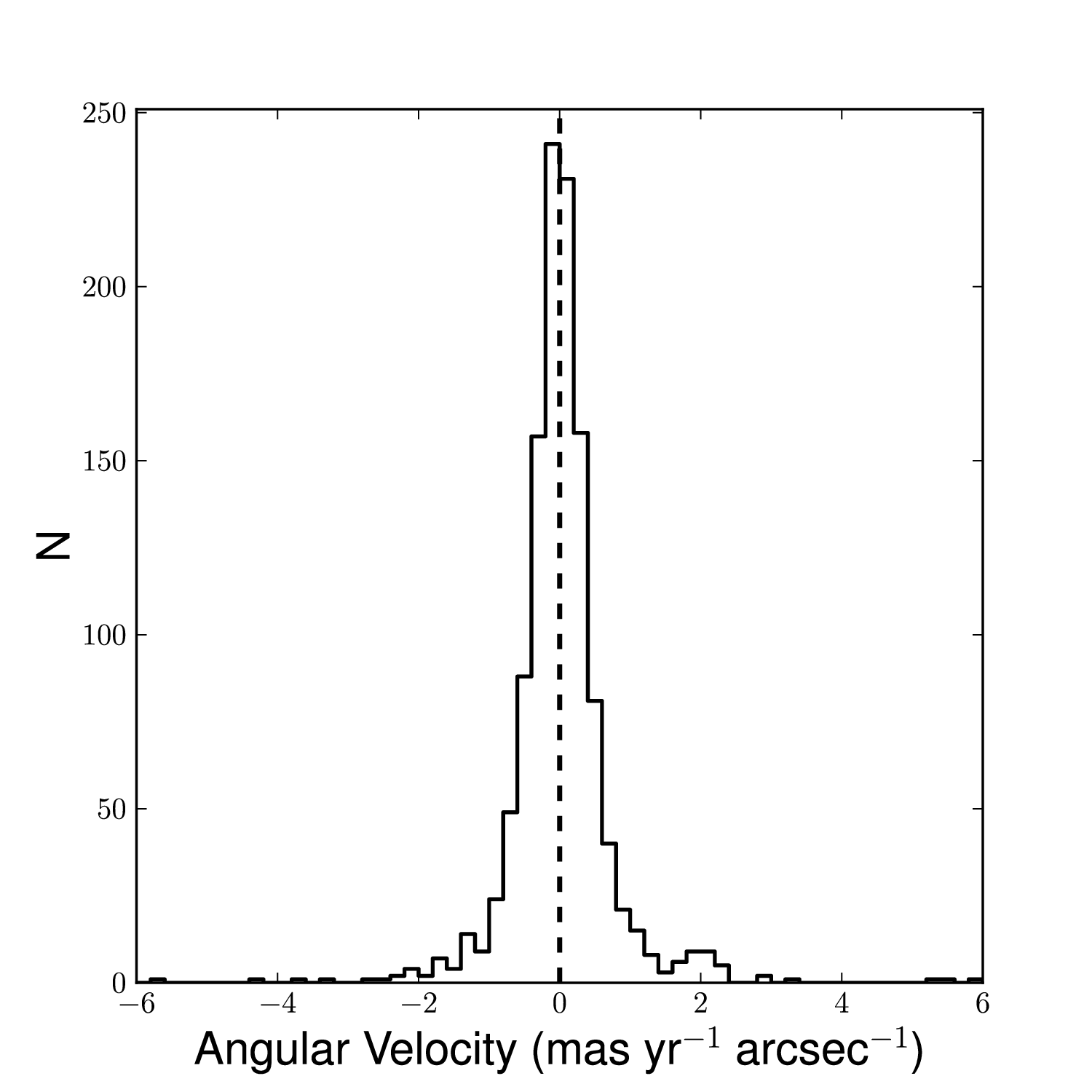}
\caption{
({\em Left}) Histogram of tangential velocities of the astrometric 
reference stars (excluding known young stars). The tangential
motion in the plane of the sky is consistent with zero mas yr$^{-1}$ 
({\em dashed line}).
({\em Right}) Histogram of angular velocities of the 
reference stars (excluding known young stars). The angular velocity
is consistent with zero mas yr$^{-1}$ ({\em dashed line}).
}
\label{fig:polarVels}
\end{figure}

\begin{figure}
\epsscale{1.0}
\plotone{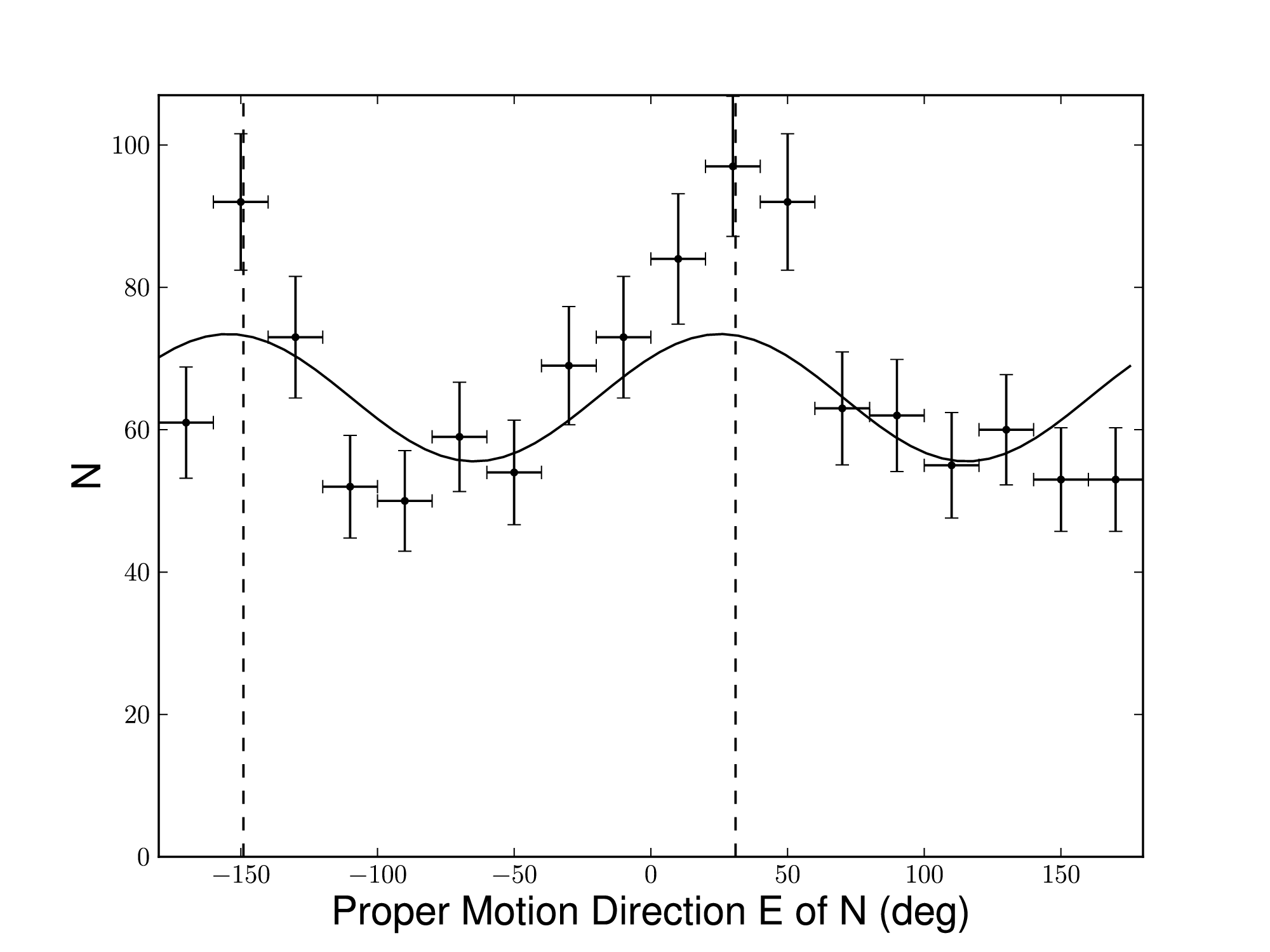}
\caption{
Histogram of proper motion direction, defined as the angle east of north, of the 
infrared astrometric standards in the central $\sim$1 pc.  The vertical error bars  
are Poisson errors, while the horizontal bars denote the width of the histogram bins.
The angle of the Galactic plane is shown as the dashed line ($\sim$31$^o$ and
180$^o$ opposite).  The data are best fit with a cosine curve which peaks at
25.4$^o$ $\pm$ 16.3$^o$, which is consistent with the angle of the Galactic
plane. 
}
\label{fig:pln_of_sky_rot}
\end{figure}

\begin{figure}
\epsscale{1.0}
\plotone{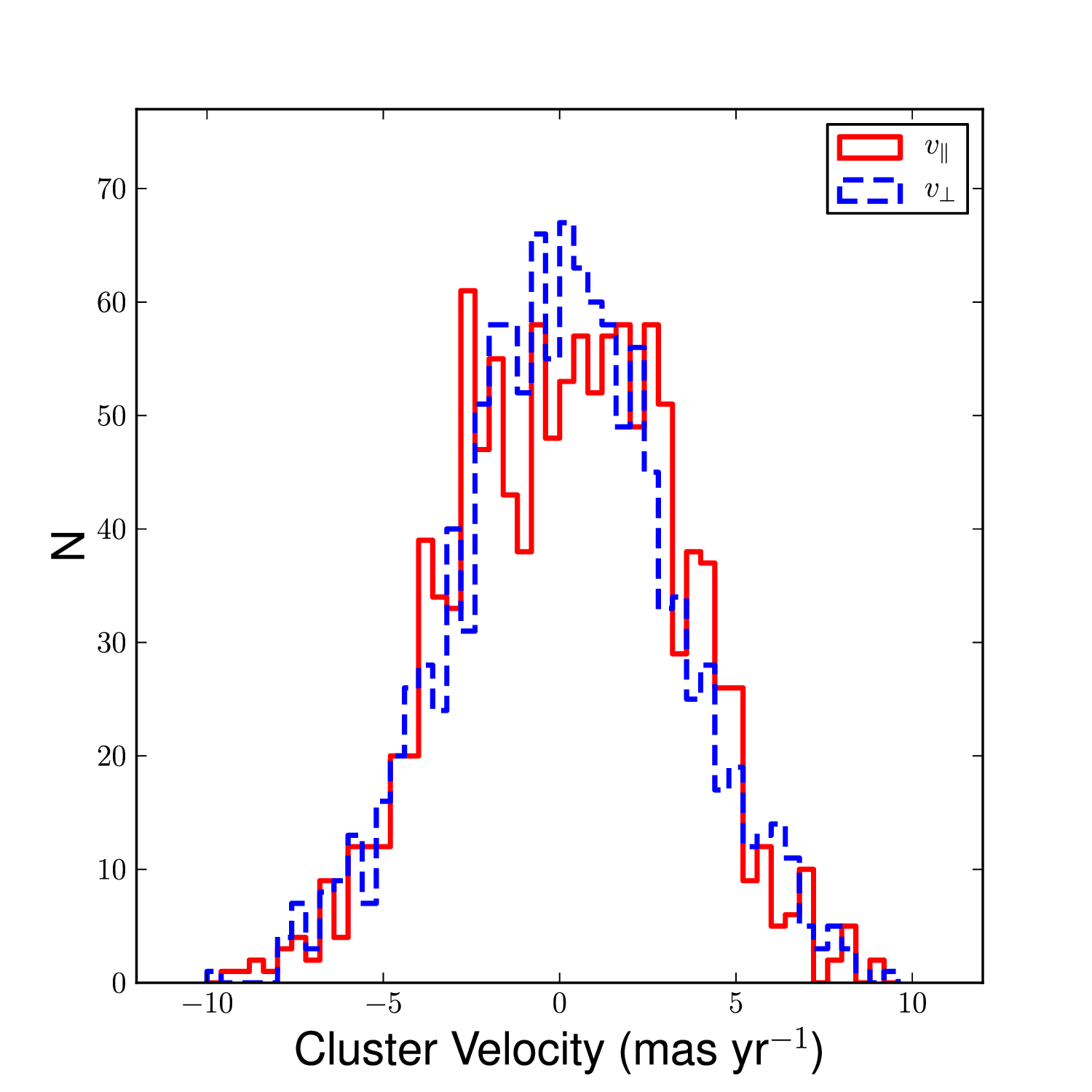}
\caption{
({\em Left}) Reference star velocities parallel ({\em solid red}) and 
perpendicular ({\em dashed red})
to the Galactic plane ($\theta$ = 31.4$^o$). The flattening of the
$v_{\parallel}$ distribution is due to the rotation of the cluster stars
along the Galactic plane.  
}
\label{fig:parPerpVels}
\end{figure}

\begin{figure}
\epsscale{1.0}
\plotone{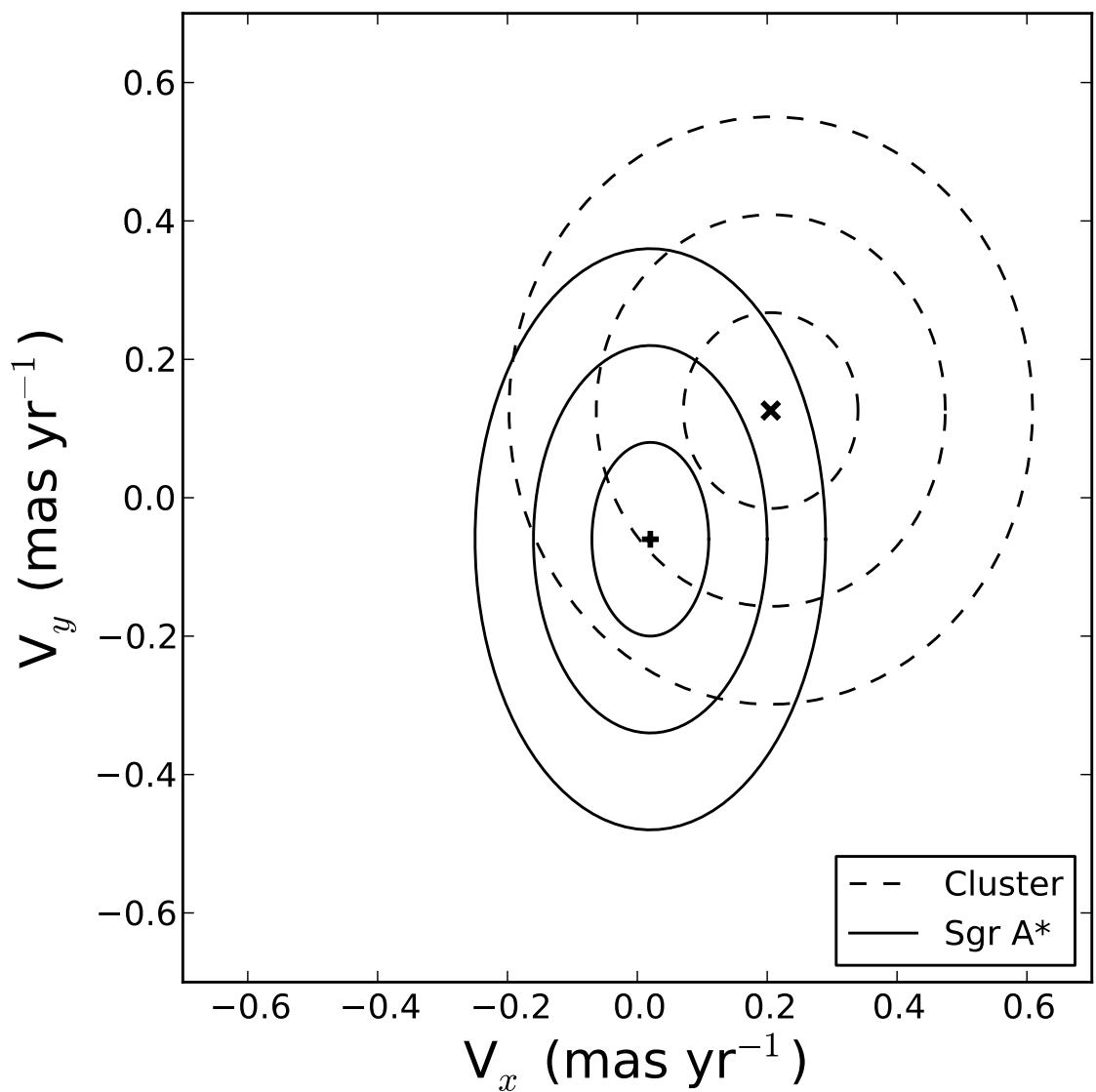}
\caption{
Velocity of Sgr A* in the infrared reference frame ({\em plus sign}) 
with 1, 2, and 3$\sigma$ contours shown ({\em solid curves}),
as compared to the cluster's weighted average velocity 
({\em cross; dashed curves}). V$_x$ and V$_y$ are defined such that positive
values are motions that increase in the east and north directions, respectively.
}
\label{fig:clusterVsSgrA}
\end{figure}

The proper motions for the secondary astrometric standards listed in Table
\ref{tab:secondary_short} also offer the first opportunity to study the kinematic
properties of the central stellar cluster directly in a Sgr A*-rest
frame. Since all previous proper motions have been made in the 
cluster-rest reference frame, any net rotation of the cluster in the plane of 
the sky is removed from these earlier measurements.
Rotationally, we find no motion in the plane of the sky in the tangential 
velocities (-0.09 $\pm$ 0.14 mas yr$^{-1}$) nor in the angular velocities 
(0.26 $\pm$ 0.36 mas yr$^{-1}$ arcsec$^{-1}$) about Sgr A* 
(Figure \ref{fig:polarVels}).  However, we do detect rotation in the plane 
of the Galaxy, as has been previously reported by both \citet{trippe08} and 
\citet{schodel09} and is shown in Figure \ref{fig:pln_of_sky_rot}, which shows
that there is a preferred angle for the proper motion vectors of 
25.4$^o \pm$ 16.3$^o$, consistent with the angle of the Galactic plane 
\citep[31.4$^o$;][]{reid04}. 
This rotation in the Galactic plane is also seen in the flattening of the 
distribution of velocities in the direction parallel to the Galactic plane
as compared to the velocities in the perpendicular direction 
(Figure \ref{fig:parPerpVels}).
Translationally, we find that the weighted average 
velocity of all the stars in the sample that are not known to be young (N=1202)
is 0.21 $\pm$ 0.13 mas yr$^{-1}$ ($\sim$8.0 $\pm$ 4.9 km s$^{-1}$ at 8 kpc) and 
0.13 $\pm$ 0.14 mas yr$^{-1}$ ($\sim$4.9 $\pm$ 5.3 km s$^{-1}$) in the X and Y
directions (where X and Y increase to the east and north), respectively. 
Figure \ref{fig:clusterVsSgrA} compares 
the mean translational motion of the nuclear stellar cluster to the motion of 
Sgr A* (as determined in \S\ref{sec:gcRefFrame}) and shows that there is
no relative motion between the cluster and the black hole.
The uncertainties quoted here are the RMS errors of the weighted average velocities 
from a bootstrap analysis with 10$^5$ trials, where in each trial
a random set of data was sampled (with replacement) from the observed cluster 
velocity distribution. 

\section{Summary \& Conclusions}
\label{sec:discuss}
We have improved upon existing geometric distortion solutions for the NIRC2 camera
at the W. M. Keck II telescope and have, for the first time, implemented 
DAR corrections to our Galactic center astrometry.  
In all tests that were performed, the new distortion solution shows an 
improvement by a factor of $\sim$2-4 over existing solutions. 
We take as our final residuals:
(${\sigma}_x$, ${\sigma}_y$) $\sim$ (0.11, 0.10) pix $\sim$ (1.1, 1.0) mas.
This is the error that is incurred when combining images taken at various offsets
and position angles.  
The transformations between the ACS/WFC and NIRC2 reference frames yield
a consistent plate scale and angle offset to that obtained using Galactic
center infrared data which are tied to the radio reference frame.  We find
an average plate scale and angle offset for the NIRC2 narrow camera of
9.952 $\pm$ 0.002 mas pix$^{-1}$ and 0.252 $\pm$ 0.009$^o$, respectively.

The new distortion solution and its associated uncertainty, in the form of 
FITS files, may be obtained at
\url{http://www.astro.ucla.edu/~ghezgroup/distortion}.
The FITS files, or look-up tables, may be fed into the IRAF 
routine {\em Drizzle} during the data reduction process.  The values in the 
look-up tables specify the shifts required to put an image into a 
"distortion-free" reference frame.  

As a result of the work presented here, Galactic center astrometry can now be 
tied to a Sgr A*-radio rest frame to better than $\sim$0.6 mas and 
$\sim$0.1 mas yr$^{-1}$ ($\sim$3.5 km s$^{-1}$ at 8 kpc) in position and velocity 
space, respectively, which is a factor of 3 improvement over earlier reported efforts.
We note that the velocity of Sgr A* along the line of sight is likely to be 
minimal ($\sim$3.5 km s$^{-1}$, 1$\sigma$) given the constraints on the motion in the 
plane of the sky. Since the cluster has been found to exhibit no net motion 
with respect to the local standard of rest (LSR) to within $\pm$5 km s$^{-1}$ 
\citep{figer03,trippe08}, this adds confidence in the estimates of the distance to
the Galactic center (R$_o$) from orbital analyses in which the black hole is
assumed to have no line of sight motion with respect to the LSR. 
With this assumption, R$_o$ estimates from comparable orbital analyses in 
\citet{ghez08}, \citet{gillessen09b} and \citet{gillessen09} 
have values of 8.4 $\pm$ 0.4 kpc, 7.7 $\pm$ 0.4 kpc, and 7.3 $\pm$ 0.5 kpc, 
respectively\footnote{The orbital analysis compared is the case of S0-2 only, 
no 2002 astrometric data, and priors only on V$_z$}.

We present a new set of infrared astrometric standards that can be used to 
define the reference frame (with Sgr A*-radio at rest, at the origin) in smaller
field of view Galactic center measurements that do not contain enough of the SiO 
radio masers and that are used for stellar orbit measurements. 
We measure the motion of the stellar cluster in a Sgr A*-rest frame and confirm
that the cluster rotates in the plane of the Galaxy.

A stable astrometric reference frame is a key requirement when using stellar
orbits to study the central supermassive black hole and its environment.
Stellar orbits have already proven to be powerful tools for measuring
the black hole's mass and distance, as well as placing limits on a black hole 
companion \citep{ghez08,gillessen09}.
In time, stellar orbital work will probe the extended mass distribution and
general relativity through the detection of prograde and retrograde precession, 
respectively \citep{rubilar01}.  \citet{weinberg05} considered the effect of
an assumed extended mass distribution within the orbit of the central arcsecond,
16-year period star, S0-2, and estimated an apocenter shift after one revolution 
of ${\Delta}s\sim$0.3 mas, which corresponds to an effect of 
$\phi\sim$0.02 mas yr$^{-1}$ 
(or an angular velocity of 0.08 mas yr$^{-1}$ arcsec$^{-1}$)\footnote{
We note that this is only an approximation as the amount of extended mass
within the orbit of S0-2 is highly unknown.}.
The prograde relativistic precession of S0-2, on the other hand,
is predicted to be $\phi\sim$0.06 mas yr$^{-1}$ \citep[or an angular velocity of 
0.27 mas yr$^{-1}$ arcsec$^{-1}$;][]{weinberg05}, thereby requiring a
reference frame that is stable to 0.02 mas yr$^{-1}$.
Detection of either the prograde or retrograde precession of the central
arcsecond sources will therefore require an extremely stable astrometric 
reference frame.

\begin{figure}
\epsscale{1.0}
\plotone{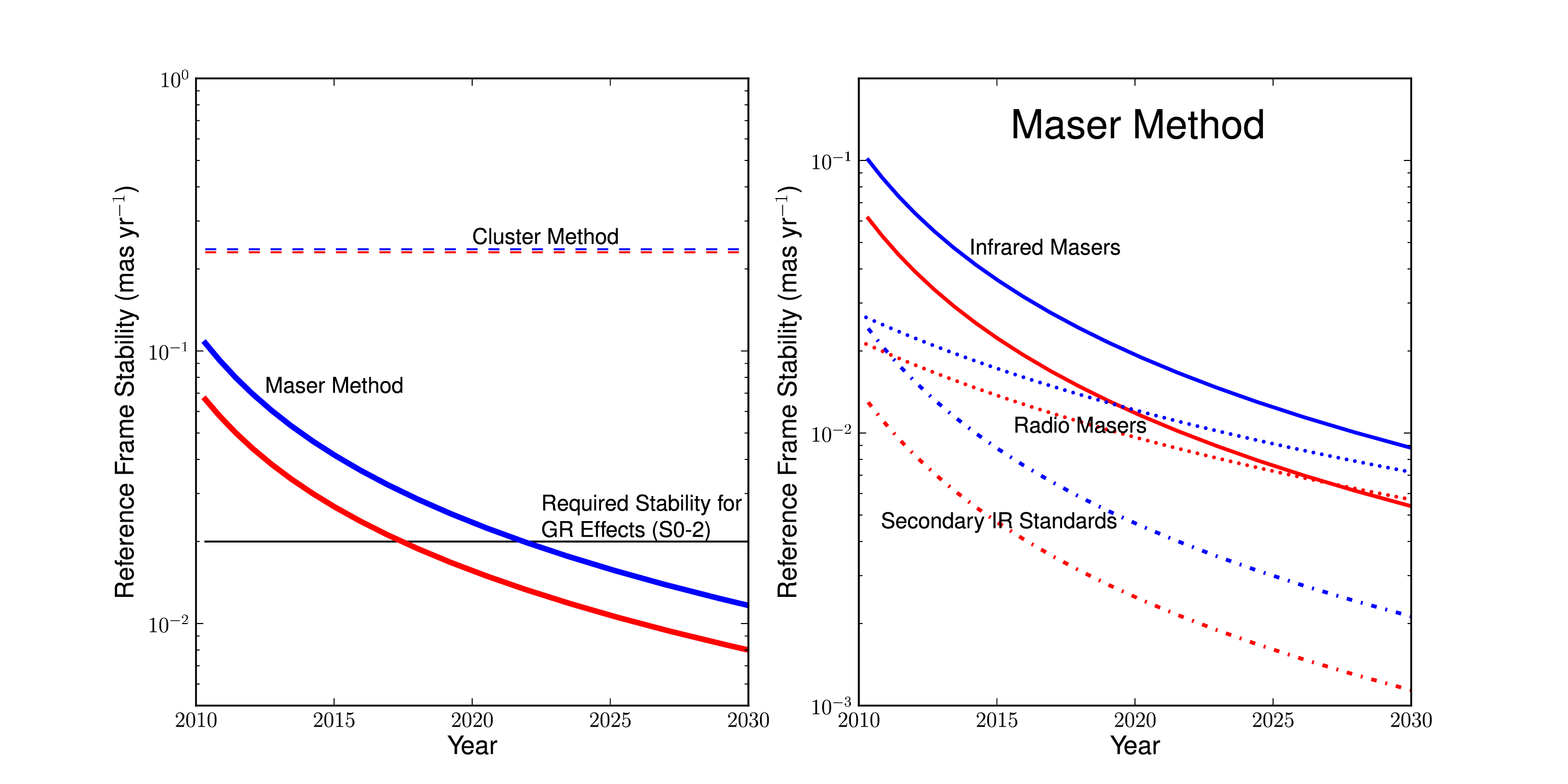}
\caption{
({\em Left}) Predicted stability in X ({\em red}) and Y ({\em blue}) of the 
reference frame with time (assuming the field of view of Keck AO images, 
10$\arcsec\times$10$\arcsec$). The maser method 
({\em solid curves}; \S\ref{sec:GC}) allows for an improvement in the stability
with time as t$^{-3/2}$, whereas the cluster method ({\em dashed lines}), 
which assumes no net motion of the stellar cluster, is fundamentally limited by
the cluster's intrinsic dispersion and therefore will not improve with time.
In order to detect the prograde relativistic precession at the 3$\sigma$ level
(neglecting the
retrograde precession due to the extended mass distribution), the reference
frame must be stable to within $\sim$0.02 mas yr$^{-1}$
({\em black line}).  Using the maser method,
a significant detection of the retrograde precession of S0-2's orbit will be 
possible beginning in the year $\sim$2022. 
({\em Right}) The three sources of error that contribute to the 
stability of the reference frame using the maser method are shown separately.
These include the radio masers ({\em dotted}), the 
infrared masers ({\em thick solid}), and the secondary astrometric standards
({\em dash-dotted}).  Note the different scaling for the Y axis in the two plots. 
}
\label{fig:stableFrame}
\end{figure}

Figure \ref{fig:stableFrame} shows the expected improvement
in the stability of the astrometric reference frame with time using the maser 
method described in \S\ref{sec:GC}.  The various contributions to the stability
of the reference frame come from measurements of the SiO masers in both the 
radio ({\em dotted lines}) and infrared ({\em dash-dotted lines}), 
as well as from the transformation of the infrared stars into the 
Sgr A*-radio rest frame ({\em thick solid lines}).  In order to detect either
the prograde relativistic precession or the retrograde precession from the
extended mass distribution, the combination of these various sources of error 
must be reduced to less than 0.02 mas yr$^{-1}$. Figure \ref{fig:stableFrame} shows
that this will be possible only starting in the year $\sim$2022, using the method 
described in \S\ref{sec:GC}.  High precision radio measurements of additional
masers within this region, such as IRS 14NE \citep{li10} or others that may be 
discovered with the Expanded Very Large Array (EVLA), would help to achieve the
required level of precision more rapidly, as the IR measurements are already 
in hand for this entire region.

\acknowledgements
We thank the staff of the Keck Observatory, especially Randy Campbell, Al Conrad,
Jim Lyke, and Hien Tran for their help in obtaining the observations.
We are grateful to Mark Reid for providing the latest VLA measurements of
the Galactic center.
We also thank Mark Morris, Quinn Konopacky, Marshall Perrin, and Leo Meyer 
for their helpful suggestions on this work and the manuscript.
Support for this work was provided by the NSF grant AST-0406816 and the 
NSF Science \& Technology Center for Adaptive Optics, managed by UCSC 
(AST-9876783). The W. M. Keck Observatory is operated as a scientific 
partnership among the California Institute of Technology, the University 
of California and the National Aeronautics and Space Administration.  
The Observatory was made possible by the generous financial support of 
the W. M. Keck Foundation.  The authors also recognize and acknowledge
the very significant cultural role and reverence that the summit of Mauna
Kea has always had within the indigenous Hawaiian community. We are most
fortunate to have the opportunity to conduct observations from this mountain.

{\it Facilities:} \facility{Keck: II (NIRC2)}

\begin{appendix}
\section{StarFinder}
\label{app:starfinder}

{\em Starfinder} iteratively determines first the PSF, from a set of 
user selected "PSF stars", and then the positions and fluxes of all stars 
in the field \citep{stf}. 
Successive iterations improve the PSF estimate by subtracting off stars 
identified in the previous pass. However, errors in the initial PSF estimate 
can lead to spurious source detections due to speckles or airy ring 
substructure 
that are incorrectly identified as stars. These errors propagate through all 
iterations and lead to increased astrometric noise from fitting an incorrect 
PSF and astrometric biases due to the detection of false sources.

To minimize the impact of these false sources on the PSF estimation and 
subsequent astrometry and photometry, we insert a step into each 
{\em StarFinder}
iteration that trims out these false sources from the list of identified 
stars before re-extracting the PSF on subsequent iterations. We define our 
valid star detection limits as a contrast curve of delta-magnitude vs. 
separation, which is computed by azimuthally averaging the PSF.
For every source, we remove all detections that are fainter than this
contrast curve. This typically removes 20\% of the originally detected 
sources, with roughly half coming from from substructure in the first 
airy wing and the other half coming from speckles in the extended PSF halo. 
We also increased our PSF box size from 1'', as used in \citet{ghez08} and
\citet{lu09}, to 2'' to improve photometric accuracy; however, this had a 
minimal impact on astrometry.

\section{Distortion Uncertainties for the IR Maser Mosaic}
\label{app:distErr_mosaic}
We generate a map of positional uncertainties that arise from uncertainties
in the distortion model for our IR maser mosaic to facilitate assignment of
this source of uncertainty.  This is simply a mosaic of the
distortion uncertainty models discussed in \S\ref{sec:newDist} and 
shown in Figure \ref{fig:monteHist}, where the residual distortion of
0.1 pix (\S\ref{sec:testNew}) was added in quadrature to each pixel.
To construct the mosaicked error map, we compute the distortion error 
contribution at each pixel as
\begin{equation}
\sigma_{dist} = \sqrt{\frac{\sigma^2_1 + \sigma^2_2 + ... + \sigma^2_N}{N}}
\end{equation}
where $\sigma_i$ is the distortion uncertainty at each individual pixel and 
N is the number of overlapping fields, which can vary between 1 and 4.
This resulted in a mosaicked map of approximately 2200$\times$2200 pix.

\section{Possible Astrometric Bias from IRS 7}
\label{app:irs7}
The linear motion modeling of the IR maser measurements in 
\S\ref{sec:gcRefFrame} have unexpectedly low reduced $\chi^2$.   
In our current analysis, alignment uncertainties are treated as 
purely random errors. If there is a systematic problem with one 
of the radio maser positions, this could create a significant 
correlated alignment error that is not captured in our present 
analysis and cause the reduced $\chi^2$ to be smaller than its 
expected value for random errors. Indeed, one possible culprit 
is the radio position of IRS 7, which, as discussed in 
\citet{reid03} is more uncertain than the other masers used for two
related reasons.  First, it is a supergiant and therefore is expected 
to have a much larger maser emission region (r $\sim$ 10 mas) than 
the other masers used, which are thought to be Mira variables 
(r $\sim$ 1 mas).  Second, the maser spot location for IRS 7 
jumped in 1998 (although the spots moved with similar proper 
motion before and after 1998), making the position of IRS 7 
harder to assess than its proper motion. The solution has been to 
take the mid-point between the positions before and after 1998, 
which amounts to applying a 10 mas offset to the post-1998 values 
and increasing the positional uncertainties to 5 mas. If we remove 
this offset from the reported position of IRS 7 in our analysis 
(prior to aligning the IR positions in each epoch), the resulting 
average $\chi^2$ of the linear motion models for the IR maser 
measurements is 0.97, which may suggest that this 
offset should not be applied.  Since this has only a minor impact 
on our current analysis, we have used the values reported in 
\S\ref{sec:gcRefFrame} and Table \ref{tab:maser_polyfit} for the 
results reported in this study.  However this may become a more 
important issue in the future as the precision of the IR maser 
measurements improves as discussed in \S\ref{sec:discuss}.

\end{appendix}

%
%

\clearpage
\begin{deluxetable}{lclcccccccccccc}
\rotate
\tabletypesize{\scriptsize}
\tablewidth{0pt}
\tablecaption{Summary of M92 Images}
\tablehead{
  \colhead{Date$\tablenotemark{a}$} & 
  \colhead{PA} & 
  \colhead{(X,Y)$_{GS}$$\tablenotemark{b}$} & 
  \colhead{$\Delta$(X,Y)$\tablenotemark{c}$} & 
  \colhead{El} & 
  \colhead{Temp} & 
  \colhead{Pressure} & 
  \colhead{RH$\tablenotemark{d}$} & 
  \colhead{$\Delta$R$\tablenotemark{e}$} & 
  \colhead{t$_{exp,i}\times$coadd} & 
  \colhead{$\langle$FWHM$\rangle$} & 
  \colhead{$\langle$Strehl$\rangle$} & 
  \colhead{N stars} & 
  \colhead{N stars} & 
  \colhead{$\langle\sigma_{pos}\rangle$\tablenotemark{f}} \\ 

  \colhead{(UT)} & 
  \colhead{(deg)} & 
  \colhead{(pix)} & 
  \colhead{(pix)} & 
  \colhead{(deg)} & 
  \colhead{(K)} & 
  \colhead{(mbar)} & 
  \colhead{\%} & 
  \colhead{(mas)} & 
  \colhead{(sec)} & 
  \colhead{(mas)} & 
  \colhead{} & 
  \colhead{detected} & 
  \colhead{used} & 
  \colhead{(pix)} 
}
\startdata
 2007 June 21 &    0 &   508, 512 &         0, 0 & 53 & 271.2 & 616.8 & 92 &  2.74 &       3.0$\times$10 & 49 & 0.55 & 105 & 69 & 0.037\\ 
              &      &            &    254, -252 & 53 & 271.1 & 616.7 & 89 &  2.77 &                     & 48 & 0.55 & 145 & 58 & 0.035\\ 
              &      &            &     251, 252 & 52 & 271.1 & 616.7 & 89 &  2.81 &                     & 48 & 0.57 & 112 & 46 & 0.045\\ 
              &      &            &   -251, -252 & 52 & 271.3 & 616.7 & 93 &  2.84 &                     & 48 & 0.59 & 176 & 56 & 0.034\\ 
              &      &            &    -253, 250 & 51 & 271.3 & 616.7 & 93 &  2.89 &                     & 49 & 0.54 & 110 & 57 & 0.037\\ 
              &      &            &       251, 0 & 51 & 271.2 & 616.7 & 95 &  2.93 &                     & 48 & 0.60 & 124 & 58 & 0.044\\ 
              &      &            &     -251, -1 & 50 & 271.2 & 616.7 & 95 &  2.97 &                     & 47 & 0.59 & 115 & 64 & 0.049\\ 
              &      &            &      2, -250 & 50 & 271.1 & 616.7 & 96 &  3.02 &                     & 47 & 0.58 & 124 & 47 & 0.063\\ 
              &      &            &       0, 253 & 49 & 271.0 & 616.6 & 97 &  3.07 &                     & 46 & 0.61 &  94 & 51 & 0.058\\ 
\\ 
 2007 July 29 &   90 &   457, 499 &         0, 0 & 67 & 272.9 & 615.3 & 11 &  2.07 &       0.8$\times$60 & 45 & 0.64 &  73 & 47 & 0.031\\ 
              &      &            &    255, -251 & 67 & 272.9 & 615.3 & 11 &  2.07 &                     & 45 & 0.72 &  53 & 38 & 0.048\\ 
              &      &            &     251, 255 & 67 & 272.9 & 615.3 & 12 &  2.07 &                     & 45 & 0.68 &  84 & 41 & 0.042\\ 
              &      &            &   -249, -251 & 67 & 272.9 & 615.2 & 12 &  2.07 &                     & 46 & 0.66 &  65 & 40 & 0.052\\ 
              &      &            &    -252, 251 & 67 & 272.9 & 615.2 & 13 &  2.07 &                     & 45 & 0.69 & 132 & 52 & 0.031\\ 
              &      &            &       254, 2 & 67 & 272.9 & 615.3 & 13 &  2.07 &                     & 45 & 0.71 &  72 & 44 & 0.021\\ 
              &      &            &      -252, 0 & 66 & 272.9 & 615.2 & 13 &  2.08 &                     & 45 & 0.72 &  93 & 48 & 0.037\\ 
              &      &            &      3, -250 & 66 & 272.8 & 615.2 & 12 &  2.09 &                     & 44 & 0.73 &  69 & 47 & 0.029\\ 
              &      &            &      -2, 253 & 66 & 272.8 & 615.2 & 12 &  2.10 &                     & 45 & 0.72 &  92 & 51 & 0.024\\ 
              &      &            &   -125, -124 & 65 & 272.8 & 615.2 & 12 &  2.11 &                     & 45 & 0.72 &  85 & 50 & 0.026\\ 
              &      &            &    128, -375 & 65 & 272.8 & 615.2 & 13 &  2.12 &                     & 45 & 0.73 &  89 & 48 & 0.030\\ 
              &      &            &     126, 129 & 65 & 272.8 & 615.2 & 12 &  2.14 &                     & 45 & 0.70 &  84 & 54 & 0.030\\ 
              &      &            &   -374, -376 & 64 & 272.8 & 615.2 & 13 &  2.16 &                     & 46 & 0.69 &  47 & 33 & 0.033\\ 
              &      &            &    -378, 126 & 64 & 272.8 & 615.2 & 12 &  2.18 &                     & 46 & 0.68 &  74 & 39 & 0.067\\ 
              &      &            &    128, -123 & 63 & 272.9 & 615.2 & 12 &  2.20 &                     & 46 & 0.67 &  77 & 49 & 0.038\\ 
              &      &            &   -374, -125 & 62 & 272.9 & 615.2 & 12 &  2.22 &                     & 46 & 0.67 &  72 & 38 & 0.032\\ 
              &      &            &   -121, -374 & 62 & 272.9 & 615.2 & 12 &  2.25 &                     & 46 & 0.67 &  51 & 38 & 0.028\\ 
              &      &            &    -127, 129 & 61 & 273.0 & 615.2 & 11 &  2.27 &                     & 46 & 0.66 &  87 & 45 & 0.045\\ 
              &      &            &    127, -121 & 60 & 273.1 & 615.1 & 11 &  2.31 &                     & 46 & 0.65 &  81 & 47 & 0.045\\ 
              &      &            &    380, -373 & 60 & 273.2 & 615.1 & 12 &  2.34 &                     & 47 & 0.63 &  52 & 33 & 0.071\\ 
              &      &            &     378, 132 & 59 & 273.1 & 615.1 & 12 &  2.38 &                     & 46 & 0.65 &  51 & 39 & 0.025\\ 
              &      &            &   -126, -373 & 58 & 273.3 & 615.1 & 12 &  2.41 &                     & 46 & 0.66 &  54 & 36 & 0.032\\ 
              &      &            &    -128, 130 & 57 & 273.3 & 615.1 & 11 &  2.46 &                     & 47 & 0.65 &  76 & 45 & 0.031\\ 
              &      &            &    381, -120 & 56 & 273.3 & 615.0 & 11 &  2.50 &                     & 47 & 0.64 &  43 & 35 & 0.040\\ 
              &      &            &   -127, -120 & 56 & 273.4 & 615.0 & 10 &  2.55 &                     & 47 & 0.61 &  64 & 43 & 0.022\\ 
              &      &            &    129, -371 & 55 & 273.3 & 615.0 & 12 &  2.60 &                     & 47 & 0.63 &  54 & 39 & 0.040\\ 
              &      &            &     124, 133 & 54 & 273.3 & 614.9 & 11 &  2.66 &                     & 48 & 0.60 &  79 & 47 & 0.036\\ 
\\ 
  2008 Apr 28 &  180 &   496, 477 &         0, 0 & 67 & 271.0 & 615.8 & 82 &  2.08 &       0.8$\times$60 & 47 & 0.56 &  31 & 20 & 0.022\\ 
              &      &            &    252, -252 & 67 & 271.0 & 615.7 & 81 &  2.08 &                     & 48 & 0.53 &  32 &  0 & -1.000\\ 
              &      &            &     248, 253 & 67 & 271.0 & 615.8 & 83 &  2.08 &                     & 48 & 0.55 &  55 & 15 & 0.061\\ 
              &      &            &    113, -375 & 63 & 271.5 & 616.0 & 70 &  2.23 &                     & 48 & 0.54 &  12 &  0 & -1.000\\ 
              &      &            &   -143, -120 & 59 & 271.0 & 616.0 & 77 &  2.39 &                     & 51 & 0.44 &  13 &  9 & 0.022\\ 
\\ 
  2008 June 3 &    0 &   776, 573 &         0, 0 & 42 & 273.3 & 616.0 & 64 &  3.85 &        1.5$\times$6 & 50 & 0.47 &  30 &  0 & -1.000\\ 
              &      &            &         4, 4 & 42 & 273.2 & 616.1 & 65 &  3.89 &                     & 51 & 0.48 &  32 &  0 & -1.000\\ 
              &      &            &        -4, 0 & 42 & 273.2 & 616.1 & 65 &  3.93 &                     & 58 & 0.29 &  29 &  0 & -1.000\\ 
              &      &            &         4, 0 & 41 & 273.2 & 616.1 & 65 &  3.97 &                     & 54 & 0.38 &  32 &  0 & -1.000\\ 
              &      &            &        -4, 3 & 41 & 273.2 & 616.1 & 65 &  4.01 &                     & 54 & 0.38 &  25 &  0 & -1.000\\ 
              &      &            &       -5, -4 & 41 & 273.2 & 616.1 & 65 &  4.05 &                     & 54 & 0.40 &  26 &  0 & -1.000\\ 
\\ 
 2008 July 24 &   45 &   173, 565 &         0, 0 & 50 & 271.6 & 617.2 & 39 &  3.03 &       2.8$\times$10 & 66 & 0.35 & 128 & 32 & 0.077\\ 
              &      &            &       0, -49 & 49 & 271.6 & 617.2 & 39 &  3.07 &                     & 63 & 0.35 & 110 & 35 & 0.068\\ 
              &      &            &      0, -100 & 49 & 271.6 & 617.1 & 39 &  3.11 &                     & 72 & 0.30 &  88 & 30 & 0.065\\ 
              &      &            &      1, -149 & 48 & 271.6 & 617.1 & 39 &  3.16 &                     & 72 & 0.29 &  89 & 30 & 0.089\\ 
              &      &            &      3, -199 & 48 & 271.6 & 616.9 & 39 &  3.21 &                     & 51 & 0.46 & 131 & 45 & 0.065\\ 
              &      &            &      2, -249 & 47 & 271.6 & 616.9 & 39 &  3.26 &                     & 86 & 0.22 &  89 & 21 & 0.067\\ 
\\ 
   2009 May 9 &    0 &   910, 668 &         0, 0 & 66 & 270.6 & 614.6 & 36 &  2.11 &       0.8$\times$60 & 52 & 0.45 &  18 & 10 & 0.072\\ 
              &      &            &      1, -153 & 66 & 270.6 & 614.6 & 36 &  2.12 &                     & 49 & 0.50 &  19 & 15 & 0.031\\ 
              &      &            &      3, -304 & 65 & 270.6 & 614.6 & 36 &  2.13 &                     & 49 & 0.47 &  19 & 16 & 0.026\\ 
              &      &            &     -154, -2 & 65 & 270.5 & 614.6 & 36 &  2.14 &                     & 51 & 0.48 &  25 & 17 & 0.036\\ 
              &      &            &   -153, -154 & 65 & 270.5 & 614.7 & 36 &  2.15 &                     & 47 & 0.58 &  30 & 25 & 0.036\\ 
              &      &            &   -152, -305 & 64 & 270.5 & 614.7 & 36 &  2.16 &                     & 50 & 0.47 &  27 & 22 & 0.038\\ 
              &      &            &     -305, -4 & 64 & 270.8 & 614.7 & 35 &  2.17 &                     & 50 & 0.51 &  23 & 19 & 0.034\\ 
              &      &            &   -305, -155 & 64 & 271.0 & 614.7 & 35 &  2.18 &                     & 56 & 0.38 &  26 & 12 & 0.041\\ 
              &      &            &   -302, -306 & 63 & 271.0 & 614.7 & 35 &  2.20 &                     & 63 & 0.30 &  22 & 13 & 0.033\\ 
\\ 
   2009 May 9 &   90 &   365, 411 &         0, 0 & 62 & 270.8 & 614.6 & 35 &  2.27 &       0.8$\times$60 & 49 & 0.49 &  25 & 19 & 0.025\\ 
              &      &            &      2, -153 & 61 & 270.8 & 614.6 & 35 &  2.29 &                     & 50 & 0.48 &  24 & 17 & 0.028\\ 
              &      &            &      0, -302 & 61 & 270.8 & 614.5 & 35 &  2.31 &                     & 50 & 0.47 &  10 &  8 & 0.031\\ 
              &      &            &     -151, -1 & 60 & 270.8 & 614.5 & 35 &  2.33 &                     & 50 & 0.48 &  28 & 19 & 0.061\\ 
              &      &            &   -152, -154 & 60 & 270.8 & 614.5 & 36 &  2.35 &                     & 47 & 0.56 &  29 & 22 & 0.035\\ 
              &      &            &   -149, -304 & 59 & 270.8 & 614.5 & 36 &  2.37 &                     & 46 & 0.67 &  26 & 20 & 0.020\\ 
              &      &            &     -302, -4 & 59 & 270.6 & 614.5 & 36 &  2.40 &                     & 46 & 0.61 &  35 & 24 & 0.023\\ 
              &      &            &   -304, -156 & 58 & 270.7 & 614.5 & 34 &  2.42 &                     & 46 & 0.64 &  29 & 26 & 0.025\\ 
              &      &            &   -301, -305 & 58 & 270.7 & 614.5 & 34 &  2.45 &                     & 48 & 0.53 &  18 & 15 & 0.030\\ 
\\ 
   2009 May 9 &  315 &   697, 499 &         0, 0 & 56 & 271.1 & 614.5 & 33 &  2.55 &       0.8$\times$60 & 45 & 0.67 &  38 & 27 & 0.055\\ 
              &      &            &      2, -152 & 55 & 271.1 & 614.5 & 33 &  2.58 &                     & 45 & 0.68 &  48 & 29 & 0.055\\ 
              &      &            &      0, -304 & 55 & 271.0 & 614.4 & 34 &  2.62 &                     & 45 & 0.67 &  58 & 30 & 0.060\\ 
              &      &            &      -152, 0 & 54 & 270.7 & 614.3 & 35 &  2.66 &                     & 47 & 0.56 &  35 & 22 & 0.044\\ 
              &      &            &   -148, -154 & 54 & 270.7 & 614.3 & 35 &  2.69 &                     & 56 & 0.32 &  21 & 13 & 0.085\\ 
              &      &            &   -149, -305 & 53 & 270.7 & 614.3 & 35 &  2.74 &                     & 50 & 0.45 &  43 & 18 & 0.078\\ 
              &      &            &     -300, -2 & 53 & 270.6 & 614.4 & 35 &  2.78 &                     & 50 & 0.45 &  25 & 18 & 0.044\\ 
              &      &            &   -301, -154 & 52 & 270.6 & 614.5 & 35 &  2.82 &                     & 59 & 0.29 &  21 & 13 & 0.063\\ 
\\ 
\enddata 
\label{tab:m92_obs_all}

\tablenotetext{a}{2008 June 3 data set taken in LGS-AO mode. All other data sets taken with NGS-AO.}
\tablenotetext{b}{Position of guide star in first image of a given epoch.}
\tablenotetext{c}{Positional offset of guide star in NIRC2 pixels relative to first pointing of epoch.}
\tablenotetext{d}{Relative Humidity.}
\tablenotetext{e}{Model of differential atmospheric refraction relative to center of image.}
\tablenotetext{f}{Images thrown out are given a value of -1.0 for the average positional uncertainty (see text for details).}

\end{deluxetable}
\clearpage

\clearpage
\begin{deluxetable}{lcccccccc}
\tabletypesize{\scriptsize}
\tablewidth{0pt}
\tablecaption{Summary of GC Maser Mosaic Images}
\tablehead{
  \colhead{Date} & 
  \colhead{Start Pos\tablenotemark{a}} & 
  \colhead{t$_{exp,i}\times$coadd} & 
  \colhead{N$_{exp}\tablenotemark{b}$} & 
  \colhead{K$'_{lim}\tablenotemark{c}$} & 
  \colhead{FWHM} & 
  \colhead{Strehl} & 
  \colhead{N$_{stars}$} & 
  \colhead{$\sigma_{pos}$} \\ 

  \colhead{(UT)} & 
  \colhead{(pix)} & 
  \colhead{(sec)} & 
  \colhead{} & 
  \colhead{(mag)} & 
  \colhead{(mas)} & 
  \colhead{} & 
  \colhead{} & 
  \colhead{mas} 
}
\startdata
 2005 June 30 & 851, 426 &    0.181$\times$60 &  2 & 15.6 & 62 & 0.25 & 1306 & 1.14\\ 
   2006 May 3 & 852, 426 &    0.181$\times$60 &  3 & 15.7 & 60 & 0.21 & 1372 & 1.13\\ 
  2007 Aug 12 & 852, 425 &    0.181$\times$60 &  3 & 15.7 & 58 & 0.23 & 1626 & 1.06\\ 
  2008 May 15 & 856, 427 &    0.181$\times$60 &  3 & 15.9 & 52 & 0.31 & 2017 & 1.04\\ 
 2009 June 28 & 855, 426 &    0.181$\times$60 &  9 & 16.2 & 63 & 0.20 & 2354 & 1.04\\ 
   2010 May 4 & 858, 428 &    0.181$\times$60 &  3 & 15.5 & 69 & 0.17 & 1174 & 1.08\\ 
\enddata 
\label{tab:maserObs}

\tablecomments{All images taken at PA=0$^o$.}
\tablenotetext{a}{The X,Y position of IRS16C in the first image of a given epoch.}
\tablenotetext{b}{The number of exposures per dither position.}
\tablenotetext{c}{K$'_{lim}$ is the magnitude at which the cumulative distribution function of the observed K' magnitudes reaches 90\% of the total sample size.}

\end{deluxetable}
\clearpage

\clearpage
\begin{singlespace} 
\begin{deluxetable}{lcccccccccccccr} 
\rotate
\tabletypesize{\scriptsize} 
\setlength{\tabcolsep}{1.5mm} 
\tablewidth{0pt} 
\tablecaption{Measurements of SiO Masers}
\tablehead{ 
  \colhead{Star Name} & 
  \multicolumn{3}{c}{$[$IR $-$ Radio$]$ Position} & 
  \multicolumn{2}{c}{Uncertainty in Radio} & 
  \multicolumn{2}{c}{Uncertainty in IR Centroid} & 
  \multicolumn{2}{c}{Uncertainty in Alignment} & 
  \multicolumn{2}{c}{Distortion Error \tablenotemark{a}} & 
  \multicolumn{2}{c}{$\sigma$ Offset} & 
  \colhead{DAR \tablenotemark{b}} \\ 
  \colhead{} & 
  \colhead{X} & 
  \colhead{Y} & 
  \colhead{Total} & 
  \colhead{X} & 
  \colhead{Y} & 
  \colhead{X} & 
  \colhead{Y} & 
  \colhead{X} & 
  \colhead{Y} & 
  \colhead{X} & 
  \colhead{Y} & 
  \colhead{X} & 
  \colhead{Y} & 
  \colhead{} \\ 
  \colhead{} & 
  \colhead{(mas)} & 
  \colhead{(mas)} & 
  \colhead{(mas)} & 
  \colhead{(mas)} & 
  \colhead{(mas)} & 
  \colhead{(mas)} & 
  \colhead{(mas)} & 
  \colhead{(mas)} & 
  \colhead{(mas)} & 
  \colhead{(mas)} & 
  \colhead{(mas)} & 
  \colhead{} & 
  \colhead{} & 
  \colhead{(mas)} 
} 
\startdata 
IRS 9         (Avg) &  -1.44 &   0.31 &   1.65 &   0.40 &   0.67 &   0.91 &   1.19 &   0.78 &   0.55 &   1.04 &   1.05 &  -0.82 &   0.13 &  -2.32 \\ 
$\;\;\;$2005 June       &  -1.54 &   0.90 &   1.78 &   0.30 &   0.50 &   0.36 &   0.62 &   0.40 &   0.60 &   1.04 &   1.05 &  -1.27 &   0.62 &  -2.95 \\ 
$\;\;\;$2006 May        &  -2.94 &   1.31 &   3.22 &   0.30 &   0.50 &   2.94 &   3.02 &   1.30 &   0.60 &   1.04 &   1.06 &  -0.87 &   0.40 &  -2.01 \\ 
$\;\;\;$2007 August     &  -1.00 &   0.17 &   1.01 &   0.40 &   0.60 &   0.59 &   1.01 &   0.50 &   0.30 &   1.04 &   1.05 &  -0.74 &   0.11 &  -1.77 \\ 
$\;\;\;$2008 May        &  -1.60 &  -0.43 &   1.66 &   0.40 &   0.70 &   0.48 &   0.62 &   0.80 &   0.70 &   1.05 &   1.05 &  -1.10 &  -0.27 &  -1.79 \\ 
$\;\;\;$2009 June       &  -0.99 &  -0.84 &   1.30 &   0.50 &   0.80 &   0.86 &   0.89 &   0.90 &   0.50 &   1.05 &   1.05 &  -0.58 &  -0.50 &  -2.84 \\ 
$\;\;\;$2010 May        &  -0.54 &   0.73 &   0.91 &   0.50 &   0.90 &   0.25 &   1.00 &   0.80 &   0.60 &   1.05 &   1.05 &  -0.38 &   0.40 &  -2.54 \\ 
 & & & & & & &  \\ 
IRS 7         (Avg) &   1.26 &  -6.61 &   6.75 &   5.02 &   5.03 &   0.32 &   0.40 &   0.28 &   1.32 &   1.09 &   1.07 &   0.24 &  -1.24 &   2.24 \\ 
$\;\;\;$2005 June       &   1.96 &  -5.27 &   5.62 &   5.00 &   5.00 &   0.20 &   0.68 &   0.40 &   0.80 &   1.09 &   1.07 &   0.38 &  -1.01 &   2.21 \\ 
$\;\;\;$2006 May        &   1.17 &  -6.85 &   6.95 &   5.00 &   5.00 &   0.51 &   0.31 &   0.30 &   1.50 &   1.09 &   1.07 &   0.23 &  -1.28 &   2.25 \\ 
$\;\;\;$2007 August     &   1.57 &  -6.35 &   6.54 &   5.00 &   5.00 &   0.35 &   0.19 &   0.30 &   1.10 &   1.09 &   1.07 &   0.31 &  -1.21 &   2.29 \\ 
$\;\;\;$2008 May        &   0.79 &  -7.89 &   7.93 &   5.00 &   5.00 &   0.28 &   0.27 &   0.20 &   1.50 &   1.09 &   1.07 &   0.15 &  -1.48 &   2.30 \\ 
$\;\;\;$2009 June       &   1.47 &  -7.08 &   7.23 &   5.00 &   5.10 &   0.35 &   0.38 &   0.30 &   1.60 &   1.09 &   1.07 &   0.29 &  -1.30 &   2.17 \\ 
$\;\;\;$2010 May        &   0.59 &  -6.22 &   6.25 &   5.10 &   5.10 &   0.22 &   0.56 &   0.20 &   1.40 &   1.08 &   1.07 &   0.11 &  -1.15 &   2.20 \\ 
 & & & & & & &  \\ 
IRS 12N       (Avg) &  -3.46 &  -3.46 &   5.17 &   0.47 &   0.50 &   3.35 &   5.29 &   1.32 &   0.95 &   1.03 &   1.02 &  -1.08 &  -0.87 &  -2.94 \\ 
$\;\;\;$2005 June       &  -1.81 &  -5.37 &   5.67 &   0.40 &   0.40 &   0.72 &   1.26 &   0.70 &   0.80 &   1.03 &   1.03 &  -1.21 &  -2.89 &  -2.53 \\ 
$\;\;\;$2006 May        &  -4.64 &  -1.39 &   4.84 &   0.40 &   0.40 &   3.81 &   1.88 &   2.00 &   1.10 &   1.04 &   1.04 &  -1.04 &  -0.57 &  -3.13 \\ 
$\;\;\;$2007 August     &  -1.30 &  -0.47 &   1.38 &   0.40 &   0.50 &   0.30 &   2.53 &   0.90 &   0.60 &   1.03 &   1.02 &  -0.89 &  -0.17 &  -3.35 \\ 
$\;\;\;$2008 May        &  -3.35 &  -5.81 &   6.71 &   0.50 &   0.50 &   1.82 &   7.26 &   1.40 &   1.20 &   1.02 &   1.01 &  -1.31 &  -0.78 &  -3.36 \\ 
$\;\;\;$2009 June       &  -3.99 &  -3.65 &   5.41 &   0.50 &   0.60 &  10.51 &  10.45 &   1.50 &   1.00 &   1.02 &   1.02 &  -0.37 &  -0.35 &  -2.54 \\ 
$\;\;\;$2010 May        &  -5.70 &  -4.07 &   7.00 &   0.60 &   0.60 &   2.95 &   8.36 &   1.40 &   1.00 &   1.04 &   1.00 &  -1.64 &  -0.48 &  -2.74 \\ 
 & & & & & & &  \\ 
IRS 28        (Avg) &  -2.03 &   6.53 &   7.23 &   0.72 &   0.63 &   2.11 &   5.56 &   0.60 &   0.62 &   1.05 &   1.04 &  -0.84 &   1.34 &  -1.91 \\ 
$\;\;\;$2005 June       &  -3.88 &   5.57 &   6.79 &   0.80 &   0.70 &   2.18 &   0.59 &   0.50 &   0.60 &   1.05 &   1.04 &  -1.49 &   3.69 &  -3.09 \\ 
$\;\;\;$2006 May        &  -0.66 &   2.32 &   2.41 &   0.70 &   0.60 &   1.79 &   6.35 &   0.90 &   0.90 &   1.06 &   1.01 &  -0.28 &   0.36 &  -1.33 \\ 
$\;\;\;$2007 August     &  -1.04 &  -0.54 &   1.17 &   0.60 &   0.50 &   2.49 &   1.77 &   0.30 &   0.40 &   1.07 &   1.03 &  -0.37 &  -0.25 &  -0.84 \\ 
$\;\;\;$2008 May        &  -2.22 &   5.73 &   6.15 &   0.60 &   0.50 &   1.17 &   2.10 &   0.60 &   0.70 &   1.06 &   1.03 &  -1.24 &   2.30 &  -0.87 \\ 
$\;\;\;$2009 June       &  -1.71 &   1.99 &   2.62 &   0.70 &   0.70 &   0.61 &   2.34 &   0.60 &   0.50 &   1.05 &   1.04 &  -1.12 &   0.74 &  -2.93 \\ 
$\;\;\;$2010 May        &  -2.66 &  24.09 &  24.24 &   0.90 &   0.80 &   4.45 &  20.22 &   0.70 &   0.60 &   1.04 &   1.10 &  -0.57 &   1.19 &  -2.37 \\ 
 & & & & & & &  \\ 
IRS 10EE      (Avg) &   0.71 &   1.75 &   2.00 &   0.32 &   0.38 &   0.39 &   0.45 &   0.40 &   0.73 &   1.11 &   1.03 &   0.57 &   1.24 &   2.03 \\ 
$\;\;\;$2005 June       &   0.75 &   1.00 &   1.25 &   0.20 &   0.30 &   0.13 &   0.90 &   0.50 &   0.50 &   1.11 &   1.03 &   0.61 &   0.67 &   1.12 \\ 
$\;\;\;$2006 May        &  -0.17 &   2.63 &   2.64 &   0.30 &   0.30 &   0.99 &   0.29 &   0.30 &   0.70 &   1.11 &   1.04 &  -0.11 &   2.00 &   2.46 \\ 
$\;\;\;$2007 August     &   0.12 &   1.94 &   1.94 &   0.30 &   0.40 &   0.43 &   0.32 &   0.40 &   0.70 &   1.10 &   1.04 &   0.09 &   1.44 &   2.89 \\ 
$\;\;\;$2008 May        &   1.19 &   1.01 &   1.56 &   0.30 &   0.40 &   0.31 &   0.14 &   0.50 &   0.80 &   1.10 &   1.04 &   0.93 &   0.73 &   2.88 \\ 
$\;\;\;$2009 June       &   1.13 &   1.72 &   2.06 &   0.40 &   0.40 &   0.33 &   0.44 &   0.40 &   0.80 &   1.11 &   1.04 &   0.88 &   1.20 &   1.20 \\ 
$\;\;\;$2010 May        &   1.27 &   2.18 &   2.52 &   0.40 &   0.50 &   0.19 &   0.60 &   0.30 &   0.90 &   1.11 &   1.03 &   1.03 &   1.38 &   1.64 \\ 
 & & & & & & &  \\ 
IRS 15NE      (Avg) &   1.93 &   4.55 &   5.06 &   0.40 &   0.50 &   0.50 &   0.68 &   0.52 &   1.58 &   1.12 &   1.11 &   1.40 &   2.10 &   4.62 \\ 
$\;\;\;$2005 June       &   2.89 &   3.25 &   4.35 &   0.30 &   0.40 &   0.41 &   0.34 &   0.60 &   1.00 &   1.11 &   1.11 &   2.12 &   2.05 &   4.42 \\ 
$\;\;\;$2006 May        &   2.41 &   2.69 &   3.61 &   0.30 &   0.40 &   0.09 &   0.62 &   0.60 &   1.80 &   1.12 &   1.11 &   1.84 &   1.20 &   4.70 \\ 
$\;\;\;$2007 August     &   1.95 &   4.11 &   4.55 &   0.40 &   0.50 &   0.49 &   0.73 &   0.50 &   1.30 &   1.11 &   1.12 &   1.42 &   2.13 &   4.85 \\ 
$\;\;\;$2008 May        &   1.80 &   5.32 &   5.62 &   0.40 &   0.50 &   0.10 &   0.23 &   0.50 &   1.80 &   1.11 &   1.11 &   1.40 &   2.43 &   4.86 \\ 
$\;\;\;$2009 June       &   1.63 &   5.66 &   5.89 &   0.50 &   0.60 &   0.61 &   0.72 &   0.50 &   1.90 &   1.12 &   1.11 &   1.12 &   2.37 &   4.37 \\ 
$\;\;\;$2010 May        &   0.91 &   6.27 &   6.34 &   0.50 &   0.60 &   1.28 &   1.44 &   0.40 &   1.70 &   1.12 &   1.11 &   0.50 &   2.45 &   4.48 \\ 
 & & & & & & &  \\ 
IRS 17        (Avg) &  -0.85 &   0.34 &   1.54 &   4.52 &   2.35 &   0.51 &   0.46 &   0.83 &   0.53 &   1.12 &   1.04 &  -0.17 &   0.21 &   2.82 \\ 
$\;\;\;$2005 June       &  -0.12 &   0.97 &   0.98 &   3.20 &   1.60 &   0.39 &   0.05 &   0.80 &   0.40 &   1.13 &   1.04 &  -0.03 &   0.50 &   1.27 \\ 
$\;\;\;$2006 May        &  -0.48 &   1.57 &   1.64 &   3.50 &   1.80 &   1.00 &   0.45 &   0.80 &   0.60 &   1.12 &   1.04 &  -0.12 &   0.71 &   3.55 \\ 
$\;\;\;$2007 August     &  -1.55 &   1.37 &   2.07 &   4.10 &   2.10 &   0.70 &   0.58 &   0.70 &   0.50 &   1.13 &   1.05 &  -0.36 &   0.56 &   4.28 \\ 
$\;\;\;$2008 May        &  -0.35 &   0.00 &   0.35 &   4.60 &   2.40 &   0.18 &   0.28 &   1.00 &   0.50 &   1.12 &   1.04 &  -0.07 &   0.00 &   4.26 \\ 
$\;\;\;$2009 June       &  -1.77 &   0.35 &   1.80 &   5.50 &   2.90 &   0.16 &   0.38 &   0.90 &   0.50 &   1.12 &   1.04 &  -0.31 &   0.11 &   1.41 \\ 
$\;\;\;$2010 May        &  -0.86 &  -2.23 &   2.39 &   6.20 &   3.30 &   0.63 &   1.01 &   0.80 &   0.70 &   1.12 &   1.04 &  -0.13 &  -0.61 &   2.16 \\ 
 & & & & & & &  \\ 
\enddata 
\tablenotetext{a}{Distortion error includes the residual distortion term described in the text.}
\tablenotetext{b}{Differential Atmospheric Refraction relative to the center of the image}
\label{tab:error_breakdown}
\end{deluxetable} 
\end{singlespace}

\begin{deluxetable}{lcc}
\tabletypesize{\scriptsize}
\tablewidth{0pt}
\tablecaption{NIRC2 Plate Scale and Orientation}
\tablehead{
  \colhead{Method} &
  \colhead{Plate Scale} &
  \colhead{Orientation} \\

  \colhead{} &
  \colhead{(mas pix$^{-1}$)} &
  \colhead{(deg)} 
}

\startdata
Calibrated w.r.t. ACS observations of M92\tablenotemark{a}        & 9.950 $\pm$ 0.003 $\pm$ 0.001 & 0.254 $\pm$ 0.014 $\pm$ 0.002 \\
$\;\;\;\;\;\;\;\;\;\;\;\;$2007 June              & 9.948 $\pm$ 0.001 $\pm$ 0.001 & 0.249 $\pm$ 0.006 $\pm$ 0.002 \\
$\;\;\;\;\;\;\;\;\;\;\;\;$2007 July              & 9.948 $\pm$ 0.001 $\pm$ 0.001 & 0.256 $\pm$ 0.006 $\pm$ 0.002 \\
$\;\;\;\;\;\;\;\;\;\;\;\;$2008 April             & 9.946 $\pm$ 0.007 $\pm$ 0.001 & 0.276 $\pm$ 0.030 $\pm$ 0.002 \\
$\;\;\;\;\;\;\;\;\;\;\;\;$2008 June              & 9.952 $\pm$ 0.003 $\pm$ 0.001 & 0.270 $\pm$ 0.009 $\pm$ 0.002 \\
$\;\;\;\;\;\;\;\;\;\;\;\;$2008 July              & 9.951 $\pm$ 0.002 $\pm$ 0.001 & 0.248 $\pm$ 0.004 $\pm$ 0.002 \\
$\;\;\;\;\;\;\;\;\;\;\;\;$2009 May               & 9.949 $\pm$ 0.002 $\pm$ 0.001 & 0.282 $\pm$ 0.013 $\pm$ 0.002 \\
\\
Calibrated w.r.t. VLA observations of GC Masers\tablenotemark{a}  & 9.953 $\pm$ 0.002 & 0.249 $\pm$ 0.012 \\
$\;\;\;\;\;\;\;\;\;\;\;\;$2005 June              & 9.952 $\pm$ 0.001 & 0.237 $\pm$ 0.003 \\
$\;\;\;\;\;\;\;\;\;\;\;\;$2006 May               & 9.955 $\pm$ 0.001 & 0.248 $\pm$ 0.007 \\
$\;\;\;\;\;\;\;\;\;\;\;\;$2007 August            & 9.950 $\pm$ 0.001 & 0.253 $\pm$ 0.003 \\
$\;\;\;\;\;\;\;\;\;\;\;\;$2008 May               & 9.954 $\pm$ 0.001 & 0.271 $\pm$ 0.005 \\
$\;\;\;\;\;\;\;\;\;\;\;\;$2009 June              & 9.952 $\pm$ 0.001 & 0.253 $\pm$ 0.005 \\
$\;\;\;\;\;\;\;\;\;\;\;\;$2010 May               & 9.955 $\pm$ 0.001 & 0.253 $\pm$ 0.005 \\
\\
Final Value\tablenotemark{b} & 9.952$\pm$0.002 & 0.252$\pm$0.009 \\
\enddata
\tablecomments{Statistical (first) and absolute (second) uncertainties are shown for the ACS observations (see \S\ref{sec:m92}).}
\tablenotetext{a}{Weighted averages are taken for the final values for each method. We use the more conservative RMS errors as the uncertainties on these values.}
\tablenotetext{b}{Average of each method's weighted average (see Note a).}
\label{tab:plateScaleAngleCmpr}

\end{deluxetable}

\begin{deluxetable}{lrccccccc}
\rotate
\tabletypesize{\scriptsize}
\tablewidth{0pt}
\tablecaption{Astrometry of SiO Masers}

\tablehead{
  \colhead{Maser} & 
  \colhead{$K$} & 
  \colhead{$\tilde{\chi}^{2}$\tablenotemark{a}} & 
  \colhead{IR $T_0$} & 
  \colhead{IR + Radio $T_0$\tablenotemark{b}} & 
  \colhead{[IR - Radio] X Position\tablenotemark{c}} & 
  \colhead{[IR - Radio] Y Position\tablenotemark{c}} & 
  \colhead{[IR - Radio] X Velocity} & 
  \colhead{[IR - Radio] Y Velocity} \\ 
  \colhead{} & 
  \colhead{(mag)} & 
  \colhead{} & 
  \colhead{(year)} & 
  \colhead{(year)} & 
  \colhead{(mas)} & 
  \colhead{(mas)} & 
  \colhead{(mas yr$^{-1}$)} & 
  \colhead{(mas yr$^{-1}$)} 
}
\startdata
           IRS 9  &  9.063  & 0.29  & 2007.7  & 2007.2  & -1.32 $\pm$  1.11 $\pm$  0.34  &  0.35 $\pm$  1.17 $\pm$  0.57  &  0.18 $\pm$ 0.18 $\pm$ 0.08  & -0.23 $\pm$ 0.26 $\pm$ 0.14 \\
           IRS 7  &  7.658  & 0.68  & 2007.7  & 2007.2  &  1.46 $\pm$  1.11 $\pm$  5.00  & -6.24 $\pm$  1.22 $\pm$  5.00  & -0.25 $\pm$ 0.09 $\pm$ 0.22  & -0.28 $\pm$ 0.31 $\pm$ 0.30 \\
         IRS 12N  &  9.538  & 0.60  & 2006.4  & 2006.4  & -1.88 $\pm$  1.26 $\pm$  0.42  & -2.74 $\pm$  1.59 $\pm$  0.47  & -0.28 $\pm$ 0.49 $\pm$ 0.05  &  1.24 $\pm$ 1.03 $\pm$ 0.06 \\
          IRS 28  &  9.328  & 1.19  & 2007.6  & 2007.6  & -2.12 $\pm$  1.34 $\pm$  0.62  &  3.93 $\pm$  1.41 $\pm$  0.52  &  0.16 $\pm$ 0.47 $\pm$ 0.22  & -0.95 $\pm$ 0.49 $\pm$ 0.24 \\
        IRS 10EE  & 11.270  & 0.64  & 2008.1  & 2007.7  &  0.62 $\pm$  1.14 $\pm$  0.29  &  1.75 $\pm$  1.13 $\pm$  0.32  &  0.17 $\pm$ 0.12 $\pm$ 0.04  & -0.01 $\pm$ 0.24 $\pm$ 0.05 \\
        IRS 15NE  & 10.198  & 0.06  & 2007.4  & 2007.1  &  2.31 $\pm$  1.16 $\pm$  0.36  &  3.90 $\pm$  1.31 $\pm$  0.47  & -0.33 $\pm$ 0.20 $\pm$ 0.05  &  0.68 $\pm$ 0.39 $\pm$ 0.06 \\
          IRS 17  &  8.910  & 0.90  & 2007.7  & 2005.1  & -0.68 $\pm$  1.22 $\pm$  3.72  &  0.80 $\pm$  1.08 $\pm$  1.91  & -0.23 $\pm$ 0.24 $\pm$ 0.99  & -0.33 $\pm$ 0.15 $\pm$ 0.54 \\
\\
\tableline
\\
Weighted Average\tablenotemark{d}  &         & 0.62  &         & 2006.9  & -0.31 $\pm$  0.55  &  1.44 $\pm$  0.59  &  0.02 $\pm$ 0.09  & -0.06 $\pm$ 0.14 \\
\enddata
\tablecomments{Infrared (first) and radio (second) formal uncertainties are reported for each maser's position and velocity. Average distortion errors ($\sigma\sim$1 mas) for each maser are added in quadrature to the infrared formal uncertainties. X and Y increase to the east and north, respectively.}
\tablenotetext{a}{$\tilde{\chi}^2$ is the average of the X and Y $\chi^2$ per degree of freedom.}
\tablenotetext{b}{Average $T_0$ from both IR and radio measurements weighted by velocity errors (see Equation \ref{eq:t0}).
}\tablenotetext{c}{Positional offsets computed for the common epoch of 2006.9.}
\tablenotetext{d}{Weighted average and error in the weighted average are reported for all columns except the $\tilde{\chi}^2$ and $T_0$ columns, where we report the average.}
\label{tab:maser_polyfit}
\end{deluxetable}

\begin{deluxetable}{lrccc}
\rotate
\tabletypesize{\scriptsize}
\tablewidth{0pt}
\tablecaption{Astrometry of SiO Masers (Polar Coordinates)}

\tablehead{
  \colhead{Maser} & 
  \colhead{$R$\tablenotemark{a}} & 
  \colhead{[IR - Radio] Rad Velocity} & 
  \colhead{[IR - Radio] Tan Velocity} & 
  \colhead{[IR - Radio] Ang Velocity} \\ 
  \colhead{} & 
  \colhead{(arcsec)} & 
  \colhead{(mas yr$^{-1}$)} & 
  \colhead{(mas yr$^{-1}$)} & 
  \colhead{(mas yr$^{-1}$ arcsec$^{-1}$)} 
}
\startdata
           IRS 9  &  8.503  &  0.29 $\pm$ 0.23 $\pm$ 0.12  &  0.02 $\pm$ 0.22 $\pm$ 0.11  & 0.00 $\pm$ 0.22 $\pm$ 0.11 \\
           IRS 7  &  5.524  & -0.28 $\pm$ 0.31 $\pm$ 0.30  & -0.25 $\pm$ 0.09 $\pm$ 0.22  &-0.04 $\pm$ 0.09 $\pm$ 0.22 \\
         IRS 12N  &  7.644  & -1.00 $\pm$ 0.96 $\pm$ 0.06  &  0.78 $\pm$ 0.62 $\pm$ 0.05  & 0.10 $\pm$ 0.62 $\pm$ 0.05 \\
          IRS 28  & 11.989  &  0.60 $\pm$ 0.47 $\pm$ 0.22  &  0.75 $\pm$ 0.48 $\pm$ 0.24  & 0.06 $\pm$ 0.48 $\pm$ 0.24 \\
        IRS 10EE  &  8.757  &  0.15 $\pm$ 0.16 $\pm$ 0.04  &  0.09 $\pm$ 0.21 $\pm$ 0.05  & 0.01 $\pm$ 0.21 $\pm$ 0.05 \\
        IRS 15NE  & 11.340  &  0.64 $\pm$ 0.39 $\pm$ 0.06  & -0.40 $\pm$ 0.20 $\pm$ 0.05  &-0.04 $\pm$ 0.20 $\pm$ 0.05 \\
          IRS 17  & 14.271  & -0.34 $\pm$ 0.23 $\pm$ 0.94  &  0.21 $\pm$ 0.17 $\pm$ 0.63  & 0.01 $\pm$ 0.17 $\pm$ 0.63 \\
\\
\tableline
\\
Weighted Average\tablenotemark{b}  &         &   0.19 $\pm$  0.12  & -0.07 $\pm$  0.11  & -0.01 $\pm$ 0.11 \\
\enddata
\tablecomments{Infrared (first) and radio (second) formal uncertainties are reported for each maser's position and velocity.}\tablenotetext{a}{Position reported for the common IR + radio epoch given in Table \ref{tab:maser_polyfit} for each maser, and using proper motions measured in the infrared.
}\tablenotetext{b}{Weighted average and error in the weighted average are reported.
}\label{tab:maser_polyfit_polar}
\end{deluxetable}

\begin{deluxetable}{lcccccccccc}
\tabletypesize{\scriptsize}
\tablewidth{0pt}
\tablecaption{Galactic Center Secondary IR Astrometric Standards}
\tablehead{
  \colhead{Name} & 
  \colhead{K'} & 
  \colhead{$T_{0,IR}$} & 
  \colhead{Radius} & 
  \colhead{$\Delta$ R.A.} & 
  \colhead{$\sigma_{R.A.}$\tablenotemark{a}} & 
  \colhead{$\Delta$ Dec.} & 
  \colhead{$\sigma_{Dec}$\tablenotemark{a}} & 
  \colhead{v$_{RA}$\tablenotemark{b}} & 
  \colhead{v$_{Dec}$\tablenotemark{b}} \\ 
  \colhead{} & 
  \colhead{(mag)} & 
  \colhead{(year)} & 
  \colhead{(arcsec)} & 
  \colhead{(arcsec)} & 
  \colhead{(mas)} & 
  \colhead{(arcsec)} & 
  \colhead{(mas)} & 
  \colhead{(mas yr$^{-1}$)} & 
  \colhead{(mas yr$^{-1}$)} 
}
\startdata
      S0-3  & 14.8  & 2008.67  & 0.36  &   0.3351  & 1.4  &   0.1189  & 1.4  &      9.1 $\pm$      0.3     &     -0.9 $\pm$      0.5\\ 
      S0-6  & 14.2  & 2008.43  & 0.36  &   0.0276  & 1.1  &  -0.3625  & 1.2  &     -5.3 $\pm$      0.1     &      3.5 $\pm$      0.4\\ 
     S0-53  & 15.5  & 2005.89  & 0.40  &   0.3484  & 1.6  &   0.2037  & 1.4  &     -8.1 $\pm$      1.1     &      7.5 $\pm$      0.5\\ 
\enddata 
\tablecomments{Table \ref{tab:secondary_short} is published in its entirety in the electronic version of this paper.}
\tablenotetext{a}{Positional errors include centroiding, alignment, and residual distortion (1 mas) errors, but do not include error in position of Sgr A* (0.55 mas, 0.59 mas in RA and Dec, respectively).}
\tablenotetext{b}{Velocity errors do not include error in velocity of Sgr A* (0.09 mas yr$^{-1}$, 0.14 mas yr$^{-1}$ in RA and Dec, respectively).}
\label{tab:secondary_short}

\end{deluxetable}
\clearpage

\clearpage

\clearpage

\clearpage

\clearpage

\clearpage

\clearpage

\clearpage

\clearpage

\clearpage

\end{document}